\documentclass[num-refs]{wiley-article}   % Enable this line and disable the
\usepackage{amssymb,amsmath,psfrag, units, booktabs,algorithm,algorithmic,url,amsfonts}
\usepackage{graphicx}      % include this line if your document contains figures
\usepackage{natbib,auto-pst-pdf}        
\usepackage{psfrag}% required for bibliography
\usepackage{multirow,stmaryrd,babel}
\newtheorem{assumption}{Assumption}
\usepackage{subfig}
\newtheorem{Theorem}{Theorem}
\def\ve#1{\mathchoice{\mbox{$\displaystyle#1$}}
{\mbox{$\textstyle#1$}}
{\mbox{$\scriptstyle#1$}}
{\mbox{$\scriptscriptstyle#1$}}}
\newcommand{\bluec}[1]{{\leavevmode#1}}
\newcommand{\greenc}[1]{{\leavevmode#1}}
\newcommand{\isep}{\mathrel{,\,}\nobreak}
\newcommand{\ind}[2]{\llbracket #1\isep #2 \rrbracket}
\papertype{Original Article}

\title{Tube-enhanced Multi-stage MPC for Flexible Robust Control of Constrained Linear Systems with Additive and Parametric Uncertainties} % Title, preferably not more
\author[1]{Sankaranarayanan Subramanian}
\author[2]{Sergio Lucia}
\author[3]{Radoslav Paulen}
\author[1]{Sebastian Engell}

\affil[1]{Process Dynamics and Operations Group, TU Dortmund, Germany.}
\affil[2]{Laboratory of Process Automation Systems, TU Dortmund, Germany.}
\affil[3]{Faculty of Chemical and Food Technology, Slovak University of Technology in Bratislava, Slovakia.}

\corraddress{Process Dynamics and Operations Group, TU Dortmund, Germany}
\corremail{sankaranarayanan.subramanian@tu-dortmund.de}

\fundinginfo{The research leading to these results has received funding from the European Commission under grant agreement number 291458 (MOBOCON). RP acknowledges the contribution of the Slovak Research and Development Agency under the project APVV 15-0007.}

\begin{document}
	\maketitle
\begin{abstract}                          % Abstract of not more than 200 words.
The trade-off between optimality and complexity has been one of the most important challenges in the field of robust Model Predictive Control (MPC). To address the challenge, we propose a flexible robust MPC scheme by synergizing the multi-stage and tube-based MPC approaches. The key idea is to exploit the non-conservatism of the multi-stage MPC and the simplicity of the tube-based MPC. The proposed scheme provides two options for the user to determine the trade-off depending on the application: the choice of the robust horizon and the classification of the uncertainties. Beyond the robust horizon, the branching of the scenario-tree employed in multi-stage MPC is avoided with the help of tubes. The growth of the problem size with respect to the number of uncertainties is  reduced by handling \emph{small} uncertainties via an invariant tube that can be computed offline. This results in linear growth of the problem size beyond the robust horizon and no growth of the problem size concerning small magnitude uncertainties.  The proposed approach helps to achieve a desired trade-off between optimality and complexity compared to existing robust MPC approaches. We show that the proposed approach is robustly asymptotically stable. Its advantages are demonstrated for a CSTR example. 
\end{abstract}

\section{Introduction}
\bluec{Robust Model Predictive Control (MPC) schemes address the presence of  uncertainties in the model with the goal to achieve constraint satisfaction and closed-loop stability}. It is desirable that the robust MPC schemes are computationally cheap and non-conservative. The dual goal of non-conservatism and low complexity is a key challenge that is being actively researched in the field of robust MPC, and often a trade-off is needed. Open-loop min-max MPC was one of the earliest robust MPC schemes proposed~\cite{campo1987}. In this approach, the worst-case cost is minimized while satisfying the constraints for all realizations of the uncertainty. The scheme, however, does not account for the presence of feedback in the predictions and predicts a single control input at every stage. Because of the lack of recourse, the robustness comes at the cost of a significant loss of performance.

Feedback min-max MPC models the presence of feedback information explicitly in the predictions and thus reduces the conservatism of  the open-loop schemes~\cite{lee1997}. A general feedback min-max MPC optimizes the worst-case value of the cost function over a sequence of control policies, leading to infinite-dimensional optimization problems. One possibility to formulate a feedback MPC method with a finite-dimensional optimization problem is to consider a tree-structure to represent the evolution of the uncertainty~\cite{scokaert1998} because, for each predicted state at every stage, the possibility is considered to adapt the inputs in the predictions. The tree structure grows exponentially with respect to the length of the prediction horizon, making the approach inapplicable in practice for long prediction horizons.

Other related robust MPC approaches optimize the expected value of the cost function \cite{bernardini2009} or a weighted sum of all the predicted scenarios, as done in multi-stage MPC~\cite{lucia2013}. The weights of the multi-stage MPC are tuning parameters that provide additional degrees of freedom to improve the closed-loop performance compared to a feedback min-max MPC scheme.

An alternative to the representation of feedback via a scenario tree consists of restricting the optimization to control policies with a fixed structure, linear policies~\cite{KOTHARE19961361},  or affine policies~\cite{lee2000robust, CHISCI20011019, lofberg2003,LANGSON2004125}. Tube-based MPC is one of the most discussed robust MPC approaches in the literature that usually considers an affine parameterization of the feedback policies~\cite{LANGSON2004125, mayne2005}. It was shown in~\cite{mayne2005} that the problem size can be kept the same as that of nominal MPC if the feedback gain is chosen offline and kept constant in the predictions. However, this comes at  the cost of performance loss. Tube-based MPC approaches that relax the structure of the control policy or that predict the tube online (as opposed to an invariant tube) can improve the performance  as shown in~\cite{rakovic2012parameterized,rakovic2012homothetic,rakovic2016elastic,VILLANUEVA2017311}. The performance advantages come at the cost of an increase in computational complexity with respect to the length of the prediction horizon. To handle parametric uncertainties, tube-based MPC based on Farkas' Lemma was proposed. The complexity of the approach grows linearly with respect to the length of the prediction horizon~\cite{Fleming2015,Munoz2015}.  Advanced tube-based schemes such as \cite{rakovic2012homothetic, rakovic2016elastic,Fleming2015, Munoz2015}  use contractive sets  for the prediction of tubes online. The number of inequalities and  the number of vertices that characterize the tube can increase rapidly with respect to the dimension of the states and this makes the approach difficult to implement for high dimensional systems. If low complexity tubes are employed as proposed in \cite{Lee2000minimal,Blanco2010minimal}, the schemes can be highly conservative.

The aim of this paper is to propose a novel scheme to achieve the dual goal of low computational cost and low conservatism. To achieve this goal, we combined the multi-stage and the tube-based MPC approaches by the classification of uncertainties in~\cite{subramanian2018nmpc}. The multi-stage MPC is employed to handle \emph{significant} uncertainties, and the tube-based MPC is used to handle \emph{small} magnitude disturbances.  We extend the scheme proposed in~\cite{subramanian2018nmpc} in this contribution in such a way that the rapid increase in problem complexity in multi-stage MPC is addressed both with respect to the number of uncertainties and with respect to the length of the prediction horizon. The proposed scheme gives the user two options that determine the trade-off depending on the requirements of an application. The two options are the choice of the robust horizon in multi-stage MPC and the classification of the uncertainties. The key aspects of the proposed approach are  as follows:

\begin{enumerate}
\item The branching of the scenario tree is stopped beyond a certain prediction step called robust horizon. An affine feedback policy is employed beyond the robust horizon with the help of tubes to achieve robust constraint satisfaction and recursive feasibility guarantees. 
\item Different scenarios are predicted in the proposed framework for any choice of robust horizon greater than or equal to $1$. The resulting increase in the number of degrees of freedom enables the employment of low complexity tubes for high dimensional systems without a significant loss of performance when compared to standard tube-based schemes.
\item In addition, the growth of the problem size with respect to the number of uncertainties is reduced by formulating an invariant tube for small disturbances by making use of the ideas proposed in~\cite{subramanian2018nmpc}.
\end{enumerate}

We investigate in detail the theoretical properties of the proposed approach and demonstrate that the proposed approach is robustly asymptotically stable. We present the resulting tube-enhanced multi-stage MPC scheme as a convex optimization problem that is solved at every time step. This is achieved by employing the tube-based formulations  from~\cite{rakovic2012homothetic,Fleming2015} and the multi-stage formulation from \cite{subramanian2018nmpc}. The advantages of the scheme are demonstrated for a CSTR example.

\section{Preliminaries}
We study discrete-time linear dynamical systems of the form: 
\begin{align}
\ve x^+&=\ve{A} \ve x+\ve{B} \ve u+\ve w,\label{lsys}
\end{align}
where $\ve x \in \mathbb{R}^{n_x}$ represents the state, $\ve u  \in \mathbb{R}^{n_u}$ represents the input, $\ve w \in \ve{\mathsf W} \subset \mathbb{R}^{n_x}$ denotes additive disturbances, the matrix $\ve{A}\in\mathbb{R}^{n_x\times n_x}$ represents the uncertain system matrix and $\ve{B}\in\mathbb{R}^{n_x\times n_u}$ denotes the uncertain input matrix of the controlled system.  The system matrix $\ve A$ and the input matrix $\ve B$ are contained in a convex polytope and can be represented as $(\ve A,\ve B)\in\mathrm{conv}(\{(\ve A_i,\,\ve B_i),\,\forall i\in\Gamma_p\})$, where $\mathrm{conv}()$ denotes the convex-hull operator and $\Gamma_p:=\{1,\,\dots,\,n_p\}$. We assume that there exists a feedback gain $\ve K$ that is stabilizing for all  $(\ve A,\ve B)\in\mathrm{conv}(\{(\ve A_i,\,\ve B_i),\,\forall i\in\Gamma_p\})$. $\ve{\mathsf W}$ is assumed to be  a convex polytope with the origin in its interior and is characterized by $n_w$ vertices. The bounds of the additive disturbances can be defined in terms of vertices of the set as $\mathsf{W}:=\{ w|w\in\mathrm{conv}(\{w_l,\,\forall l \in\Gamma_w\})\}$, where $\Gamma_w=\{1,\,\dots,\,n_w\}$. The following definitions of invariant sets adapted from~\cite{rawlings2009b,blanchini2008set} will be used throughout the paper.
% \times <sergio.lucia@tu-berlin.de> 2018-11-06T19:03:32.578Z:
% 
% > The extreme realizations (vertices) of the additive disturbances are defined as $\mathsf{W}:=
%  SL: here you are not defining the the vertices but the set W, isn't it?
% 
% ^.
\begin{definition}
A set $\ve{\mathsf S}$ is said to be robust positively invariant (RPI) for the system $\ve x^+=(\ve A_i+\ve B_i \ve K)\ve x+\ve w$, if $\forall \ve x\in\ve{\mathsf S}$, $\ve x^+\in\ve{\mathsf S},\,\forall\ve w\in\ve{\mathsf W},\forall \ve i \in\Gamma_p$. 
\end{definition}
\begin{definition}
A set $\ve{\mathsf S}_{\mathrm{min}}$ is said to be the minimal robust positively invariant set (mRPI) if $\ve{\mathsf S}_{\mathrm{min}}$ is contained in every closed robust positively invariant set.
\end{definition}
\begin{definition}
A set $\ve{\mathsf S}_{\mathrm{max}}$ is said to be the maximal robust positively invariant set (MRPI) if $\ve{\mathsf S}_{\mathrm{max}}$ contains every closed robust positively invariant set.
\end{definition}

\subsection{Multi-stage MPC}
\begin{figure}
\begin{center}
\includegraphics[width=0.84\columnwidth]{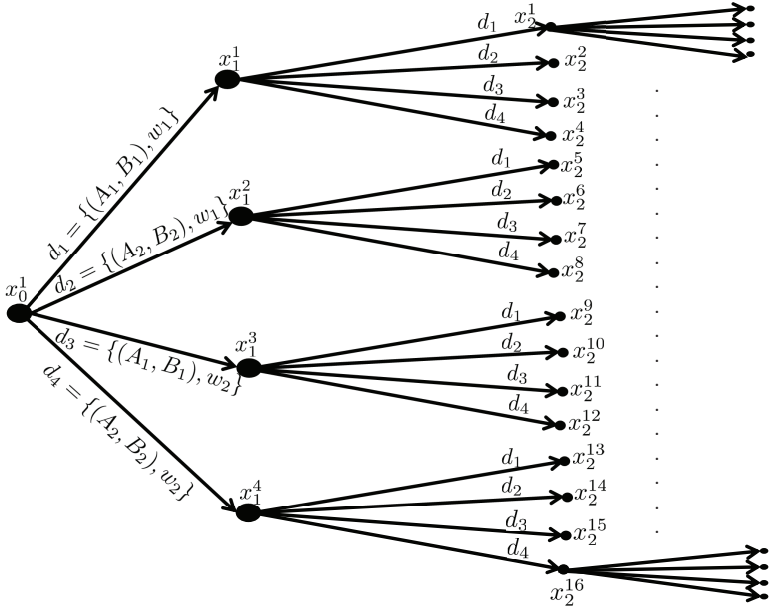}
\caption{Scenario tree representation of the uncertainty evolution for multi-stage MPC for the system with additive and parametric uncertainties.  The vertex matrices are given by $\{(A_1,B_1),(A_2,B_2)\}$ and the vertices of the additive disturbances are given by $w_1$ and $w_2$.}  % width is 8.4 cm.
%SL: Where are the supporting tubes that you mention in the caption?
%SS: Ok, deleted
\label{fig_scenario_tree}                                 % Size the figures
\end{center}                                 % accordingly.
\end{figure}

The robustness of  multi-stage MPC is achieved by modeling the future evolution of the system by a scenario tree as shown in Fig.~\ref{fig_scenario_tree}.
Each branch of the tree denotes a realization of the uncertainties. Each node denotes a predicted state at the corresponding point in time. If all the extreme values of the uncertainties are realized in the predictions, the predicted states form the convex hull of all the possible trajectories in the future until the end of the prediction horizon. \bluec{Realizations of the uncertainties that are not extreme  can also be included to improve the resulting closed-loop performance}. The tree branches until the end of the prediction horizon for each predicted node. The availability of feedback information in the predictions is explicitly modeled in the tree structure without restricting the structure of the feedback policy. This makes the approach less conservative than those which assume linear or affine feedback policies, but exponentially complex \cite{hadjiyiannis2011}. The optimization problem that is  solved at every time step is given as follows:
\begin{subequations} \label{mpc_general0}
\begin{align}
% \min_{\stackrel{\ve y_k^j,\ve x_k^j,\ve u_k^j}{\forall (j,k)\in \ve I}}
% \min_{\stackrel{\ve y_{k+1}^j,\ve x_{k+1}^j,\ve u_k^j}{\forall (j,k)\in \ve I}}
% \;\;\;\tilde{\mathcal J}(\ve y_{k+1}^j,\ve u_k^j)
\min_{\ve u_k^j, \forall (j,k)\in \ve I_{\ind{0}{N_p-1}}}
\;\;\; \sum_{k=0}^{N_p-1}\sum_{j=1}^{{n_d}^k}\omega_k^j
\ell(\ve x_{k}^j,\ve u_k^j)+\sum_{j=1}^{{n_d}^{N_p}}V_f (\ve x_{N_p}^j)
\end{align}
subject to:
\begin{align}
 &\ve x_{k+1}^c=\ve A_i\ve x_k^j+\ve B_i \ve u_k^j+w_l, &&\forall\, (j, k) \in \ve I_{\ind{0}{N_p-1}},\,\forall i\in\Gamma_p,\, \forall l\in\Gamma_w,\label{eq_sys1}\\
 &\ve x_{k}^j\in  \mathbb{X},\,\ve u_k^j\in \mathbb{U},\label{eq_nc0} &&\forall \, (j, k) \in \ve I_{\ind{0}{N_p-1}},\\
  &\ve x_{N_p}^j\in  \mathbb{X}_f,\; \label{eq_init_term0}
 &&\forall \, (j, N_p) \in \ve I_{N_p},
% \ve u_k^j &= \ve u_k^l \mathrm{ if } \ve z_k^{p(j)}=\ve z_k^{p(l)},
%  \;\forall \, (j,k), (l,k) \in \ve I,\label{eq_nonanti}
\end{align}
\end{subequations}
\greenc{where the set of all indices $(j,k)$ in the scenario tree is denoted as $I$ and the set of indices occurring  from a  stage $k_1$ until a certain stage $k_2$   is denoted by $I_{\ind{k_1}{k_2}}$, where $0\leq k_1\leq N_p$ and $k_1\leq k_2\leq N_p$. Also, the set of indices occurring at a stage at $k$ is denoted by $I_k \triangleq I_{\ind{k}{k}}$, where $0\leq k\leq N_p$. $I_{\ind{k_1}{k_2} }\triangleq \emptyset$, if $k_2 <k_1$. } Each state $\ve x_{k+1}^c$ predicted at the time step $k+1$ is the child node in the scenario tree obtained from the node $\ve x_k^j$, the input $\ve u_k^j$, the realization $i\in\Gamma_p$ of the parametric uncertainties $(A_i,B_i)$, and the realization $l\in\Gamma_w$ of the additive disturbances $w_l$.  The weighted sum of the stage costs $\ell(x,u)$ along the prediction horizon and the terminal penalty function $V_f (\ve x)$ constitute the overall objective   function. The number of branches at every predicted node is given by $n_d=n_p\times n_w$.

The state and the input bounds, and the bounds on the terminal state are enforced via~\eqref{eq_nc0} using polytopic sets $\mathbb{X}$ and $\mathbb{U}$, and $\mathbb{X}_f$, respectively.

\bluec{The control input for a particular node must be the same for all the branches to enforce the causality of the control policy. I.e., $u_k^j=u_k^l$ if $z_k^j=z_k^l$ for all $ (j,k), (l,k) \in \ve I_{\ind{0}{N_p-1}}$}. However, the future inputs at different nodes can be different as  measurement information will be available at the next stages. \bluec{I.e., $u_k^j$ can be different from $u_k^l$  if $z_k^j\neq z_k^l$ for all $ (j,k), (l,k) \in \ve I_{\ind{0}{N_p-1}}$.} The optimal input at the fist prediction step $u_0^{1*}(x)$ obtained by solving the optimization problem~\eqref{mpc_general0} is applied to the plant. The terminal region $\mathbb{X}_f$ is chosen as the maximal RPI set for a stabilizing control law $K_f x$. The problem size grows rapidly with respect to the number of uncertainties and the length of the prediction horizon. Therefore we investigate solutions of reduced complexity that approximately realize the performance of the multi-stage scheme. This will be discussed in detail in the rest of the paper.

In the linear case considered here, since all the extreme realizations of the uncertainties are used in the predictions, the scenario tree predicts the reachable set of state trajectories. Every node of the scenario tree denotes the vertices of the polytope that forms the reachable set of the system for the predicted control policy.

\section{Tube-enhanced multi-stage MPC}
\begin{figure}
	\centering
	\includegraphics[width=0.84\columnwidth]{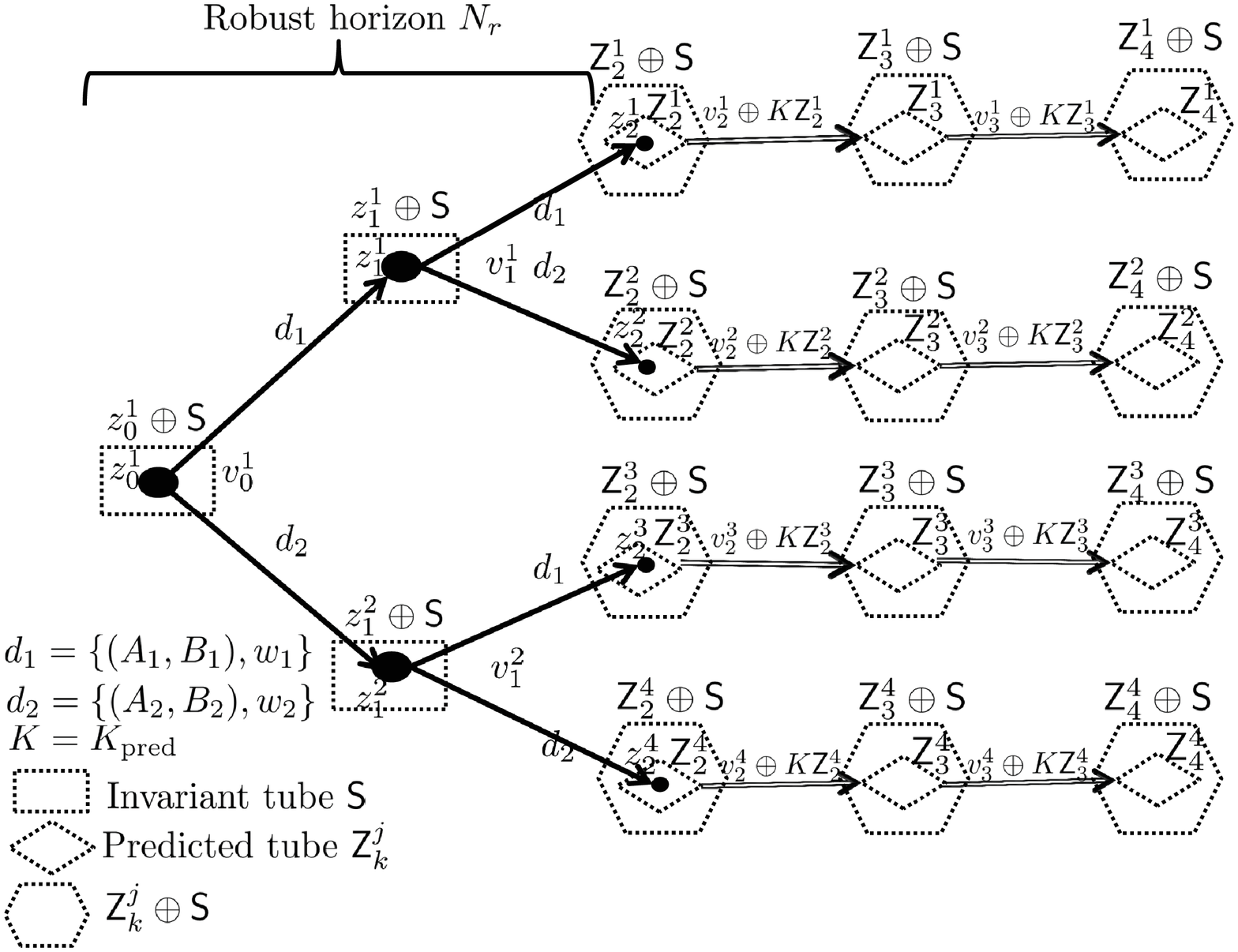}
	\caption{Scenario tree representation of the evolution of the uncertainties for tube-enhanced multi-stage MPC with robust horizon $N_r=2$.}  % width is 8.4 cm.
	
	\label{tree_RH}                                 % Size the figures
	% accordingly.
\end{figure}

 The problem defined in~\eqref{mpc_general0} suffers from rapid growth with respect to the number of realizations of the uncertainties and the length of the prediction horizon. We propose to employ  two kinds of tubes to deal with the growth in problem complexity as described below: 
\begin{enumerate}
    \item An invariant tube using an affine feedback policy is employed to handle small-magnitude disturbances.
    \begin{itemize}
    \item The invariant tube is obtained offline and hence, the complexity of the optimization problem does not grow with respect to the number of small disturbances.
    \item The invariant tube is employed only for small disturbances and hence the method does not introduce a large conservatism.
    \end{itemize}
    \item  Different tubes for each scenario are introduced to handle the significant uncertainties after a predefined horizon (robust horizon) in the prediction, instead of further branching of the scenario tree (see Figure~\ref{tree_RH}).
    \begin{itemize}
    \item The problem complexity grows only linearly with respect to the prediction horizon beyond the robust horizon.
    \item The formulation  optimizes for different feed-forward terms of the  predicted tubes at every stage that belong to different scenarios beyond the robust horizon (in addition to modeling full recourse until the robust horizon) and hence the approach is less conservative when compared to a pure tube-based scheme. Low complexity tubes can also be employed for less conservatism making the approach applicable to high dimensional systems.
    \end{itemize}
\end{enumerate}
The following subsections will elaborate the  key points discussed above to obtain an improved trade-off between optimality and complexity.

\subsection{Handling small disturbances using an invariant tube}
%The non-conservatism of the robust multi-stage MPC arises because the recourse is modeled in the predictions without restricting the structure of the feedback policy. This results in an rapid growth in problem complexity with respect to the number of realizations of the uncertainties and the length of the prediction horizon. Unstructured feedback policies can result in significant performance advantages when compared to affine policies for large magnitude uncertainties, but the advantages obtained for small magnitude uncertainties are small. The small improvement in the performance for a  rapid growth in the problem size for small uncertainties makes the approach not implementable in real-time. For this reason, we propose to classify the uncertainties into two categories: small-magnitude and large-magnitude uncertainties. The classification is based on the influence that the uncertainty has on the state trajectory and not on the actual range of the uncertainty.

Because of the multiplicative nature of the parametric uncertainties, they will have a large influence on the state trajectory far away from the origin. Therefore, we classify all parametric uncertainties as large magnitude uncertainties. The additive disturbances can be large or small depending on the application. To reduce the computational complexity, the set of additive disturbances $\mathsf{W}$ is decomposed into two polytopes $\overline{\mathsf{W}}$ and $\underline{\mathsf{W}}$ that contain the origin in their interiors such that $\mathsf{W}\subseteq\overline{\mathsf{W}}\oplus\underline{\mathsf{W}}$, where $\overline{\mathsf{W}},\, \underline{\mathsf{W}} \subset\mathbb{R}^{n_x}$ .

The set $\overline{\mathsf{W}}$ denotes large magnitude disturbances and the set $\underline{\mathsf{W}}$ denotes small magnitude uncertainties. It is recommended that the uncertain set $\mathsf{W}$ is decomposed such that the number of vertices of the large uncertainties $ \overline{\mathsf{W}}$ is small while $\underline{\mathsf{W}}$ is of smaller volume.  The model that accounts for large uncertainties is defined as follows: 
\begin{align}
\ve z^+&=\ve{A}_i \ve z+\ve{B}_i \ve v+\ve w,\label{lnom}
\end{align}
for all $i\in\Gamma_p$ and for all $\ve w\in\overline{\mathsf{W}}$, where $\ve z\in\mathbb{R}^{n_x}$ is the state and $\ve v\in\mathbb{R}^{n_u}$ is the input of the model~\eqref{lnom}. The number of vertices of the additive uncertainty set $\overline{\mathsf{W}}$ is denoted by $n_{\overline{w}}$ and the set $\overline{\mathsf{W}}$ is defined as $\overline{\mathsf{W}}:=\{w|w\in\mathrm{conv}(\{w_l,\,\forall l\in\Gamma_{\overline{w}})\}$, where $\Gamma_{\overline{w}}=\{1,\,\dots,\,n_{\overline{w}}\}$. The large uncertainties are considered in the predictions using the multi-stage approach. To handle small disturbance set $\underline{\mathsf{W}}$, an affine feedback policy is employed as follows:
\begin{align}\label{ch3_eq_affine_policy}
\ve u=\ve v+\ve K_{\mathrm{inv}}(\ve x-\ve z),
\end{align}
where $\ve K_{\mathrm{inv}}$  denotes the feedback gain associated with the invariant tube and is chosen such that the parameter-varying closed-loop system $\{(\ve A_i+\ve B_i\ve K_{\mathrm{inv}}),\,\forall i\in\Gamma_p\}$ is asymptotically stable. For the system given in~\eqref{lsys}, the state $\ve x$ and the control law \eqref{ch3_eq_affine_policy}, a set $\ve{\mathsf S}$ is defined as small disturbance invariant if $(\ve A_i+\ve B_i\ve K_{\mathrm{inv}})\ve{\mathsf S}\oplus\ve{\underline{\mathsf{W}}}\subseteq\ve{\mathsf S}$ for all $i\in\Gamma_p$.  This disturbance invariant set $\ve{\mathsf S}$ can be chosen as the convex RPI over-approximation of the minimal RPI set of the model \eqref{lnom}.  A convex outer-approximation ($\ve{\mathsf S}$) of the minimal RPI  set $\ve{\mathsf S}$ can be  obtained from the algorithm given in~\cite{Kouramas2005}. This problem is solved offline and hence does not affect the online computation time of the proposed algorithm. In the implementation section, we propose a novel convex optimization problem to over-approximate the minimal RPI set that is disturbance invariant using a linear programming problem. Since the small magnitude uncertainties are handled using an affine feedback policy, they will not be considered in the online optimization problem and hence do not affect the problem complexity.
\subsection{Handling long prediction horizons using predicted tubes}
 For the large uncertainties considered in the scenario tree, the problem complexity increases exponentially with respect to the prediction horizon $N_p$.  To reduce the problem complexity, the branching of the tree can be stopped beyond a certain prediction step called the robust horizon $N_r$.  Beyond the robust horizon, the affine policies ($v+K_{\mathrm{pred}}z$) are employed to handle all large uncertainties considered in the scenario tree. Here $K_{\mathrm{pred}}$ denotes the feedback gain associated with the tubes predicted online. The dynamics of the system beyond $N_r$ can be described using the following set recursion:
\begin{align}\label{eq_set_rec}
 \mathsf{Z}^+\supseteq(\ve A_i+\ve B_i K_{\mathrm{pred}})\mathsf{Z}\oplus\{B_i v\}\oplus\overline{\mathsf{W}},
\end{align}
for all $i\in\Gamma_p$ and  $\mathsf{Z}$ denotes the tube of states at the current time step and $\mathsf{Z}^+$ denotes the tube at the successor time step.
The idea is illustrated in Fig.~\ref{tree_RH} for a robust horizon $N_r=2$. It can be seen that the tubes replace branches beyond $N_r$ leading to a linear growth of the complexity of the scenario tree with respect to the prediction horizon. The feed-forward terms associated with the affine control law can be different for different scenarios. This can improve the performance of the controller when compared to the tube-based scheme~\cite{Fleming2015}, where the problem formulation considers one feed-forward term per prediction step. The invariant tube is shown on top of the scenario tree for representational purposes. The invariant tube $\mathsf{S}$ is used  to obtain a suitable back-off that guarantees satisfaction of the original constraints for all possible values of the small magnitude uncertainties $\underline{\mathsf{W}}$ and does not contribute additional costs to the online optimization problem.
\bluec{\begin{remark}
The idea of robust horizon has been already proposed for nonlinear systems in~\cite{lucia2013} by assuming that the uncertainty remains constant beyond the robust horizon. However, no rigorous study on the recursive feasibility and the stability of the closed-loop system was performed. In \cite{LuciaPhd}, it was proposed that a terminal set must be employed at the end of robust horizon to achieve stability but this is clearly restrictive.  In this work, we propose a rigorous solution using a robust horizon in the multi-stage MPC framework enhanced by a tube-based formulation. We make use of affine control  policies  beyond the robust horizon and formulate the problem such that the closed-loop system is stable. The proposed formulation is not as restrictive as in~\cite{LuciaPhd}, the terminal constraints are enforced at the end of the prediction horizon as in conventional MPC schemes and the proposed scheme does not assume that the uncertainties remain constant beyond $N_r$. 
\end{remark}}
%The optimization problem with the robust horizon is formulated as follows:

\subsection{Problem Formulation}
The optimization problem  $\mathtt{P}_{N_p}(\ve x)$ that is solved at every time step is given by
\begin{subequations} \label{mpc_generalm}
\begin{align}
% \min_{\stackrel{\ve y_k^j,\ve x_k^j,\ve u_k^j}{\forall (j,k)\in \ve I}}
% \min_{\stackrel{\ve y_{k+1}^j,\ve x_{k+1}^j,\ve u_k^j}{\forall (j,k)\in \ve I}}
% \;\;\;\tilde{\mathcal J}(\ve y_{k+1}^j,\ve u_k^j)
%&\min_{\ve,z_0^1,\ve v_k^j \forall (j,k)\in \ve I}\bigg(\sum_{k=0}^{N_r} \sum_{j=1}^{{n_{d}}^k}\omega_k^j \mathcal \ell(\ve z_{k}^j,\ve v_k^j)+\nonumber\\&       \max_{\tilde{z}_k^j\, \forall (j,k\geq N_r)\in \ve I}\big(\sum_{k=N_r+1}^{N_p-1}\sum_{j=1}^{{n_{d}}^{N_r}}\omega_k^j  \ell(\tilde{z}_{k}^j,\ve v_k^j+K \tilde{z}_{k}^j+\nonumber\\&   \sum_{j=1}^{{n_{d}}^{N_p}}V_f (\tilde{z}_{N_p}^j)\big)\bigg)
\min_{\ve z_0^1,\ve v_k^j \forall (j,k)\in \ve I_{\ind{0}{N_p-1}}}\bigg(\sum_{k=0}^{N_r-1} J^{\mathrm{MS}}_k+ \sum_{k=N_r}^{N_p-1} J^{\mathrm{tube}}_k+J_{N_p}^{\mathrm{term}}\bigg)
\end{align}
subject to:
\begin{align}
 &\ve z_{k+1}^c=\ve A_i\ve z_k^j+\ve B_i \ve v_k^j+\ve w_l, &&\forall\, (j, k) \in \ve I_{\ind{0}{N_r-1}}, \forall i\in\Gamma_{P}, \forall l\in\Gamma_{\overline{w}},\label{eq_sys1m}\\
  &\ve z_{k}^j\in  \mathbb{Z},\,\ve v_k^j\in \mathbb V, \label{eq_nc1m} &&\forall  (j, k) \in \ve I_{\ind{0}{N_r-1}},\\
&\ve z_{N_r}^j\in  \mathsf{Z}_{N_r}^j\subseteq\mathbb{Z},\;\label{rh_initm}&&\forall (j, N_r) \in \ve I_{N_r},\\
  &(\ve A_i+\ve B_iK_{\mathrm{pred}})\mathsf{Z}_k^j\oplus\{B_i v_k^j\}\oplus\overline{\mathsf{W}}\subseteq\mathsf{Z}_{k+1}^j,\label{eq_nc2m} &&\forall (j, k) \in \ve I_{\ind{N_r}{N_p-1}}, \forall i\in\Gamma_p,\\
 &\mathsf{Z}_{k}^j\subseteq \mathbb{Z}, \tilde{z}_k^j\in \mathsf{Z}_k^j,\,\{v_k^j\}\oplus\ve K_{\mathrm{pred}}\mathsf{Z}_{k}^j\subseteq \mathbb{V},\label{eq_nc3m} &&\forall (j, k) \in \ve I_{\ind{N_r}{N_p-1}},\\
 &\mathsf{Z}_{N_p}^j\subseteq  \mathbb{Z}_f,\,\label{final_termm}&&\forall (j, N_p) \in \ve I_{N_p},\\
   &\ve x\in\{ z_1^0\}\oplus\mathsf{S},\, \label{init_termm}
 \end{align}
  where
\begin{align}
 J^{\mathrm{MS}}_k&= \sum_{j=1}^{{n_{d}}^k}\omega_k^j \ell(\ve z_{k}^j,\ve v_k^j),\\
 J^{\mathrm{tube}}_k&=\max_{\tilde{z}_k^j,\,\forall(j,k)\in I_{\ind{N_r}{N_p-1}}}\sum_{j=1}^{{n_{d}}^{N_r}}\omega_k^j \ell(\tilde{z}_{k}^j,\ve v_k^j+K_{\mathrm{pred}} \tilde{z}_{k}^j),\\
 J_{N_p}^{\mathrm{term}}&=\max_{\tilde{z}_{N_p}^j,\,\forall(j,N_p)\in I_{N_p}}\sum_{j=1}^{{n_{d}}^{N_r}}V_f (\tilde{z}_{N_p}^j).
\end{align}
\end{subequations}
There are several differences in both the objective and the constraints in~\eqref{mpc_generalm} compared to that of the standard multi-stage MPC~\eqref{mpc_general0}. In the optimization problem~\eqref{mpc_generalm}, the objective can be divided into three parts: a multi-stage part $J_k^{\mathrm{MS}}$, a tube-based part $J_k^{\mathrm{tube}}$ and the terminal penalty part $J_k^{\mathrm{term}}$. The multi-stage part of the objective function $J_k^{\mathrm{MS}}$ is applied until $N_r-1$, and beyond $N_r-1$, the tube-based part of the objective function $J_k^{\mathrm{tube}}$ is applied. As always, the terminal penalty $J_k^{\mathrm{term}}$ is applied at the last prediction step $N_p$. The multi-stage part of the objective function $J_k^{\mathrm{MS}}$ is the same as in~\eqref{mpc_general0}. The tube-based part of the objective function $J_k^{\mathrm{tube}}$ and the terminal part of the objective function $J_k^{\mathrm{term}}$ have maximization terms associated with them. For the tube-based part of the optimization problem, the worst-case cost associated with the predicted tubes $\mathsf{Z}_k^j$ are obtained at every prediction step. This helps in establishing the optimal value function as a Lyapunov function for the proposed scheme.

The state inside the tube that maximizes the objective function is defined as $\tilde{z}_k^j$ and is constrained by $\tilde{z}_k^j\in \mathsf{Z}_k^j$ as given in~\eqref{eq_nc3m}. The variables $\tilde{z}_k^j$ are formulated as decision variables for all $(j,k)\in I_{\ind{N_r}{N_p-1}}$. Equations \eqref{eq_nc2m}--\eqref{final_termm} represent the constraints beyond the robust horizon $N_r$.  Equation~\eqref{eq_nc2m} guarantees the recursive bounding of the state trajectories for the chosen control law $v+K_{\mathrm{pred}}z$. It can also be seen that the affine term $v_k^j$ can be freely chosen for all $(j,k)\in I_{\ind{0}{N_p-1}}$. This can improve the resulting solution even for a robust horizon of $N_r=1$ when compared to a pure tube-based scheme. Equations.~\eqref{eq_nc3m}--\eqref{final_termm} denote the state, input and the terminal constraints. The set of all $x\in\mathbb{X}$ for which there exists a feasible feedback policy is denoted as $\mathtt{X}_{N_p}$. Equation~\eqref{rh_initm} is formulated at the robust horizon to establish continuity between the predicted scenarios until the robust horizon and the tubes predicted beyond the robust horizon.

The initial state of the scenario tree $\ve z^{1}_0$ is a function of the current state $\ve x$ as defined in~\eqref{init_termm} and is a decision variable of the optimization problem~\eqref{mpc_generalm}. The optimization problem is solved with tightened constraints $\mathbb{Z}=\mathbb{X}\ominus\mathsf{S}$ and $\mathbb{V}=\mathbb{U}\ominus K_{\mathrm{inv}}\mathsf{S}$ in~\eqref{eq_nc1m} and~\eqref{eq_nc3m}. The number of branches at every node is given by $n_{d}=n_p\times n_{\overline{w}}$ until the robust horizon $N_r$.

Note that the number of branches can be reduced dramatically if $n_{\overline{w}}\ll n_w$ when compared to the consideration of all uncertainties in a scenario tree. The control input $\ve u$ applied to the system is given by $\ve u=\ve v^{1*}_0+\ve K_{\mathrm{inv}}(\ve x-\ve z^{1*}_0 (\ve x))$, where $\ve v^{1*}_0$ is the first element of the optimal control input sequence obtained by solving \eqref{mpc_generalm}. Since the polytope $\overline{\mathsf{W}}$ has smaller number of vertices compared to $\mathsf{W}$, the problem size solved using~\eqref{mpc_generalm} is reduced.
% Rp: ``The problem size is reduced'' compared to what?
Despite the reduced complexity when compared to a standard multi-stage problem, the proposed controller can often achieve a performance comparable to that of multi-stage MPC for a fraction of the computational complexity by the choice of $N_r$ and $\overline{\mathsf{W}}$. 
\begin{remark}
The proposed scheme is flexible and includes options to further improve it which we do not analyze in this work. Two of the possible modifications are listed below:
\begin{enumerate}
    \item The feedback gain is denoted as a constant $K_{\mathrm{pred}}$ for the predicted tubes. This is done only to simplify the presentation. The feedback gains can be different for different scenarios. The only necessary condition is that the gain must be stabilizing and can be chosen freely for different scenarios to improve the performance of the closed-loop.
    \item The number of tubes at the robust horizon is formulated equal to the number of nodes predicted until that stage using the scenario tree in \eqref{mpc_generalm}. The constraint \eqref{rh_initm} represents the continuity equation. However, different nodes can be bundled together in one tube and the number of tubes can be smaller than the number of nodes predicted until $N_r$ by modifying the continuity constraint \eqref{rh_initm}. This can help to reduce problem complexity further.
\end{enumerate}
\end{remark}
\subsection{Stabilizing objective function and choice of weights}
Since a persisting disturbance $w\in\mathsf{W}$ is assumed, convergence to the origin cannot be established. Instead, as described in~\cite{kerrigan2004feedback}, a robust positively invariant set $\mathsf{T}$ will be shown to be asymptotically stable using the proposed robust model predictive control scheme. To achieve stability, we assume that the terminal set, the proposed stage cost and the terminal penalty function satisfy the following properties:
\begin{enumerate}\label{stage_prop}
\item The terminal set $\mathbb{Z}_f:=\mathsf{T}$ is a robust positively invariant set for a stabilizing control law $K_fz$.
\item The stage cost $\ell(z,K_fz)=0$ and the terminal penalty $V_f(z)=0$, $\forall z\in \mathsf{T}$.
\end{enumerate}
A stage cost with these properties as proposed in~\cite{kerrigan2004feedback} is given as follows:
\begin{align}\label{stage_cost}
\ell(z,v)=\min_{y\in\mathbb{Z}_f} \vert\vert Q(z-y)\vert\vert_p+\vert\vert R(v-K_fz)\vert\vert_p,
\end{align}
where $Q$ and $R$ are positive semi-definite matrices. The terminal penalty function can be simply set to zero i.e. $V_f(z)=0,\,\forall z\in\mathbb{R}^{n_x}$. The choice of the stage cost is different from the nominal cost $\ell_{\mathrm{nom}}(z,v)=\vert\vert Qz\vert\vert_p+\vert\vert Rv\vert\vert_p$ which is generally used in the case of nominal MPC. The nominal stage cost $\ell_{\mathrm{nom}}$ penalizes the distance to the origin and the control effort depending on the choice of tuning matrices $Q$ and $R$. The stage cost $\ell(z,v)$ penalizes the distance to the set $\mathbb{Z}_f$ and the deviations from the control law $K_fz$. \greenc{The terminal gain $K_f$ is chosen equal to the gain of the predicted tubes  $K_{\mathrm{pred}}$. The gain $K_f$ can be chosen freely when there is no tube-based part of the scheme. This is further discussed in Section \ref{sec_gain} and in Section \ref{sec_theory}.}

Each realization of the uncertain matrices $(\ve A_i,\,\ve B_i)$ and of $w_l$ has a fixed positive weight $\omega_{i,l}>0$ associated with it for all $i\in\Gamma_p,\, l\in\Gamma_{\overline{w}}$.  Appropriate weights can be chosen depending on the applications. However, to establish stability the weights must follow certain rules which are formalized in the following assumption. The weights associated with the vertex matrices will be assigned to the nodes that result from them in the predictions. For example, if $z_1^1$ is realized because of $(\ve A_1,\,\ve B_1)$, the weight associated with the node $z_1^1$ will be equal to $\omega_1$ (the corresponding weight of $(\ve A_1,\,\ve B_1)$). The requirement of weights associated with each node in the scenario tree is given in the following assumption.

\begin{assumption}\label{as:weights}
The weight of the root node $\omega_0^1$ is smaller than or equal to the minimum of all the weights (i.e. $\omega_0^1\le\min\{\omega_0,\omega_1,\,\dots,\,\omega_{n_{d}}\}$) and it must be positive $\omega_0^1>0$. The weights $\omega_k^j$ associated with the other nodes $z_k^j$ are equal to the weights associated with the realization of the uncertainty from which they are obtained for all $(j,k)\in I_{\ind{1}{N_r-1}}$. The weights associated with tubes are chosen as	  \greenc{$\omega_k^j=n_d^{k-N_r}\omega_{\mathrm{tube}},\,\forall (j,k)\in I_{\ind{N_r}{N_p-1}}$} where  $\omega_{\mathrm{tube}}\ge\max\{\omega_0,\omega_1,\,\dots,\,\omega_{n_{d}}\}$ and is  bounded.
\end{assumption}
\bluec{
	\begin{remark}
		The choice of the weights is important in establishing the stability properties of the proposed approach. The weights affect the objective function  and thus the value function. If the weights are chosen as per Assumption \ref{as:weights}  stability  can be proven, see Lemma~\ref{lem:stab}. In the multi-stage part of the scheme, the weights are chosen the same in each stage  for the same realization of the uncertainty. Then feasible values of the stage costs for the succeeding step can be obtained by the convex combination of the stage costs that are realized at the current time step. Since the branching stops beyond the robust horizon, the weights must be updated to account for the receding horizon implementation of the MPC scheme. Hence the weights of the tube-based part of the scheme are employed as   $\omega_k^j=n_d^{k-N_r}\omega_{\mathrm{tube}},\,\forall (j,k)\in I_{\ind{N_r}{N_p-1}}.$
		\end{remark}}
\section{Implementation details}
We employ the tube-based MPC approach based on Farkas' Lemma as proposed in~\cite{Fleming2015} to account for the disturbances after the robust horizon. The proposed approach works for any choice of robust horizon and in principle can be chosen equal to $N_r=0$ in which case, the scheme reduces to the approach in~\cite{Fleming2015} if $\underline{\mathsf{W}}=\{0\}$.

The tightened state constraint set $\mathbb{Z}$ is defined as $\mathbb{Z}:=\{z|Fz\le\textbf{1}\}$, where $F$ is a $n_c\times n_x$ dimensional matrix, $n_c$ denotes the number of state constraints, and $\textbf{1}$ denotes a vector with all elements $1$ of an appropriate dimension. The tightened input constraint set $\mathbb{V}$ is defined as $\mathbb{V}:=\{v|Gv\leq\textbf{1}\}$, where $G$ is a $n_m\times n_u$ matrix and $n_m$ denotes the number of input constraints.
In the following, we discuss three types of tubes: general complexity tube, homothetic tube and low complexity tube. We show that how the proposed formulation \eqref{mpc_generalm} can be implemented as convex optimization problem. In addition, we also discuss the subtleties in the reformulation with respect to the proposed approach.
{\subsection{General complexity tube}
The complexity of the tubes that are employed beyond the robust horizon can be fixed and defined as $$\mathsf{Z}:=\{z|Tz\leq\tau\},$$ where $T\in\mathbb{R}^{n_r\times n_x}$ is fixed for all prediction steps and $\tau\in\mathbb{R}^{n_r}$ is a decision variable that is chosen online. Here $n_r$ denotes the number of inequalities that describes the tube $\mathsf{Z}$ and it is typically larger than $2n_x$.  The matrix $T$ is chosen such that the set $\Lambda:=\{z\mid Tz\leq\textbf{1}\}$ is $\lambda-$contractive for $\lambda\in[\lambda_\mathrm{min},1)$, where $\lambda_\mathrm{min}$ is the joint spectral radius of the closed-loop uncertain system matrices for the chosen feedback gain $K_{\mathrm{pred}}$.
The details on algorithm for obtaining  the largest $\lambda-$contractive set can be found in~\cite{blanchini2008set} (pp. 171-184). One can also use off-the-shelf toolboxes such as Multi-Parametric Toolbox~\cite{MPT3} to obtain $\lambda-$contractive sets.

Now, let us have a look at the problem formulation~\eqref{mpc_generalm}, where the eqs.~\eqref{eq_nc2m}--\eqref{eq_nc3m} are in the form of set operations. Farkas' Lemma can be employed to convert them from set operations to a set of linear equalities and inequalities. This reformulation will help us formulate the optimization problem as a convex optimization problem that does not require any set operations online. Equation~\eqref{eq_nc2m} bounds the error dynamics recursively as $ (A_i+B_i K_{\mathrm{pred}})\mathsf{Z}_k^j\oplus\mathsf{W}\subseteq\mathsf{Z}_{k+1}^j$ for all $i\in\Gamma_p$. Let $\mathsf{Z}_k^j$ be represented as
\begin{align}\label{eq_tube_ineq}
	\mathsf{Z}_k^j&=\{z|T z\leq \tau_k^j\},
\end{align}
where the variables $\tau_k^j,\,\forall (j,k)\in I_{\ind{N_r}{N_p}}$ can be formulated as decision variables in the optimization problem. Since an affine control law $v+K_{\mathrm{pred}}z$ is employed beyond the robust horizon $N_r$, $z^+\in\{(A_i+B_i K_{\mathrm{pred}})z+B_iv\}\oplus\overline{\mathsf{W}}$ for a given $i\in\Gamma_p$ and the tube at the next prediction step that bounds all the trajectories for the predicted input and the arbitrary realizations of the uncertainties for all $(j,k)\in I_{\ind{N_r}{N_p-1}}$ is given by 
\begin{align}\label{eq_tube1} 
\mathsf{Z}_{k+1}^{j}:=\{z|T((A_i+B_iK_{\mathrm{pred}})z+B_iv_k^j+w_l)\leq \tau_{k+1}^{j},\,\forall i\in\Gamma_p\,\forall l\in\Gamma_{\overline{w}},\,\forall z\in\mathsf{Z}_k^j\}.
\end{align}

To achieve the set recursion, Farkas' Lemma is employed and it is given below \cite{blanchini2008set,Fleming2015} :

\begin{lemma}\label{Farkas}
	Given two non-empty sets $\mathsf{X}_1:=\{x|T_1x\leq \tau_1\}$, $\mathsf{X}_2:=\{x|T_2x\leq \tau_2\}$, $\mathsf{X}_1\subseteq\mathsf{X}_2$ holds iff there exists a non-negative matrix $P$ that satisfies the equality $PT_1=T_2$ and the inequality $P\tau_1\leq \tau_2$.
\end{lemma}
Using Lemma~\ref{Farkas}, we can employ non-negative matrices $P_i,\,\forall i\in\Gamma_p,$ and $\ve (A_i+B_i K_{\mathrm{pred}})\mathsf{Z}_k^j\oplus\overline{\mathsf{W}}\subseteq\mathsf{Z}_{k+1}^j$ holds for all $i\in\Gamma_p$ iff:
\begin{subequations}\label{eq_Farka1}
	\begin{align}
	&P_iT=T(A_i+B_i K_{\mathrm{pred}}),&&\forall i\in\Gamma_p,\\
	&P_i\tau_k^j+T B_iv_k^j+T w_l\leq \tau_{k+1}^j, &&\forall i\in\Gamma_{p},\, \forall l\in\Gamma_{\overline{w}}, \,(j,k)\in I_{\ind{N_r}{N_p-1}}.\label{eq_Farka1_ineq}
	\end{align} 
\end{subequations}
The set~\eqref{eq_tube1} is reformulated using linear equalities and inequalities as shown in~\eqref{eq_Farka1}. Similarly, the set operations \eqref{eq_nc3m}--\eqref{final_termm} can be reformulated using Lemma~\ref{Farkas}. The state and input constraints can be formulated using non-negative matrices $P_x,\, P_u,\,P_T$ as follows:
\begin{subequations}\label{eq_Farka2}
	\begin{align}
	&P_x T=F,\label{Farka2_eq}\\
	&P_x\tau_k^j\leq\textbf{1},&&\forall(j,k)\in I_{\ind{N_r}{N_p-1}},\label{Farka2_ineq}\\
	&P_u T=GK_{\mathrm{pred}},\label{Farka3_eq}\\
	&Gv_k^j+P_u\tau_k^j\leq\textbf{1}, &&\forall(j,k)\in I_{\ind{N_r}{N_p-1}},\label{Farka3_ineq}
	\end{align} 
\end{subequations}
 The choice of non-negative matrices $P_i,\,\forall i\in\Gamma_p,\,P_x,\,P_u$ if obtained online, results in a non-convex optimization problem. Hence as proposed in~\cite{Fleming2015}, the non-negative matrices can be obtained offline such that the equality constraints are satisfied. This reduces the computational load of the online problem and convexifies it. The matrices $P_i,\,\forall i\in\Gamma_p$ can be obtained by solving the following problem for a given $i$:
\begin{align} \label{LP1}
P_i=\arg\min_{\tilde{P}_i}\vert\vert  \tilde{P}_i\vert\vert_\infty,\mathrm{s.t. } \tilde{P}_iT=T(A_i+B_i K_{\mathrm{pred}}).
\end{align}
Similar to~\eqref{LP1}, the matrices $P_x,\,P_u,\,P_T$ can be obtained by formulating linear programming problems that satisfy the equality constraints~\eqref{Farka2_eq}, and \eqref{Farka3_eq}. 
\bluec{The terminal constraints are implemented as follows:
\begin{align}
&{P}_i\tau_{N_p}^j+T w_l\leq\tau_{N_p}^j,\;\forall i\in\Gamma_{p},\, \forall l\in\Gamma_{\overline{w}} ,\forall(j,N_p)\in I_{N_p} ,\label{Farka4_ineq}
\end{align}
where the terminal set is defined as
\begin{align}\label{eq_term_MS}
\mathbb{Z}_f\triangleq\{z\mid Tz\leq\tau,{P}_iT=T(A_i+B_i K_{\mathrm{pred}}),{P}_i\tau+T w_l\leq\tau ,\forall i\in\Gamma_{p},\, \forall l\in\Gamma_{\overline{w}} \},
\end{align}
and $\mathbb{Z}_f\subseteq\mathbb{Z}$  and $K_f\mathbb{Z}_f\subseteq \mathbb{V}, K_f=K_{\mathrm{pred}}$ hold.  The terminal gain $K_f$ is chosen as the gain of the predicted gain to simplify the implementation and the discussion that follows. Further discussions on the computation of the terminal set can be found in Section~\ref{sec_gain}.
}
The minimal RPI set can be obtained using the methods discussed in~\cite{Kouramas2005}. Another possible over-approximation of the mRPI set can be obtained as $\mathsf{S}:=\{z|\hat{T}z\leq\tau\}$. Here the matrix $\hat{T}$ is chosen such that $\{z\mid \hat{T}z\leq\textbf{1}\}$ is $\lambda-$contractive for the chosen feedback gain $K_{\mathrm{inv}}$. The right hand side of the inequality that defines $\mathsf{S}$, $\tau$ can be obtained as follows:
\begin{subequations} \label{LP2}
	\begin{align}
	\min_{\ve \tau}\vert\vert\tau\vert\vert_p
	\end{align}
	subject to:
	\begin{align}
	\hat{P}_i\tau+ \hat{T} w_l\le\tau,\forall i\in\Gamma_p,\, \forall l \in \Gamma_{\overline{w}}.
	\end{align}
\end{subequations}
The objective can be chosen as $1-$norm or $\infty-$norm so that the optimization problem is an LP. The LP~\eqref{LP2} guarantees that the set $S:=\{z|\hat{T}z\leq\tau\}$ is robust positively invariant. The non-negative matrices $\hat{P}_i,$ for all $i\in\Gamma_p$ can be chosen as discussed  in \eqref{eq_Farka2}. Since $\hat{T}$ and $\hat{P}_i,\,\forall i\in\Gamma_p$ are fixed, the set $\mathsf{S}$ can be a conservative over-approximation. The advantage however is that the LP~\eqref{LP2} can be solved much faster compared to the algorithm given in~\cite{Kouramas2005}.

In addition to the reformulation of the set operations in the constraints, the objective function can be simplified by removing the maximization part with the help of slack variables as shown in~\citep{scokaert19981,kerrigan2004feedback}.  If the stage cost~\eqref{stage_cost} is used with $p=1$, $\min_{v\in\mathbb{V}}\max_{\tilde{z}}\ell(\tilde{z},v+K_{\mathrm{pred}}\tilde{z})$ can be obtained as follows:
\begin{subequations}\label{objective}
	\begin{align}
	\min_{v\in\mathbb{V}}\max_{\tilde{z}\in\mathsf{Z}}\ell(\tilde{z},v+K_{\mathrm{pred}}\tilde{z})=\min_{v,\tilde{z},y,\mu,\eta,\gamma} \gamma
	\end{align}
	subject to:
	\begin{align}
	-\mu\leq Q(\tilde{z}-y)\leq\mu ,\,\forall \tilde{z}\in\mathsf{Z}, y\in\mathbb{Z}_f,\label{state_bound} \\
	-\eta\leq Rv+K_{\mathrm{pred}}\tilde{z}-K_f\tilde{z}\leq\eta ,\,\forall \tilde{z}\in\mathsf{Z},\label{input_bound}\\
	\textbf{1}^T\mu+\textbf{1}^T\eta\leq \gamma.\label{sum_ub}
	\end{align}
\end{subequations}
Note that $\tilde{z}$ is not known a priori and has been added as a decision variable in~\eqref{objective}. The constraint \eqref{input_bound} simplifies to $	-\eta\leq Rv\leq\eta $ because the terminal gain $K_f$ is chosen equal to the gain employed in the predicted tubes (i.e., $K_f=K_{\mathrm{pred}}$). The constraints of the state are infinite dimensional. However, a simplification is possible by reformulating the bounds on the state  objective~\eqref{state_bound} with the help of a non-negative matrix $P_Q$ as discussed earlier and the constraint~\eqref{state_bound} can be satisfied for any given $\tilde z\in\mathsf{Z}$ if all $z\in\mathsf{Z}$ satisfy~\eqref{state_bound} as proposed in \cite{lu2019robust}. Using Lemma~\ref{Farkas}, we can define a  non-negative matrix  $P_Q$  such that the following equations hold to satisfy the constraints~\eqref{state_bound} and the constraints \eqref{state_bound}$-$\eqref{sum_ub} can be rewritten as follows for all $(j,k)\in I_{\ind{N_r}{N_p-1}}$.:
\begin{subequations}\label{eq_obj_reformulate1}
	\begin{align}
	&P_Q T=Q,\label{obj_Fark}\\
	&-\mu_k^j\leq P_Q\tau_k^j-Qy_k^j\leq \mu_k^j,&&y_k^j\in\mathbb{Z}_f,\,\forall(j,k)\in I_{\ind{N_r}{N_p-1}},\label{obj_Fark2}\\
%	&P_R T=R (K_{\mathrm{pred}}-K_f),\label{obj_Fark3}\\
	&-\eta_k^j\leq Rv_k^j\leq \eta_k^j, &&\forall(j,k)\in I_{\ind{N_r}{N_p-1}},\label{obj_Fark4}\\
	&\textbf{1}^T\mu_k^j+\textbf{1}^T\eta_k^j\leq \gamma_k^j,&&\forall(j,k)\in I_{\ind{N_r}{N_p-1}}.\label{obj_combi}
	\end{align}
\end{subequations}
The equality constraint~\eqref{obj_Fark} can be obtained offline and the inequality constraints~\eqref{obj_Fark2} and \eqref{obj_Fark4} can be included as the constraints in the optimization problem online. However, this reformulation of the inequality constraints does not provide a tight upper bound which means that the obtained cost is not the same as the one obtained using the original inner maximization in (36a).  The formulation, however, retains the theoretical properties of recursive feasibility and stability of the original formulation \eqref{mpc_generalm}.
Combining all the reformulations, the resulting optimization problem $\mathtt{P}_{N_p}^G(x)$ can be formulated as
   \bluec{                           
	\begin{align}\label{MS_implement}
	% \min_{\stackrel{\ve y_k^j,\ve x_k^j,\ve u_k^j}{\forall (j,k)\in \ve I}}
	% \min_{\stackrel{\ve y_{k+1}^j,\ve x_{k+1}^j,\ve u_k^j}{\forall (j,k)\in \ve I}}
	% \;\;\;\tilde{\mathcal J}(\ve y_{k+1}^j,\ve u_k^j)
	\min_{\stackrel{\ve z_0^1,\ve v_k^j, y_k^j,\,\forall (j,k)\in \ve I_{\ind{0}{N_p-1}}}{\tau_k^j,\gamma_k^j,\eta_k^j,\mu_k^j,\, \forall (j,k)\in I_{\ind{N_r}{N_p-1}}}}&\sum_{k=0}^{N_r-1} \sum_{j=1}^{n_ {d}^k}\omega_k^j  \ell(\ve z_{k}^j,\ve v_k^j)+\sum_{k=N_r}^{N_p-1}\sum_{j=1}^{n_{d}^{N_r}}\omega_k^j \gamma_k^j
	\end{align}
	subject to: \\\hspace*{2.5cm}\eqref{eq_sys1m}, \eqref{eq_nc1m}, \eqref{init_termm}, \eqref{eq_Farka1_ineq}, \eqref{Farka2_ineq}, \eqref{Farka3_ineq}, \eqref{Farka4_ineq}, \eqref{obj_Fark2}, \eqref{obj_Fark4}, \eqref{obj_combi}, 	$T z_{N_r}^j\leq\tau_{N_r}^j,\,\forall(j,N_r)\in I_{N_r}$.
}
%	&\ve z_{k+1}^c=\ve A_i\ve z_k^j+\ve B_i \ve v_k^j+\ve w_l,\label{eq_sys1i}
%	&&\forall\, (j, k\leq N_r) \in \ve I, \forall i\in\Gamma_{p},\, \forall l\in\Gamma_{\overline{w}},\\
%	&\ve F \ve z_{k+1}^j\leq\textbf{1},\,\ve G v_k^j\leq\textbf{1}, \label{eq_nc1i} &&\forall \, (j, k\leq N_r) \in \ve I,\\
%	&\ve P_i \ve \tau_k^j+B_iv_k^j+w_i\leq \tau_{k+1}^j,\label{eq_nc2i} &&\forall \, (j, k\geq N_r) \in \ve I, \forall i\in\Gamma_{p}, \forall l\in\Gamma_{\overline{w}}\\
%	&\ve P_x \tau_k^j\leq \textbf{1}\label{eq_nc3i}&& \forall \, (j, k\geq N_r) \in \ve I,\\
%	& \ve v_k^j+\ve P_u \tau_k^j\leq \textbf{1},\label{eq_nc4i}&& \forall \, (j, k\geq N_r) \in \ve I,\\
%	&P_T\ve \tau_{N_p}^j\leq\textbf{1},\;\label{final_termi}
%	&&\forall \, (j, N_p) \in \ve I,\\
%	& -\mu_k^j\leq Q(\hat{z}_k^j+\alpha_k^j \vee_r-y)\leq\mu_k^j,&&\forall \, (j, k\geq N_r) \in \ve I,\, \forall r\in\{1,\,\dots,\,n_v\}\label{objmi}\\
%	&-\eta_k^j\leq R(v_k^j+(K_{\mathrm{pred}}-K_f)(\hat{z}_k^j+\alpha_k^j \vee_r)) \leq\eta_k^j,&&\forall \, (j, k\geq N_r) \in \ve I,\, \forall r\in\{1,\,\dots,\,n_v\}\label{objmi1}\\
%	&\textbf{1}^T\mu_k^j+\textbf{1}^T\eta_k^j=\gamma_k^j,&&\forall \, (j, k\geq N_r) \in \ve I,\label{objmi2}\\
%	&\ve x  \in \ve \{z_0^1(\ve x)\}\oplus\ve{\mathsf S}\label{eq_init_termi}.
	% \ve u_k^j &= \ve u_k^l \mathrm{ if } \ve z_k^{p(j)}=\ve z_k^{p(l)},
	%  \;\forall \, (j,k), (l,k) \in \ve I,\label{eq_nonanti}

In \eqref{MS_implement}, it can be seen that the set operations are replaced by inequality constraints with the help of the non-negative matrices $P_i,\forall i\in\Gamma_p,\,P_x,\,P_u,P_Q$ that are obtained offline. The constraints \eqref{obj_Fark2}$-$\eqref{obj_combi} bound the objective from above using the slack variables $\gamma_k^j\in I_{\ind{N_r}{N_p-1}}$. The set of all $x\in\mathbb{X}$ for which there exists a feasible solution for the optimization problem \eqref{MS_implement} is denoted as $\mathtt{X}_{N_p}^G$. The optimal value function obtained by solving the  optimization problem \eqref{MS_implement} is denoted as $V_{N_p}^{G*}(x)$. The stability properties of the implementation \eqref{MS_implement} is proven in Section \ref{sec_theory}. 
%\begin{remark}
%If $N_r=N_p$. the terminal set $\mathbb{Z}_f$ can be chosen as the maximal RPI set for the system 
%\end{remark}
%\begin{remark}
%	The terminal feedback gain$K_f$ is chosen as the prediction gain $K_{\mathrm{pred}}$ to simplify the implementation. If the terminal gain $K_f$ is chosen different from the prediction gain $K_{\mathrm{pred}}$, the terminal set can be defined as 
%	\begin{align}\label{eq_term_MS_remark}
%	\mathbb{Z}_f\triangleq\{z\mid& Tz\leq\tau,{P}_iT=T(A_i+B_i K_{\mathrm{pred}}),{P}_i\tau+T w_l\leq\tau ,Tz\leq\tau,\nonumber\\&\hat{P}_iT=T(A_i+B_i K_f),\hat{P}_i\tau+T w_l\leq\tau\forall i\in\Gamma_{p},\, \forall l\in\Gamma_{\overline{w}} \}
%	\end{align}
%\end{remark}
\subsection{Homothetic tube}
Since the tubes are characterized using inequalities in the general complexity tube, explicit values of the vertices are not available online. Hence, the tight upper bound for the stage costs of the tubes are difficult to obtain. By formulating the predicted tubes as homothetic tubes (i.e. by fixing the shape of the sets  and only varying the scaling variable), the characterizations of the vertices can be obtained  as follows:
\begin{subequations}
	\begin{align}
	\mathsf{Z}_k^j&=\{z|T (z-\hat{z}_k^j)\leq \alpha_k^j\textbf{1}\},\\
	&=\hat{z}_k^j\oplus\alpha_k^j\Lambda,\\
	&=\hat{z}_k^j\oplus \alpha_k^j\mathrm{conv}\{\vee_1,\vee_2,\,\dots,\,\vee_{n_v}\},
	\end{align}
\end{subequations}
where $\alpha_k^j\geq 0$ is a non-negative scalar and the $\lambda-$contractive set $\Lambda$ is characterized by  the vertices $\vee_1,\vee_2,\,\dots,\,\vee_{n_v}$, where $n_v$ denotes the number of vertices.  The constraints for the homothetic tube can be rewritten as follows using Lemma \ref{Farkas}.
\bluec{
\begin{subequations}
\begin{align}
&P_i(\alpha_k^j\textbf{1}+T\hat{z}_{k}^j)+T B_iv_k^j+T w_l\leq T\hat{z}_{k+1}^j+\alpha_{k+1}^j\textbf{1}, &&\forall i\in\Gamma_{p},\, \forall l\in\Gamma_{\overline{w}}, \,(j,k)\in I_{\ind{N_r}{N_p-1}}\label{eq_homo_ineq1},\\
&P_x(\alpha_k^j\textbf{1}+T\hat{z}_{k}^j)\leq\textbf{1},&&\forall(j,k)\in I_{\ind{N_r}{N_p-1}},\label{homo_ineq2}\\
&Gv_k^j+P_u(\alpha_k^j\textbf{1}+T\hat{z}_{k}^j)\leq\textbf{1}, &&\forall(j,k)\in I_{\ind{N_r}{N_p-1}},\label{homo_ineq3}\\
&{P}_i(\alpha_{N_p}^j\textbf{1}+T\hat{z}_{N_p}^j)+T w_l\leq T\hat{z}_{N_p}^j+\alpha_{N_p}^j\textbf{1},&&\forall i\in\Gamma_{p},\, \forall l\in\Gamma_{\overline{w}} ,\forall(j,N_p)\in I_{N_p} .\label{homo_ineq4}
\end{align}
\end{subequations}
Note that the terminal feedback gain is chosen as the gain of the predicted tubes $K_{\mathrm{pred}}$ as in the general complexity tube case. The resulting terminal set is defined as
\begin{align}\label{eq_term_MS2}
\mathbb{Z}_f\triangleq\{z\mid T(z-\hat{z})\leq\alpha\textbf{1},{P}_iT=T(A_i+B_i K_{\mathrm{pred}}),{P}_i(T\hat{z}+\alpha\textbf{1})+T w_l\leq T\hat{z}+\alpha\textbf{1} ,\forall i\in\Gamma_{p},\, \forall l\in\Gamma_{\overline{w}} \}.
\end{align}}
The homothetic tube enables us to formulate the tube and obtain a tight upper bound on the extreme stage costs of the tubes as follows:
\begin{subequations}\label{objective2}
	\begin{align}
	\sum_{k=N_r}^{N_p} J^{\mathrm{tube}}_k&=\min_{\alpha_k^j,\gamma_k^j,\eta_k^j,\mu_k^j \forall (j,k)\in I_{\ind{N_r}{N_p-1}}}\sum_{k=N_r}^{N_p}\sum_{j=1}^{n_{d}^{N_r}}\omega_k^j \gamma_k^j
	\end{align}
	subject to
	\begin{align}
	&-\mu_k^j\leq Q(\hat{z}_k^j+\alpha \vee_r-y_k^{j,r})\leq\mu_k^j, && \forall r\in\{1,\,\dots,\,n_v\}, (j,k)\in I_{\ind{N_r}{N_p-1}}, \label{state_bound2} \\
	&-\eta_k^j\leq Rv_k^j\leq\eta_k^j ,&&\forall r\in\{1,\,\dots,\,n_v\},\,(j,k)\in I_{\ind{N_r}{N_p-1}}, \label{input_bound2}\\
	& y_k^{j,r}\in\mathbb{Z}_f,&&\forall(j,k)\in I_{\ind{N_r}{N_p-1}},\\
	&\textbf{1}^T\mu_k^j+\textbf{1}^T\eta_k^j\leq \gamma_k^j&&\forall(j,k)\in I_{\ind{N_r}{N_p-1}},\label{obj_combi2}
	\end{align}
\end{subequations}
where $(j,r)$ denotes the indices associated with the vertices of the tube $\mathsf{Z}_k^j$ for all $(j,k)\in I_{\ind{N_r}{N_p-1}}$. 
\bluec{
 The optimization formulation in the case of homothetic tubes $\mathtt{P}_{N_p}^H(x)$  results as follows:
	\begin{align}\label{MS_implement2}
% \min_{\stackrel{\ve y_k^j,\ve x_k^j,\ve u_k^j}{\forall (j,k)\in \ve I}}
% \min_{\stackrel{\ve y_{k+1}^j,\ve x_{k+1}^j,\ve u_k^j}{\forall (j,k)\in \ve I}}
% \;\;\;\tilde{\mathcal J}(\ve y_{k+1}^j,\ve u_k^j)
\min_{\stackrel{\ve z_0^1,\ve v_k^j, y_k^j,\,\forall (j,k)\in \ve I_{\ind{0}{N_p-1}}}{\alpha_k^j,\gamma_k^j,\eta_k^j,\mu_k^j \forall (j,k)\in I_{\ind{N_r}{N_p-1}}}}&\sum_{k=0}^{N_r-1} \sum_{j=1}^{n_ {d}^k}\omega_k^j  \ell(\ve z_{k}^j,\ve v_k^j)+\sum_{k=N_r}^{N_p-1}\sum_{j=1}^{n_{d}^{N_r}}\omega_k^j \gamma_k^j
\end{align}
subject to: \\\hspace*{4cm}\eqref{eq_sys1m}, \eqref{eq_nc1m}, \eqref{init_termm}, \eqref{eq_homo_ineq1}$-$\eqref{homo_ineq4}, \eqref{state_bound2}$-$\eqref{obj_combi2}, $T z_{N_r}^j\leq\alpha_{N_r}^j\textbf{1},\,\forall(j,N_r)\in I_{N_r}$.
}

As in the general complexity tube case, the non-negative matrices $P_i,\forall i\in\Gamma_p,\,P_x,\,P_u$ are obtained offline as defined in \eqref{eq_Farka1}, \eqref{Farka2_eq}, \eqref{Farka3_eq}.  The set of all $x\in\mathbb{X}$ for which there exists a feasible solution for the optimization problem \eqref{MS_implement} is denoted as $\mathtt{X}_{N_p}^H$. The value function obtained by solving the optimization problem \eqref{MS_implement2} at every time step is denoted as $V_{N_p}^{H*}(x)$.
\begin{remark}
	The homothetic tubes can also be implemented using the vertices of the tube as proposed in \cite{rakovic2012homothetic}. However, the formulation proposed here using the approach in \cite{Fleming2015} can be advantageous in terms of reduced computational complexity for a small conservatism. For example,  a low complexity tube for a system of $n_x$ dimensions require $2n_x$ inequalities only, while $2^{n_x}$ vertices are required to represent the same tube (an exponential increase against a linear increase). By restricting the vertices to get a tight upper bound on the stage cost and not computing the reachable sets, applications to high dimensional systems can be achieved. This is the motivation for the proposed formulation of the tube-enhanced multi-stage MPC.
\end{remark}

\bluec{\subsection{Low complexity tube}
To reduce the computational complexity of the scheme, low complexity tubes can be employed. A polytopic tube for the low complexity tube for a given $(j,k)\in I_{\ind{N_r}{N_p-1}}$ can be defined as follows:
\begin{align}\label{eq_low_tube}
\mathsf{Z}_k^j&=\{z|\underline{\tau}_k^j\leq \bar{T}z\leq \overline{\tau}_k^j\},
\end{align}
where $\bar{T}\in\mathbb{R}^{n_x\times n_x}$ and $\underline{\tau}_k^j,\overline{\tau}_k^j\in\mathbb{R}^{n_x}$. Notice that the matrix $\bar{T}$ is a square matrix with $n_x$ rows and columns. Because of the reduced number of inequalities describing the tube, the complexity can be reduced significantly. In addition, there exists efficient implementations of the tubes (See \cite{lee2000robust,kouvaritakis2015model}) to improve the online computational time. The low complexity tubes can, however, lead to conservatism of the resulting scheme. In the proposed formulation, because of the presence of more degrees of freedom, the conservatism of low complexity tubes can be mitigated to a certain extent while reducing the computational time. A more detailed discussion is presented in Section~\ref{sec:Results}.
\subsection{Computation of feedback gains and invariant sets}\label{sec_gain}
The feedback gains  $K_{\mathrm{inv}}$  and $K_{\mathrm{pred}}$ should be chosen offline such that the joint spectral radii of the systems $\{(A_i+B_iK_{\mathrm{inv}}),i\in\Gamma_{p}\}$  and $\{(A_i+B_iK_{\mathrm{pred}}),i\in\Gamma_{p}\}$  are in the unit circle of the complex plane.  They can be chosen such that they optimize a performance measure: for example, the feedback gain $K_{\mathrm{inv}}$  can be chosen such that the volume of the invariant tube $\mathsf{S}$ is as small as possible and the prediction gain $K_{\mathrm{pred}}$ can be chosen as the optimal gain for the linear quadratic regulator either in the nominal  or in the worst-case. As we point out in Remark 2,  instead of choosing a prediction gain, one can choose different gains for different scenarios. However, a systematic way to obtain multiple prediction gains is an issue for further investigations.  If $N_r=N_p$, the terminal gain $K_f$ can be chosen to maximize the volume of the terminal set  to improve the volume of the feasible domain for a chosen prediction horizon $N_p$.

 The implementation of the terminal ingredients is simplified and the complexity is reduced by using \eqref{eq_term_MS} and \eqref{eq_term_MS2} because these formulations can be implemented directly in the optimization problem without computing additional sets. As pointed in \cite{Fleming2015, kouvaritakis2015model}, this simplifies the implementation but can  lead to a decrease in the feasible domain (for a fixed prediction horizon). In \cite{Fleming2015,kouvaritakis2015model}, mode 2 dynamics are introduced to find a trade-off between improved feasible domain and computational complexity. Another option  is to choose the terminal set  as large as possible as described in \cite{kouvaritakis2015model} (pp. 216-219). Since the proposed scheme offers additional feed-forward terms, the proposed implementation of the terminal set is not as restrictive as in the standard tube-based MPC schemes. The disturbance invariant set $\mathsf{S}$ can be obtained by solving the optimization problem defined in \eqref{LP2}.
}}
\section{Recursive feasibility and stability properties}\label{sec_theory}
We formulate the fundamental assumptions to establish the theoretical properties of the proposed scheme as follows.

\begin{assumption}\label{as:1}
A convex compact disturbance invariant polytopic set $\ve{\mathsf S}$  is available for the system~\eqref{lnom} if $\underline{\mathsf{W}}\neq \{0\}$, where $\ve{\mathsf S}\subset \mathbb{X}$ and  $ \ve K_{\mathrm{inv}}\ve{\mathsf S}\subset\mathbb{U}$. If $\underline{\mathsf{W}}=\{0\}$, $\mathsf{S}=\{0\}$.
\end{assumption}
\bluec{Assumption~\ref{as:1} is required to obtain a non-empty feasible domain. The proposed formulation~\eqref{mpc_generalm} offers flexibility to decompose the given uncertainty set $\mathsf{W}$. Only the disturbances considered in $\underline{\mathsf{W}}$ is used to build the set $\mathsf{S}$. Hence, satisfying the assumptions is always possible because $ \underline{\mathsf{W}}=\{0\}$ is always a possible choice. If $ \underline{\mathsf{W}}=\{0\}$, there is no tightening of the constraints required, but this can result in increased computational complexity.}
\begin{assumption}\label{as:2}
 An RPI polytopic set $\mathbb{Z}_f\subseteq\mathbb{X}\ominus\ve{\mathsf S}$ that contains the origin is available for the system~\eqref{lnom} for the feedback gain  \greenc{$K_f=K_{\mathrm{pred}} $ that satisfies $\ve K_f \mathbb{Z}_f\subseteq\mathbb{U}\ominus\ve K_{\mathrm{inv}} \ve{\mathsf S}$} such that for all  	\greenc{$z\in\mathsf{Z}\subseteq\mathbb{Z}_f$, $(A_i+B_i K_f )\mathsf{Z}\oplus\overline{W}\subseteq\mathsf{Z}^+\subseteq\mathbb{Z}_f,\,\forall i\in\Gamma_p$} holds, where $\mathsf{Z}$ and $\mathsf{Z}^+$ represent the employed tubes (that over-approximates the reachable set) in the optimization problem \eqref{mpc_generalm}.
\end{assumption}
\bluec{ If $N_r=N_p$, there is no tube-based part in the predictions. Hence it is sufficient that the terminal set is robustly invariant with respect to the system \eqref{lnom} for a terminal feedback gain $K_f$. In this case, the tube $\mathsf{Z}$ in Assumption \ref{as:2} is defined as a singleton $\mathsf{Z}=\{z\}$ and $\mathsf{Z}^+=\{(A_i+B_iKf)z+w_l, i\in\Gamma_p,l\in\Gamma_{\overline{w}}\}$. However , if $N_r<N_p$, the tube-based part of the scheme requires that the set recursion \eqref{eq_set_rec} employed by the tube is also robustly invariant.  I.e., it is also necessary that the terminal set in robust positively invariant with the employed tubes that over-approximate the reachable sets of the system at every time step for the feedback gain $K_{\mathrm{pred}}$.
\begin{remark}
	We propose to keep the terminal feedback gain $K_f$ the same as that of  gain of the predicted  tubes. If  $K_f$ is chosen different from $K_{\mathrm{pred}}$, then the requirement in Assumption~\ref{as:2} should be modified as follows to guarantee recursive feasibility: An RPI polytopic set $\mathbb{Z}_f\subseteq\mathbb{X}\ominus\ve{\mathsf S}$ that contains the origin is available for the system~\eqref{lnom} for the control law $\ve K_{\mathrm{pred}} \mathbb{Z}_f\subseteq\mathbb{U}\ominus\ve K_{\mathrm{inv}} \ve{\mathsf S}$ and $K_f \mathbb{Z}_f\subseteq\mathbb{U}\ominus\ve K_{\mathrm{inv}} \ve{\mathsf S}$ such that for all  	$\mathsf{Z}\subseteq\mathbb{Z}_f$, $(A_i+B_i K_{\mathrm{pred}} )\mathsf{Z}\oplus\overline{W}\subseteq\mathsf{Z}^+\subseteq\mathbb{Z}_f,\,\forall i\in\Gamma_p$ and 	$(A_i+B_i K_f )\mathsf{Z}_f\oplus\overline{W}\subseteq\mathbb{Z}_f,\,\forall i\in\Gamma_p$ hold. This leads to additional complexity in the terminal set and it is not clear if it leads to advantages in terms of performance. Keeping $K_f=K_{\mathrm{pred}}$ simplifies the requirement and is consistent with the tube-based schemes proposed in \cite{Fleming2015,kouvaritakis2015model}.
\end{remark}} 
\begin{assumption}\label{as:3}
	The stage cost $\ell(z,v)$ and the terminal penalty $V_f(z)$ are convex and positive definite functions and satisfy the following relationships:
	\begin{enumerate}
		\item $\ell(z,v)\geq c\vert z\vert_{\mathbb{Z}_f},\,\forall z\in\mathbb{Z}\setminus\mathbb{Z}_f$,    $\ell(z,K_fz)=0,\,\forall z\in\mathbb{Z}_f$, $V_f(z)=0,\,\forall z\in\mathbb{Z}_f$, where $\vert z\vert_{\mathbb{Z}_f}:= \min_{y\in\mathbb{Z}_f} \|z-y\|_p$ and $c$ is a positive constant.
		\item \greenc{$\max_{\tilde{z}\in\mathsf{Z}}\ell(\tilde{z},v+K_{\mathrm{pred}}\tilde{z})\geq c'\vert \tilde{z}\vert_{\mathbb{Z}_f},\,\forall \mathsf{Z}\subseteq\mathbb{Z}\setminus\mathbb{Z}_f$, $\max_{\tilde{z}\in\mathsf{Z}}\ell(\tilde{z},K_f\tilde{z})=0,\,\forall\mathsf{Z}\subseteq\mathbb{Z}_f$, $\max_{\tilde{z}\in\mathsf{Z}} V_f(\tilde{z})=0, \forall \mathsf{Z}\subseteq\mathbb{Z}_f$, where $c'$ is a positive constant.}

	\end{enumerate}
\end{assumption}

\begin{lemma}\label{lem:rf}
Suppose Assumptions~\ref{as:1} and~\ref{as:2} hold and $x\in\mathtt{X}_{N_p}$ such that $\mathtt{P}_{N_p}(\ve x)$~\eqref{mpc_generalm} has a feasible solution, then $\mathtt{P}_{N_p}(\ve x^+)$ is feasible for all $\ve x^+\in \mathrm{conv}(\{\ve A_i\ve x+\ve B_i\ve u\})\oplus\ve{\mathsf W},\,\forall i\in\Gamma_p$ if the control input applied to the system~\eqref{lsys} follows the control policy $\ve u=\ve v+\ve K_{\mathrm{inv}}(\ve x-\ve z)$.
\end{lemma}
\textbf{Proof.} Let the sequence of optimal control inputs obtained by solving the problem~\eqref{mpc_generalm} be defined as
\begin{equation}
\textbf{v*}=\{v_0^{1*},\,\,\dots,\,v_{N_r}^{j*},\,v_{N_r+1}^{j*}+K_{\mathrm{pred}}z_{N_r+1}^{j*},\,\,\dots,\,v_{N_p-1}^{j*}+K_{\mathrm{pred}}z_{N_p-1}^{j*},\, \forall (j,k)\in I_{\ind{0}{N_p-1}}\}. 
\end{equation}
The root node of the scenario tree is a decision variable as defined in~\eqref{final_termm}. Let the optimal value be denoted as $z=z_0^{1*}$. The first element in the input sequence sequence $v=v_0^{1*}$ and the optimal root node $z=z_0^{1*}$ are used in the control law $u=v+K_{\mathrm{inv}}(x-z)$, where $x$ is the  current state of the plant. The input $u$ is then applied to the plant. The plant evolves from the current state $x$ to the state $x^+$ for the applied input $u$ and the realizations of the uncertainties $w \in\mathsf{W}$ and the system matrices $(A,B)\in\mathrm{conv}\{(\{A_i,B_i,\,\forall i\in\Gamma_p\})\}$. $x^+$ satisfies the constraints because, the additive disturbances $w\in\overline{\mathsf{W}}$ and the vertex matrices $\{(A_i,B_i),\,\forall i\in \Gamma_p\}$ are explicitly considered in the scenario tree in the predictions and the invariant set $\mathsf{S}$ accounts for the disturbances $w\in\underline{\mathsf{W}}$. At the next time step, the optimization problem~\eqref{mpc_generalm} is solved again for the realized state $x^+$. Since $\mathsf{S}$ is invariant with respect to the additive disturbances $w\in\underline{\mathsf{W}}$, there exists a $z^+\in\mathrm{conv}(\{z_1^{j*},\,\forall(j,1)\in I_1\})\subseteq\mathbb{Z}=\mathbb{X}\ominus\mathsf{S}$ from Assumption~\ref{as:1}. There exists a feasible input sequence for the optimization problem $\mathtt{P}_{N_p}(\ve x^+)$ that is in the convex hull of inputs predicted in the previous time step for the optimization problem $\mathtt{P}_{N_p}(\ve x)$ for all prediction steps until $N_p-1$. For the last prediction step, there exists a control law $K_fz$ for all $z\in\mathbb{Z}_f$ from Assumption~\ref{as:2}. A feasible input sequence for the next time step can be obtained as the convex combination of the predicted inputs as follows:
\begin{align}\label{conv_input}
\textbf{v}(x^+)&=\bigg\{\sum_{(j,1)\in I_1}\lambda_1^jv_1^{j*},\,\dots,\,\sum_{(j,N_r)\in I_{N_r}}\lambda_{N_r}^jv_{N_r}^{j*}, \sum_{(j,N_r+1)\in I_{N_r+1}}\lambda_{N_r+1}^jv_{N_r+1}^{j*}+K_{\mathrm{pred}}z_{N_r+1}^{j*},\,\dots,\, \sum_{(j,N_p)\in I_{N_p}}\lambda_{N_p}^{j*}K_fz_{N_p}^{j*}\bigg\},
\end{align}
where $\lambda_k^j$ denote the associated convex weights for all $(j,k)\in I$ ($\lambda_k^j\geq 0,$ and $\sum_{j=1}^{n_d}\lambda_k^j=1$). Since there exists a feasible root node $z^+$ and  a feasible input sequence $\textbf{v}(x^+)$, problem  $\mathtt{P}_{N_p}(\ve x^+)$ is feasible for all $\ve x^+\in \{\ve A_i\ve x+\ve B_i\ve u\}\oplus\ve{\mathsf W}$ for all $i\in\Gamma_p$ if $\mathtt{P}_{N_p}(\ve x)$ is feasible.\qed  

We now show that the recursve feasibility property is retained if the convex optimization problems \eqref{MS_implement} and \eqref{MS_implement2} related to the formulation \eqref{mpc_generalm} are solved.

\bluec{\begin{corollary}\label{cor:1}
Suppose Assumptions~\ref{as:1} and~\ref{as:2} hold and $x\in\mathtt{X}_{N_p}$ such that $\mathtt{P}_{N_p}^{\mathrm{G}}(\ve x)$~\eqref{MS_implement} has a feasible solution for the tube $\mathsf{Z}_k^j=\{z|T z\leq \tau_k^j\}, \forall (j,k)\in I_{\ind{N_r}{N_p-1}}$, then $\mathtt{P}_{N_p}^{\mathrm{G}}(\ve x^+)$ is feasible for all $\ve x^+\in \mathrm{conv}(\{\ve A_i\ve x+\ve B_i\ve u\})\oplus\ve{\mathsf W},\,\forall i\in\Gamma_p$ if the control input applied to the system~\eqref{lsys} follows the control policy $\ve u=\ve v+\ve K_{\mathrm{inv}}(\ve x-\ve z)$.
\end{corollary}
\textbf{Proof.}  The constraints of the optimization problems \eqref{mpc_generalm} and \eqref{MS_implement}  can be compared one to one. The constraints \eqref{eq_sys1m}, \eqref{eq_nc1m}, \eqref{init_termm} are retained in optimization problem \eqref{MS_implement}. The remaining constraints are direct results of  Lemma \ref{Farkas} which establishes sufficient conditions for the set recursion and guaranteeing that a set is a subset of another. Hence, the feasibility arguments discussed in Lemma \ref{lem:rf} directly applies to the formulation \eqref{MS_implement}. Hence,   $\mathtt{P}_{N_p}^{\mathrm{G}}(\ve x^+)$ is feasible for all $\ve x^+\in \mathrm{conv}(\{\ve A_i\ve x+\ve B_i\ve u\})\oplus\ve{\mathsf W},\,\forall i\in\Gamma_p$ if the control input applied to the system~\eqref{lsys} follows the control policy $\ve u=\ve v+\ve K_{\mathrm{inv}}(\ve x-\ve z)$.\qed
\begin{corollary}
	Suppose Assumptions~\ref{as:1} and~\ref{as:2} hold and $x\in\mathtt{X}_{N_p}$ such that $\mathtt{P}_{N_p}^{\mathrm{H}}(\ve x)$~\eqref{MS_implement2} has a feasible solution for the tube $\mathsf{Z}_k^j=\{z|T (z-\hat{z}_k^j)\leq \alpha_k^j\textbf{1}\}, \forall (j,k)\in I_{\ind{N_r}{N_p-1}}$, then $\mathtt{P}_{N_p}^{\mathrm{H}}(\ve x^+)$ is feasible for all $\ve x^+\in \mathrm{conv}(\{\ve A_i\ve x+\ve B_i\ve u\})\oplus\ve{\mathsf W},\,\forall i\in\Gamma_p$ if the control input applied to the system~\eqref{lsys} follows the control policy $\ve u=\ve v+\ve K_{\mathrm{inv}}(\ve x-\ve z)$.
\end{corollary}
\textbf{Proof.}  The same arguments in Corollary \ref{cor:1} directly apply here as well. \qed\\\\
}
\begin{lemma}\label{lem:mono}
If Assumptions~\ref{as:weights},~\ref{as:1},~\ref{as:2} and~\ref{as:3} hold, then $V^*_{N_p+1}(\ve x) \le V^*_{N_p}(\ve x),\,\forall \ve x\in\mathtt{X}_{N_p}$, where $V^*_{N_p} (x)$ is the optimal value function of \eqref{mpc_generalm} with the length of the prediction horizon $N_p$.
% RP: Function V^*_{N_p} is not defined.
\end{lemma}
\textbf{Proof.} Since the stage costs $\ell(z,K_fz)=0$  and $\max_{\tilde{z}\in\mathsf{Z}}\ell(z,K_fz)=0$ in the terminal set and the control law $K_fz$ is feasible, the additional prediction step does not add any cost to the optimal value function $V^*_{N_p+1}(\ve x)$ and hence $V^*_{N_p+1}(\ve x) \le V^*_{N_p}(\ve x),\,\forall \ve x\in\mathtt{X}_{N_p}$.\qed 
%\begin{align}
%V^*_{N_p}(\ve x)&=\min_{\ve z_0^1,\ve v_k^j, \ve y_k^j\forall (j,k) \ve I}\sum_{k=0}^{N_p-1}\sum_{j=1}^{p^k}\omega_k^j\mathcal \ell(\ve z_{k}^j,\ve v_k^j)\\
%&=  \mathcal \ell(\ve z_{0}^{1*},\ve v_0^{1*})+\sum_{j=1}^{p}\mathcal \ell(\ve z_{1}^{j*},\ve v_1^{j*})+\cdots\nonumber\\&+\sum_{j=1}^{p^{N_p-1}} \mathcal \ell(\ve z_{N_p-1}^{j*},\ve v_{N_p-1}^{j*})\label{eq_mono1}
%\end{align}
%To obtain $V^*_{N_p+1}(\ve x)$, we can retain the optimal decision variables from~\eqref{eq_mono1} until $N_p$ and for the extra time step in the prediction, linear control law $\ve K \ve{z}_{N_p}^{j*}$ can be applied. This choice helps us to obtain a feasible sequence of control inputs and to compare the optimal value functions. The choice of the control inputs is not necessarily optimal but feasible. The value function  $V_{N_p+1}(\ve x)$ can be obtained as follows:
%\begin{align}\label{eq_mono2}
%V_{N_p+1}(\ve x)&=V_{N_p}^*+\sum_{j=1}^{p^{N_p}} \mathcal \ell(\ve z_{N_p}^{j*},\ve K \ve z_{N_p}^{j*})
%\end{align}
%Since $z_{N_p}^{j*}\in\mathbb{Z}_f$, $\ell(\ve z_{N_p}^{j*},\ve K \ve z_{N_p}^{j*})=0$. The optimal value function satisfies the property $V^*_{N_p+1}(\ve x)\le V_{N_p+1}(\ve x)$. From \eqref{eq_mono2}, it follows that
%$V^*_{N_p+1}(\ve x)\leq V^*_{N_p}(\ve x),\, \forall x\in\mathtt{X}_{N_p}$.
%\qed
\begin{lemma}\label{lem:stab}
If Assumptions~\ref{as:weights},~\ref{as:1},~\ref{as:2} and~\ref{as:3} hold and $\mathtt{X}_{N_p}$ is compact, then the optimal
value function fulfills the following properties
\begin{align}
V_{N_p}^*(\ve x)	&=0,\,\forall \ve x \in \mathbb{Z}_f\oplus\ve{\mathsf{S}},\label{stable4}\\
V_{N_p}^*(\ve x)&\ge c_1\vert z\vert_{\mathbb{Z}_f},\,\forall \ve x \in \mathtt{X}_{N_p},\label{stable1}\\
V_{N_p}^*(\ve x^+)&\le V_{N_p}^*(\ve x)- c_1\vert z\vert_{\mathbb{Z}_f},\,\forall \ve x \in \mathtt{X}_{N_p},\label{stable3}
\end{align}
where $c_1$ is a positive constant.
\end{lemma}
\textbf{Proof.} Since $\ve x \in \mathbb{Z}_f\oplus\ve{\mathsf{S}}$ implies that $\ve z_0^1 (\ve x)\in\mathbb{Z}_f$ is a feasible point, the stage cost $\ell(z,K_fz)=0$ from Assumption~\ref{as:3} and the control law $K_fz$ keeps the state $z$ in the terminal set $\mathbb{Z}_f$ from Assumption~\ref{as:2}. Since $V_f(z)=0$ and all the stage costs are zero $V_{N_p}^*(\ve x)=0$ for all $x\in\mathbb{Z}_f\oplus\mathsf{S}$. This proves~\eqref{stable4}. 

From Assumption~\ref{as:3}, $\ell(z,v)\ge k\vert z\vert_{\mathbb{Z}_f}$ and the optimal value function satisfies the property $V_{N_p}^*(\ve x)\ge \omega_0^1 \ell(z_0^{1*},v_0^{1*})$. Since $\omega_0^1 \ge 0$, we have $V_{N_p}^*(\ve x)\ge c_1\vert z\vert_{\mathbb{Z}_f}$ for all $z\in\mathbb{Z}\setminus\mathbb{Z}_f$, where $c_1=k\omega_0^1$ a positive constant. Note that $V_{N_p}^*(\ve x)	=0$ for all $x\in\mathbb{Z}_f\oplus\mathsf{S}$ and $\vert z\vert_{\mathbb{Z}_f}=0$ for all $z\in\mathbb{Z}_f$. Hence $V_{N_p}^*(\ve x)\ge c_1\vert z\vert_{\mathbb{Z}_f}$ for all $\ve x \in\ \mathtt{X}_{N_p}$.

In the following, the descent property of the optimal value function will be proven. This property establishes the optimal value function as a Lyapunov function and is an important contribution of this paper.
\begin{figure}
\begin{center}
\includegraphics[width=0.8\columnwidth]{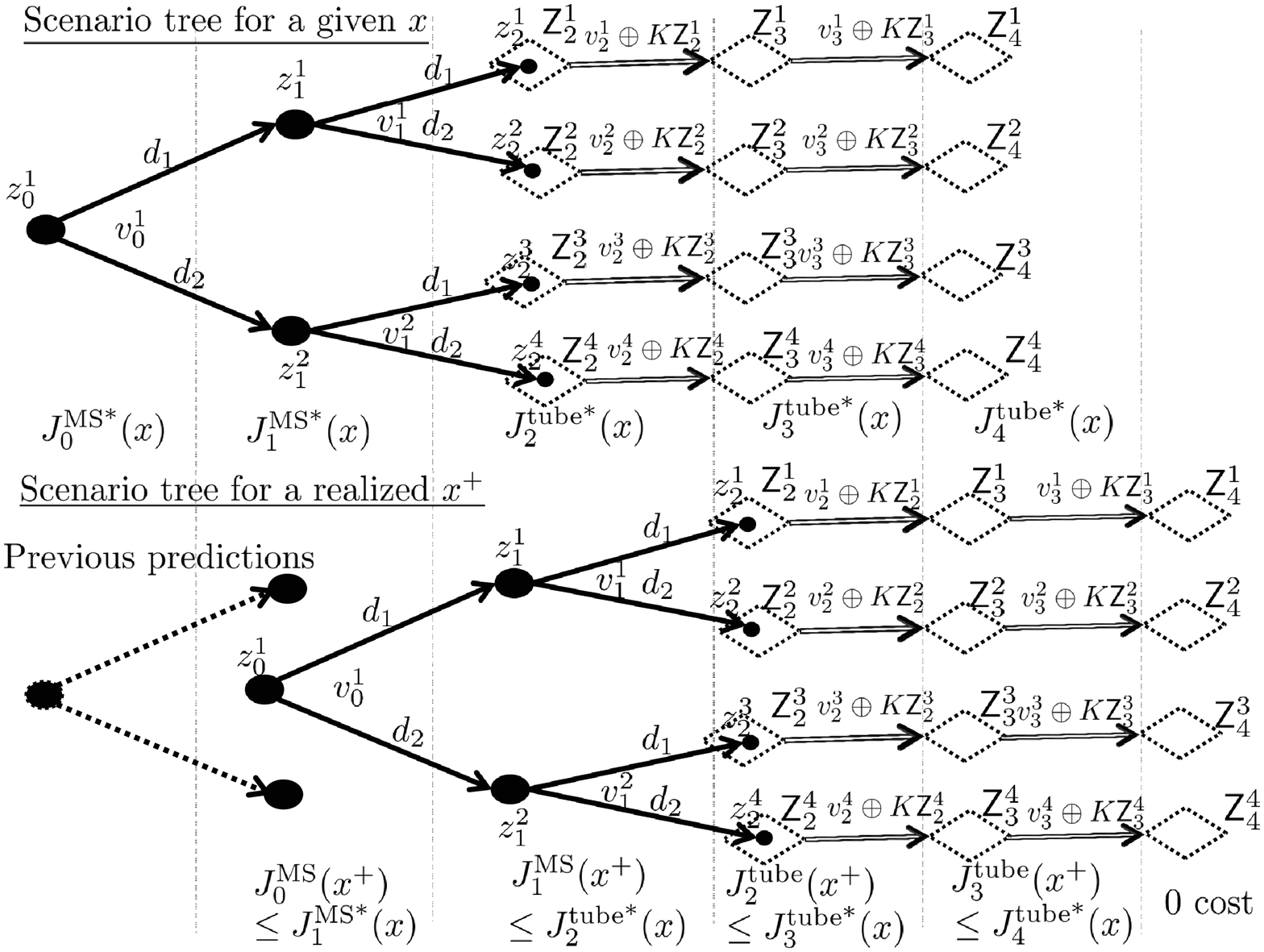}
\caption{Illustration of the descent property of the proposed scheme.}  % width is 8.4 cm.

\label{tree_RH_stability}                                 % Size the figures
\end{center}                                 % accordingly.
\end{figure}

% SL [IMPORTANT] : You are not referring to the figure within the text? It should be referred in the proof...
% SL: In the tube-part of the figure, what does actually v_2^1 \oplus KZ_2^1 mean along the branch? the meaning of that is for me not very clear. It looks like you are propagating sets but you are not really propagating them, isnt it? 
%SS: They are tubes for the inputs. It is a function of z and the tube predicted at the next stage is a function of KZ.

% To simplify the description of the proof, we consider three possible choices of the robust horizon $N_r$. 
%\begin{enumerate}
%\item $N_r=0\rightarrow$ The scheme simplifies to a tube-based MPC scheme
%\item $N_r=N_p-1\rightarrow$ The scheme simplifies to a multi-stage MPC scheme
%\item $0<N_r<N_p-1\rightarrow$ The proposed scheme
%\end{enumerate}
 As shown in Lemma 1, the initial state $\ve z_{0}^{1*}(\ve x)$ and
the control policy given in~\eqref{conv_input} are feasible for all $x \in\mathtt{X}_{N_p}$. Since $\ve z^+\in\mathrm{conv}(\{z_{1}^{j*},\,\forall(j,1)\in\ve I\})$, the optimal value function $V_{N_p}^*(x)$ and a feasible value function at the next time step $(V_{N_p}(\ve x^+))$ can be written as:
\begin{subequations}\label{desc0}
\begin{align}
&V_{N_p}^*(x)=\sum_{k=0}^{N_r-1} J^{\mathrm{MS*}}_k(x)+ J^{\mathrm{tube*}}_{N_r}(x)+\sum_{k=N_r+1}^{N_p-1} J^{\mathrm{tube*}}_k(x),\\
&V_{N_p}(x^+)=\sum_{k=0}^{N_r-1} J^{\mathrm{MS}}_k(x^+)+ \sum_{k=N_r}^{N_p-1} J^{\mathrm{tube}}_k(x^+).
\end{align}
\end{subequations}
The proof will be accomplished in two parts. Initially the multi-stage part of the value function $V_{N_p}(x^+)$ will be compared to the optimal value function $V_{N_p}^*(x)$ obtained one step before. Because of the receding horizon nature of the MPC, the comparison of the multi-stage part of  $V_{N_p}(x^+)$ will be performed with the multi-stage part and the one-step tube-based part ($J^{\mathrm{MS*}}_k(x)+ J^{\mathrm{tube*}}_{N_r}(x)$) of the optimal value function $V_{N_p}^*(x)$. It will be shown that $\sum_{k=0}^{N_r-1} J^{\mathrm{MS}}_k(x^+)\leq\sum_{k=0}^{N_r-1} J^{\mathrm{MS*}}_k(x)+ J^{\mathrm{tube*}}_{N_r}(x)-c_1\vert z\vert_{\mathbb{Z}_f}$. The multi-stage part of the value function  $\sum_{k=0}^{N_r-1} J^{\mathrm{MS}}_k(x^+)$ can be written as (refer to Figure~\ref{tree_RH_stability}):
\begin{align}\label{desc1}
\sum_{k=0}^{N_r-1} J^{\mathrm{MS}}_k(x^+)=&\, \omega_0^1  \ell\bigg(\sum_{j=1}^{n_{d}}\lambda_1^j z_{1}^{j*},\sum_{j=1}^{n_{d}}\lambda_1^j v_1^{j*}\bigg)+\cdots+\sum_{j=1}^{n_{d}^{N_r-2}}\omega_{N_r-2}^j\,\, \ell\bigg( \sum_{j=1}^{n_{d}}\lambda_{N_r-1}^j z_{N_r}^{j*}, \sum_{j=1}^{n_{d}}\lambda_{N_r-1}^j v_{N_r}^{j*}\bigg)+\nonumber\\&+\sum_{j=1}^{n_{d}^{N_r-1}}\omega_{N_r-1}^j \,\, \ell\bigg( \sum_{j=1}^{n_{d}}\lambda_{N_r}^j z_{N_r+1}^{j+}, \sum_{j=1}^{n_{d}}\lambda_{N_r}^j (v_{N_r}^{j*}+K_{\mathrm{pred}} z_{N_r}^{j+})\bigg),
\end{align}
where $z_{N_r}^{j+}\in\mathrm{conv}(\{\mathsf{Z}_{N_r}^j(x)\},\,\forall (j,N_r)\in I_{N_r})$.
%\begin{align}\label{desc1}
%&V_{N_p}(x^+)= \omega_0^1  L\bigg(\sum_{j=1}^{p}\lambda_1^j z_{1}^{j*}(x),\sum_{j=1}^{p}\lambda_1^j v_1^{j*}(x)\bigg)+\cdots\nonumber\\&+\sum_{j=1}^{N_r-1}\omega_{N_r}^j  L\bigg( \sum_{j=1}^{p}\lambda_{N_r}^j z_{N_r}^{j*}(x), \sum_{j=1}^{p}\lambda_{N_r}^j v_{N_r}^{j*}+\bigg)+\cdots\nonumber\\&+\sum_{j=1}^{N_r}\omega_{N_r+1}^j  L\bigg( \sum_{j=1}^{p}\lambda_{N_r}^j z_{N_r+1}^{j+}(x), \sum_{j=1}^{p}\lambda_{N_r}^j v_{N_r}^{j*}+Kz_{N_r+1}^{j+}\bigg)&+\max_{\tilde{z}_{N_r}^j,\forall(j,N_r)\in I}\sum_{j=1}^{p^{N_r}} L\bigg( \sum_{j=1}^{p}\lambda_{N_r}^j z_{N_r+1}^{j*}(x), \sum_{j=1}^{p}\lambda_{N_r}^j v_{N_r}^{j*}\bigg)\sum_{j=1}^{p^{N_p-1}}\omega_{N_p-1}^j  L\bigg( \sum_{j=1}^{p}\lambda_{N_p}^j z_{N_p}^{j*}(x), \sum_{j=1}^{p}\lambda_{N_p}^j  K z_{N_p}^{j*}\bigg)
%\end{align}
Since the proposed stage cost is convex from Assumption~\ref{as:3}, the following inequality holds for the stage cost of the root node of the tree at $\ve x^+$.	
\begin{align}\label{desc2}
  \ell\bigg(\sum_{j=1}^{n_{d}}\lambda_1^j z_{1}^{j*},\sum_{j=1}^{n_{d}}\lambda_1^j v_1^{j*}\bigg)&\le \sum_{j=1}^{n_{d}}\lambda_1^j \ell(\ve z_{1}^{j*},\ve v_1^{j*}).
\end{align}
Multiplying both sides with $\omega_0^1$, we get
\begin{align}\label{desc3}
 \omega_0^1 \ell\bigg(\sum_{j=1}^{n_{d}}\lambda_1^j z_{1}^{j*},\sum_{j=1}^{n_{d}}\lambda_1^j v_1^{j*}\bigg)&\le \omega_0^1\sum_{j=1}^{n_{d}}\lambda_1^j \ell(\ve z_{1}^{j*},\ve v_1^{j*}).
\end{align}
From Assumption~\ref{as:weights}, $\omega_0^1\le\min\{\omega_1,\,\dots,\,\omega_p\}$, so $\omega_0^1\sum_{j=1}^{n_{d}}\lambda_1^j \ell(\ve z_{1}^{j*},\ve v_1^{j*})\le \sum_{j=1}^{n_{d}}\lambda_1^j\omega_1^j \ell(\ve z_{1}^{j*},\ve v_1^{j*})$ and $\sum_{j=1}^{n_{d}}\lambda_1^j=1$, which results in:
\begin{align}\label{desc4}
 \omega_0^1 \ell\bigg(\sum_{j=1}^{n_{d}}\lambda_1^j z_{1}^{j*},\sum_{j=1}^{n_{d}}\lambda_1^j v_1^{j*}\bigg)&\le\sum_{j=1}^{n_{d}}\omega_1^j\lambda_1^j \ell(\ve z_{1}^{j*},\ve v_1^{j*}).
\end{align}
Similarly the stage costs for the prediction steps until $N_r-2$ of the value function $V_{N_p}(x^+)$  can be compared for all $(j,k)\in \ve I_{\ind{0}{N_r-2}}$ by grouping the nodes that are established for a particular realization of the uncertainty $i\in\Gamma_p$ and $l\in\Gamma_{\overline{w}}$ together. Note that, if $N_r=0$, there is no multi-stage part in the objective. If $N_r=1$ only the root node belongs to the multi-stage part and it will be compared with the tube-based part (first-stage) of the optimal value function in the previous time-step $V_{N_p}^*(x)$ as will be discussed later.  If $N_r=2$, the root node will be compared to the multi-stage part as in~\eqref{desc4} and for the comparison of the further stages onwards one can skip to~\eqref{desc7}. The following relationship for the case is discussed whenever $N_r>2$ until $N_r-2$ of the value function $V_{N_p}(x^+)$. 
%
%\begin{align}\label{desc5}
%  \ell\bigg(\sum_{j=1}^{n_{d}}\lambda_k^j z_{k}^{j*},\sum_{j=1}^{n_{d}}\lambda_k^j v_k^{j*}\bigg)&\le \sum_{j=1}^{n_{d}}\lambda_k^j \ell(\ve z_{k}^{j*},\ve v_k^{j*}).
%\end{align}
Since the weights $\omega_k^j$ associated with each node $z_k^j$ are the same as the weights associated with a realized uncertainty until $N_r-1$ of the problem $\mathtt{P}_{N_p}(x^+)$, we have
\begin{align}\label{desce6}
 \omega_k^j \ell\bigg(\sum_{j=1}^{n_{d}}\lambda_k^j z_{k}^{j*},\sum_{j=1}^{n_{d}}\lambda_k^j v_k^{j*}\bigg)&\le \sum_{j=1}^{n_{d}}\omega_{k+1}^j\lambda_k^j \ell(\ve z_{k}^{j*},\ve v_k^{j*}).
\end{align}
The stage cost at the prediction step $N_r-1$ of the problem  $\mathtt{P}_{N_p}(x^+)$  can be compared with the tube-based part of the optimal value function $V_{N_p}^*(x)$. Again using convexity, the following relationship can be established.
\begin{align}\label{desc7}
  \ell\bigg( \sum_{j=1}^{n_{d}}\lambda_{N_r}^j z_{N_r}^{j+}, &\sum_{j=1}^{n_{d}}\lambda_{N_r}^j (v_{N_r}^{j*}+K_{\mathrm{pred}}z_{N_r}^{j+})\bigg)\leq \ell(\tilde{z}_{N_r}^{j*},v_k^{j*}+K_{\mathrm{pred}}\tilde{z}_{N_r}^{j*}),
\end{align}
where $z_{N_r}^{j+}$ belongs to the convex hull of the tubes $\mathsf{Z}_{N_r}^{j*}$ predicted in the problem $\mathtt{P}_{N_p}(x)$. Note that the stage cost of the tube-based part of the optimal value function $V_{N_p}^*(x)$ contains the maximization objective, the optimal states $\tilde{z}_{N_r}^{j*}$ give the worst-case cost for all $(j,N_r)\in I_{N_r}$. Since the weight associated with $J^{\mathrm{tube*}}_{N_r}(x)$  is greater than the weights associated with all the realizations of the uncertainty from Assumption~\ref{as:weights}, we have
\begin{align}\label{desc8}
\omega^j_{N_r-1}& \ell\bigg( \sum_{j=1}^{n_{d}}\lambda_{N_r}^j z_{N_r}^{j+}, \sum_{j=1}^{n_{d}}\lambda_{N_r}^j (v_{N_r}^{j*}+K_{\mathrm{pred}}z_{N_r}^{j+})\bigg) \leq\sum_{j=1}^{n_{d}} \omega_{\mathrm{tube}}\lambda_{N_r}^j\ell(\tilde{z}_{N_r}^{j*},v_k^{j*}+K_{\mathrm{pred}}\tilde{z}_{N_r}^{j*}),
\end{align}
Substituting \eqref{desc4},~\eqref{desce6},~\eqref{desc8} in \eqref{desc1}, we get 
\begin{align}\label{desc9a}
\sum_{k=0}^{N_r-1} J^{\mathrm{MS}}_k(x^+)&\leq \sum_{k=1}^{N_r-1} J^{\mathrm{MS}}_k(x)+J^{\mathrm{tube*}}_{N_r}(x),\\
&\leq \sum_{k=0}^{N_r-1} J^{\mathrm{MS}}_k(x)+J^{\mathrm{tube*}}_{N_r}(x)-\omega_0^1{\ell}(\ve z_{0}^{1*},\ve v_0^{1*}).\label{desc9}
\end{align}
The tube-based part of the optimal value function $V_{N_p}(x^+)$ can be compared with the tube-based part of $V_{N_p}^*(x)$. Since the predicted tubes of the problem $\mathtt{P}_{N_p}(x^+)$ from the prediction step $N_r$ until $N_p-2$ belong to the convex hull of $\mathtt{P}_{N_p}(x)$, we have
\begin{align}
\sum_{k=N_r}^{N_p-2} J^{\mathrm{tube}}_k(x^+)\leq\sum_{k=N_r+1}^{N_p-1} J^{\mathrm{tube*}}_k(x).
\end{align}
The stage costs at the prediction step $N_p-1$ are $ \ell\bigg( \sum_{j=1}^{n_d}\lambda_{N_p}^j \tilde{z}_{N_p}^{j*}(x), \,\sum_{j=1}^{n_d}\lambda_{N_p}^j  K_f \tilde{z}_{N_p}^{j*}\bigg) =0$ by definition. Hence  the following relationship holds:
\begin{align}\label{desc10}
\sum_{k=N_r}^{N_p-1} J^{\mathrm{tube}}_k(x^+)\leq\sum_{k=N_r+1}^{N_p-1} J^{\mathrm{tube*}}_k(x).
\end{align}
  From~\eqref{desc9} and~\eqref{desc10}, the value function $V_{N_p}(x^+)$ can then be written in terms of $V_{N_p}(x)$ as follows:
\begin{align}
V_{N_p}(x^+)\le V_{N_p}^*(x)- \omega_0^1{\ell}(\ve z_{0}^{1*},\ve v_0^{1*}) .
\end{align}
Since $V^*_{N_p}(x^+)\le V_{N_p}(x^+)$, we have
\begin{align}
V_{N_p}^*(\ve x^+)&\le V_{N_p}^*(\ve x)-\omega_0^1{\ell}(\ve z_{0}^{1*},\ve v_0^{1*}) ,\\
 V_{N_p}^*(\ve x^+)&\le V_{N_p}^*(\ve x)-c_1\vert z\vert_{\mathbb{Z}_f},\forall \ve x \in \mathtt{X}_{N_p}.
\end{align}
This proves~\eqref{stable3} and with this Lemma 3 is established.\qed
%
%In Lemma~\ref{lem:stab}, the optimal value function is established as a Lyapunov function and this implies that the origin is exponentially stable for the system~\eqref{lnom}. In
%the presence of additive disturbances, we state in the following theorem that the set
%$\ve{\mathsf S}$ is robustly exponentially stable for the proposed methodology. 
\begin{remark}
The proof of descent~\eqref{stable3} is valid for all values of robust horizon in the range $N_r\in [0,N_p]$. If $N_r=0$, then the scheme is simplified into a tube-based MPC scheme enhanced by an invariant tube. If $N_r=N_p$, the scheme simplifies into a multi-stage MPC scheme enhanced by an invariant tube. The proof was established for a generic case where both the multi-stage and tube components are present. If one of the components is absent, it can be shown that the proof still holds by removing the corresponding elements in the proof.
\end{remark}
\bluec{\begin{lemma}\label{lem:stab_1}
If Assumptions~\ref{as:weights},~\ref{as:1},~\ref{as:2} and~\ref{as:3} hold and $\mathtt{X}_{N_p}^G$ is compact, the optimal value function $V_{N_p}^{G*}(\ve x)$ of the optimization problem \eqref{MS_implement} fulfills the following properties for the system \eqref{lsys}
	\begin{align}
	V_{N_p}^{G*}(\ve x)	&=0,\,\forall \ve x \in \mathbb{Z}_f\oplus\ve{\mathsf{S}},\label{stable4_1}\\
	V_{N_p}^{G*}(\ve x)&\ge c_2\vert z\vert_{\mathbb{Z}_f},\,\forall \ve x \in \mathtt{X}_{N_p}^G,\label{stable1_1}\\
	V_{N_p}^{G*}(\ve x^+)&\le V_{N_p}^{G*}(\ve x)- c_2\vert z\vert_{\mathbb{Z}_f},\,\forall \ve x \in \mathtt{X}_{N_p}^G,\label{stable_1}
	\end{align}
	where $c_2$ is a positive constant.
\end{lemma}
\textbf{Proof.}  Because of the choice of the stage costs, the proofs are \eqref{stable4_1} and \eqref{stable1_1} are  similar to  Lemma \ref{lem:stab}. In here, we show that the reformulation \eqref{MS_implement} retains these properties.

To prove \eqref{stable4_1}, it is sufficient  to show that the stage costs are $0$ for a feasible solution for all $x\in\mathbb{Z}_f\oplus\mathsf{S}$. For all   $x\in\mathbb{Z}_f\oplus\mathsf{S}$, it can be seen that $z\in\mathbb{Z}_f$ is a feasible point. The terminal control law $K_fz$ keeps the trajectory of the primary controller in the terminal set $\mathbb{Z}_f$. The stage costs associated with the multi-stage part of the scheme are $0$. I.e., $\ell(z_k^j,v_k^j)=0$ for all $z_k^j\in\mathbb{Z}_f$, for all $(j,k)\in I_{\ind{0}{N_r-1}}$. For the tube-based part of the scheme, the stage cost is reformulated as in \eqref{obj_Fark}$-$\eqref{obj_Fark4}. Equivalently, the stage cost can be represented as
\begin{subequations}\label{eq_imp_1}
\begin{align}
\max_{\tilde{z}_k^j\in\mathsf{Z}_k^j}\ell(\tilde z_k^j,v_k^j)=&\min_{y_k^j\in\mathbb{Z}_f,\eta_k^j,\mu_k^j,\gamma_k^j,\forall (j,k)\in I_{\ind{N_r}{N_p-1}}.} \gamma_k^j\\
&\text{subject to:} \nonumber\\
&-\mu_k^j\leq ( P_Q\tau_k^j-Qy_k^j)\leq \mu_k^j, (j,k)\in I_{\ind{N_r}{N_p-1}},\label{const_1}\\
& -\eta_k^j\leq Rv_k^j\leq \eta_k^j, (j,k)\in I_{\ind{N_r}{N_p-1}},\label{const_2}\\
& \textbf{1}^T\mu_k^j+\textbf{1}^T\eta_k^j\leq \gamma_k^j,(j,k)\in I_{\ind{N_r}{N_p-1}}.\label{const_3}
%&\vert\mu_k^j\vert+\vert\eta_k^j\vert\leq \gamma_k^j,(j,k)\in I_{\ind{N_r}{N_p-1}}.
\end{align}
\end{subequations}
For all $x\in\mathbb{Z}_f\oplus \mathsf{S}$ and $N_r=0$,  $\mathsf{Z}_{0}^j=\mathbb{Z}_f$ is a feasible set for all $(j,0)\in I_0$. If $N_r\geq 1$, the terminal control law keeps the state in the terminal set and hence $z_{N_r}^j\in\mathbb{Z}_f,\,\forall(j,N_r)\in I_{N_r}$. Hence,  $\mathsf{Z}_{N_r}^j=\mathbb{Z}_f$ is a feasible set for all $(j,N_r)\in I_{N_r}$. Since the set $\mathbb{Z}_f$ is robustly positive invariant for the control law $K_f \mathsf Z_k^j$, $v_k^j=0$ is a feasible control law for the tube-based part of the scheme. Substituting $v_k^j=0$, and $Q=P_QT$ from \eqref{obj_Fark}, the constraints \eqref{const_1} and \eqref{const_2} can be rewritten as follows:
\begin{subequations}
\begin{align}
&-\mu_k^j\leq P_Q(\tau_k^j-Ty_k^j)\leq \mu_k^j, (j,k)\in I_{\ind{N_r}{N_p-1}},\label{const_11}\\
 &-\eta_k^j\leq Rv_k^j\leq \eta_k^j, (j,k)\in I_{\ind{N_r}{N_p-1}},\label{const_12}
\end{align}
\end{subequations}

% \begin{subequations}\label{eq_imp_2}
%	\begin{align}
%\max_{\tilde{z}_k^j\in\mathsf{Z}_k^j}	\ell(\tilde z_k^j,v_k^j)=&\min_{y_k^j\in\mathbb{Z}_f,\eta_k^j,\mu_k^j,\gamma_k^j, (j,k)\in I_{\ind{N_r}{N_p-1}}} \gamma_k^j\\
%	&\text{subject to} \nonumber\\
%	&\vert P_Q(\tau_k^j-Ty_k^j)\vert\leq \mu_k^j, (j,k)\in I_{\ind{N_r}{N_p-1}},\\
%	&\vert 0\vert\leq \eta_k^j, (j,k)\in I_{\ind{N_r}{N_p-1}},\\
%	& \textbf{1}^T\mu_k^j+\textbf{1}^T\eta_k^j\leq \gamma_k^j, (j,k) \in I_{\ind{N_r}{N_p-1}},
%	\end{align}
%\end{subequations}
Since $\mathsf{Z}_k^j=\mathbb{Z}_f$ is a feasible solution, there exists a feasible  solution $y_k^j=z_k^j$   such that $Tz_k^j=\tau_k^j$  holds for any $(j,k)\in I_{\ind{N_r}{N_p-1}}$. Hence, $\mu_k^j=\eta_k^j=\gamma_k^j=0$ is a feasible solution for all $(j,k)\in I_{\ind{N_r}{N_p-1}}$, if $\mathsf{Z}_k^j\subseteq\mathbb{Z}_f$.  This implies that the stage costs  remain $0$ for all $x\in\mathbb{Z}_f\oplus\mathsf{S}$ for the tube-based part of the scheme in addition to the multi-stage part of the scheme for the formulation \eqref{MS_implement}. Hence 	$V_{N_p}^{G*}(\ve x)	=0,\,\forall \ve x \in \mathbb{Z}_f\oplus\ve{\mathsf{S}}$.

The proof of \eqref{stable1_1} is straight forward as shown in Lemma \ref{lem:stab}. Because of the choice of $Q$ and $R$ matrices as positive definite and that the tube-based part of the scheme over-approximates the cost, $V_{N_p}^{G*}(\ve x)\ge c_2\vert z\vert_{\mathbb{Z}_f},\,\forall \ve x \in \mathtt{X}_{N_p}^G$, where $c_2\geq c_1$.

To prove \eqref{stable_1},  the multi-stage part of the scheme follows the same arguments given in the proof of Lemma \ref{lem:stab}.  For the tube-based part, because of the reformulation \eqref{eq_obj_reformulate1} and choosing the non-negative matrices offline, we can only establish a sufficient condition online. Hence $J_k^{\mathrm{tube}*}$ solved using \eqref{MS_implement} will always over-approximate the true solution obtained using the formulation \eqref{mpc_generalm}.  From this, we see that the following inequality holds:
\begin{align}\label{desc_G_1}
\sum_{k=0}^{N_r-1} J^{\mathrm{MS}}_k(x^+)&\leq \sum_{k=0}^{N_r-1} J^{\mathrm{MS}}_k(x)+J^{\mathrm{tube*}}_{N_r}(x)-\omega_0^1{\ell}(\ve z_{0}^{1*},\ve v_0^{1*}).
\end{align}
To compare the tube-based part of the scheme, we must establish the following inequality:
\begin{align}
\sum_{k=N_r}^{N_p-1} J^{\mathrm{tube}}_k(x^+)\leq\sum_{k=N_r+1}^{N_p-1} J^{\mathrm{tube*}}_k(x).
\end{align}
for all $(j,k)\in I_{\ind{N_r}{N_p-1}}$.  To compare the components of the stage costs at consecutive time steps, let us look at the constraint \eqref{const_1} in the optimization problem \eqref{eq_imp_1}. Since the predicted tubes at the next time step can be given as the convex combination of the tubes predicted at the current time step, the following holds  for all $(j,k)\in I_{\ind{N_r}{N_p-2}}$. 
\begin{subequations}\label{first_part}
\begin{align}
\vert P_Q \tau_k^j(x^+)-Qy_k^j(x^+)\vert&=\vert P_Q\sum_{j=1}^{n_d^{N_r}}\lambda_{k+1}^j \tau_{k+1}^{j*}(x)-Q\sum_{j=1}^{n_d^{N_r}}\lambda_{k+1}^j y_{k+1}^{j*}(x)\vert,\label{eq_1st}\\
&\leq \sum_{j=1}^{n_d^{N_r}}\lambda_{k+1}^j (\vert P_Q \tau_{k+1}^{j*}(x)-Qy_{k+1}^{j*}(x)\vert),\label{eq_2nd}\\
&\leq \sum_{j=1}^{n_d^{N_r}}\lambda_{k+1}^j (\vert \mu_{k+1}^{j*}(x)\vert),\label{eq_3rd}
\end{align}
\end{subequations}
where $\lambda_k^j$ denotes the convex weights associated with the tubes for all $(j,k)\in I_{\ind{N_r}{N_p-1}}$. Minkowski's inequality leads to  \eqref{eq_2nd}  from \eqref{eq_1st} and \eqref{eq_3rd} follows from \eqref{const_1}. Similarly, the following holds for the constraint \eqref{const_2} in the optimization problem \eqref{eq_imp_1} for all $(j,k)\in I_{\ind{N_r}{N_p-2}}$. 
\begin{subequations}\label{sec_part}
	\begin{align}
\vert Rv_k^j(x^+)\vert &=\vert R\sum_{j=1}^{n_d^{N_r}}\lambda_{k+1}^j v_{k+1}^{j*}(x)\vert\label{eq_1st2}\\
	&\leq \sum_{j=1}^{n_d^{N_r}}\lambda_{k+1}^j (\vert Rv_{k+1}^{j*}(x)\vert),\label{eq_2nd2}\\
	&\leq \sum_{j=1}^{n_d^{N_r}}\lambda_{k+1}^j (\vert \eta_{k+1}^{j*}(x)\vert),\label{eq_3rd2}
	\end{align}
\end{subequations}
Substituting \eqref{first_part} and \eqref{sec_part} in \eqref{const_3}, there exist a feasible $\gamma_k^j(x^+)$  for all $(j,k)\in I_{\ind{N_r}{N_p-2}}$ such that the following holds:
\begin{align}\label{eq_2nd1}
	\gamma_k^j(x^+)\leq\sum_{j=1}^{n_d^{N_r}}\lambda_{k+1}^j\gamma_{k+1}^{j*}(x)
\end{align} 
%\begin{align}
%\vert P_Q \tau_k^j(x^+)-Qy_k^j(x^+)\vert+\vert Rv_k^j(x^+)\vert &=(\vert P_Q\sum_{j=1}^{n_d^{N_r}}\lambda_k^j \tau_k^{j*}(x)-Q\sum_{j=1}^{n_d^{N_r}}\lambda_k^j y_k^{j*}(x)\vert+\vert R\sum_{j=1}^{n_d^{N_r}}\lambda_k^j v_k^{j*}(x)\vert),\label{eq_1st}\\
%&\leq \sum_{j=1}^{n_d^{N_r}}\lambda_k^j (\vert P_Q \tau_k^{j*}(x)-Qy_k^{j*}(x)\vert+\vert Rv_k^{j*}(x)\vert),\label{eq_2nd}
%\end{align}
Combining \eqref{eq_2nd1} with Assumption \ref{as:weights}, we have
\begin{align}
\sum_{j=1}^{n_d^{N_r}}\omega_k^j\gamma_k^j(x^+)\leq\sum_{j=1}^{n_d^{N_r}}\omega_{k+1}^j\sum_{j=1}^{n_d^{N_r}}\lambda_{k+1}^j\gamma_{k+1}^{j*}(x)
\end{align} 
for all $k\in\{N_r,\,\dots,\, N_p-2\}$. As shown in \eqref{eq_imp_1}, the stage costs remain zero for the additional step. Summing up across the horizon for the tube-based part of the scheme, we get, 
\begin{align}
\sum_{k=N_r}^{N_p-1} J^{\mathrm{tube}}_k(x^+)\leq\sum_{k=N_r+1}^{N_p-1} J^{\mathrm{tube*}}_k(x).
\end{align}
This leads to the condition 	$V_{N_p}^{G*}(\ve x^+)\le V_{N_p}^{G*}(\ve x)- c_2\vert z\vert_{\mathbb{Z}_f},\,\forall \ve x \in \mathtt{X}_{N_p}^G$. This proves \eqref{stable_1}.\qed
%	\mathsf{Z}_k^j=\sum_{i=1}^{n_d^N_r}
%\end{align}
\begin{lemma}\label{lem:stab_2}
If Assumptions~\ref{as:weights},~\ref{as:1},~\ref{as:2} and~\ref{as:3} hold and $\mathtt{X}_{N_p}^H$ is compact, the optimal value function  $V_{N_p}^{H*}(\ve x)$ fulfills the following properties
	\begin{align}
	V_{N_p}^{H*}(\ve x)	&=0,\,\forall \ve x \in \mathbb{Z}_f\oplus\ve{\mathsf{S}},\label{stable5_1}\\
	V_{N_p}^{H*}(\ve x)&\ge c_3\vert z\vert_{\mathbb{Z}_f},\,\forall \ve x \in \mathtt{X}_{N_p}^H,\label{stable5_2}\\
	V_{N_p}^{H*}(\ve x^+)&\le V_{N_p}^{H*}(\ve x)- c_3\vert z\vert_{\mathbb{Z}_f},\,\forall \ve x \in \mathtt{X}_{N_p}^H,\label{stable5_3}
	\end{align}
	where $c_3$ is a positive constant.
\end{lemma}
\textbf{Proof.}  
First, we prove \eqref{stable5_1}. The multi-stage part of the scheme inherits the same properties discussed in Lemma \ref{lem:stab} and \ref{lem:stab_1} and the stage cost $\ell(z_k^j,v_k^j)=0,\,\forall (j,K)\in I_{\ind{0}{N_r-1}}$, if $x\in\mathbb{Z}_f\oplus\mathsf{S}$ .  If $N_r=0$, for all $x\in\mathbb{Z}_f\oplus\mathsf{S}$, $\mathsf{Z}_0^j=\mathbb{Z}_f$ is a feasible set for all $(j,0)\in I_0$. Also, if $N_r\geq 1$, if $z_{N_r}^j\in\mathbb{Z}_f, \,\mathsf{Z}_{N_r}^j=\mathbb{Z}_f$ is a feasible set for each $(j,N_r)\in I_{N_r}$. Since the set $\mathbb{Z}_f$ is robustly positive invariant for the control law $K_{\mathrm{pred}}\mathsf Z_k^j$ , $v_k^j=0$ is a feasible control law for the tube-based part of the scheme. All the vertex points of the tube $\hat{z}_k^j+\alpha \vee_r$  is contained in the set $\mathbb{Z}_f$. For each $\hat{z}_k^j+\alpha \vee_r$, there exists a feasible $y_k^{j,r}\in\mathbb{Z}_f$. Hence, the upper bound of   $\vert Q(\hat{z}_k^j+\alpha \vee_r-y_k^{j,r})\vert$ is $0$ in \eqref{objective2}. Also $\vert Rv_k^j\vert\leq 0$ is a feasible solution. Hence, $\gamma_k^j=0, (j,k)\in I_{\ind{N_r}{N_p-1}}$ is feasible in \eqref{objective2}. Since the optimal cost is smaller than or equal to the feasible value, the upper bound of the stage cost is $0$. From the definition of the terminal set $(A+BK_{\mathrm{pred}})\mathbb{Z}_f\subseteq\mathbb{Z}_f$, implying $\mathsf{Z}_{k+1}^j\subseteq\mathsf{Z}_k^j$. This leads to the case where the stage cost remains $0$ until $N_p$.  Hence if $x\in\mathbb{Z}_f\oplus\mathsf{S}$, $	V_{N_p}^{H*}(\ve x)	=0$. 

The proof of \eqref{stable5_2} follows directly from the choice of the stage cost.  Since $Q$ and $R$ are positive definite, \eqref{stable5_2} holds.

To prove \eqref{stable5_3}, the proof of the multi-stage part of the scheme follows the same arguments in Lemma \ref{lem:stab}. Hence the following inequality holds:
\begin{align}\label{desc_H_1}
\sum_{k=0}^{N_r-1} J^{\mathrm{MS}}_k(x^+)&\leq \sum_{k=0}^{N_r-1} J^{\mathrm{MS}}_k(x)+J^{\mathrm{tube*}}_{N_r}(x)-\omega_0^1{\ell}(\ve z_{0}^{1*},\ve v_0^{1*}).
\end{align}
For the tube-based part of the scheme, the tubes at the next step can be represented as the convex combination of tubes predicted at the previous time step. Hence the following relationship holds $(j,k)\in I_{\ind{N_r}{N_p-2}}$. 
\begin{subequations}\label{1part}
\begin{align}
\vert Q (\hat{z}_k^j (x^+)+\alpha_k^j \vee_r-y_k^{j,r}(x^+))\vert &=\vert Q \sum_{j=1}^{n_d^{N_r}}\lambda_{k+1}^j \hat{z}_{k+1}^{j*} (x)+\sum_{j=1}^{n_d^{N_r}}\lambda_{k+1}^j \alpha_{k+1}^{j*} \vee_r-\sum_{j=1}^{n_d^{N_r}}\lambda_{k+1}^j y_{k+1}^{j,r*}(x)\vert, \, r\in\{1,\dots,\,n_v\},\\
&\leq \sum_{j=1}^{n_d^{N_r}}\lambda_{k+1}^j\vert Q ( \hat{z}_{k+1}^{j*} (x)+\alpha_{k+1}^{j*} \vee_r-y_{k+1}^{j,r*}(x)\vert) , \, r\in\{1,\dots,\,n_v\},\\
&\leq \sum_{j=1}^{n_d^{N_r}}\lambda_{k+1}^j \vert\mu_{k+1}^{j*}\vert,\, \forall  r\in\{1,\dots,\,n_v\},
\end{align}
\end{subequations}
here $\lambda_k^j$ denotes the convex weights associated with the predicted tubes for all $(j,k)\in I_{\ind{N_r}{N_p-1}}$. Following the same arguments for the inputs, we arrive at the following inequality
\begin{align}\label{2part}
\vert Rv_k^j(x^+)\vert \leq \sum_{j=1}^{n_d^{N_r}}\lambda_{k+1}^j \vert\eta_{k+1}^{j*}\vert,
\end{align}
for all $(j,k)\in I_{\ind{N_r}{N_p-2}}$. Combining \eqref{1part} and \eqref{2part}  in \eqref{obj_combi2}, we can see that there exists a feasible $\gamma_k^j(x^+)$ for all $(j,k)\in I_{\ind{N_r}{N_p-2}}$ such that the following inequality holds as in Lemma~\ref{lem:stab_1}:
\begin{align}\label{eq_2nd12}
\gamma_k^j(x^+)\leq\sum_{j=1}^{n_d^{N_r}}\lambda_{k+1}^j\gamma_{k+1}^{j*}(x)
\end{align} 
%\begin{align}
%\vert P_Q \tau_k^j(x^+)-Qy_k^j(x^+)\vert+\vert Rv_k^j(x^+)\vert &=(\vert P_Q\sum_{j=1}^{n_d^{N_r}}\lambda_k^j \tau_k^{j*}(x)-Q\sum_{j=1}^{n_d^{N_r}}\lambda_k^j y_k^{j*}(x)\vert+\vert R\sum_{j=1}^{n_d^{N_r}}\lambda_k^j v_k^{j*}(x)\vert),\label{eq_1st}\\
%&\leq \sum_{j=1}^{n_d^{N_r}}\lambda_k^j (\vert P_Q \tau_k^{j*}(x)-Qy_k^{j*}(x)\vert+\vert Rv_k^{j*}(x)\vert),\label{eq_2nd}
%\end{align}
Combining \eqref{eq_2nd12} with Assumption \ref{as:weights}, we have
\begin{align}
\sum_{j=1}^{n_d^{N_r}}\omega_k^j\gamma_k^j(x^+)\leq\sum_{j=1}^{n_d^{N_r}}\omega_{k+1}^j\sum_{j=1}^{n_d^{N_r}}\lambda_{k+1}^j\gamma_{k+1}^{j*}(x)
\end{align} 
%\begin{align}
%\sum_{j=1}^{n_d^{N_r}}\omega_k^j	\max_{r\in\{1,\,\dots,\,n_v\}}(\vert Q (\hat{z}_k^j (x^+)&+\alpha_k^j V_r-y_k^{j,r}(x^+))\vert) +\vert R v_k^j(x)\vert =\nonumber\\
%&\leq \sum_{j=1}^{n_d^{N_r}}\omega_{k+1}^j\max_{r\in\{1,\,\dots,\,n_v\}}\sum_{j=1}^{n_d^{N_r}}\lambda_k^j(\vert Q ( \hat{z}_k^j (x)+\alpha_k^j V_r-y_k^{j,r}(x)\vert) +\vert R v_k^j(x)\vert),
%\end{align} 
%for all $k\in\{N_r,\,\dots,\, N_p-1\}$. 
As shown in \eqref{stable5_1}, the stage costs remain zero for the additional step, because it is contained in the terminal set. Summing up across the horizon for the tube-based part of the scheme, we get, 
\begin{align}
\sum_{k=N_r}^{N_p-1} J^{\mathrm{tube}}_k(x^+)\leq\sum_{k=N_r+1}^{N_p-1} J^{\mathrm{tube*}}_k(x).
\end{align}
This leads to the condition 	$V_{N_p}^{H*}(\ve x^+)\le V_{N_p}^{H*}(\ve x)- c_3\vert z\vert_{\mathbb{Z}_f},\,\forall \ve x \in \mathtt{X}_{N_p}^H$. This proves \eqref{stable5_3}.\qed
%\begin{subequations}\label{stage_cost_H}
%	\begin{align}
%	\max_{\tilde{z}_k^j\in\mathsf{Z}_k^j}&=\min_{\alpha_k^j,\gamma_k^j,\eta_k^j,\mu_k^j \forall (j,k\geq N_r)\in I} \gamma_k^j
%	\end{align}
%	subject to
%	\begin{align}
%	&-\mu_k^j\leq Q(\hat{z}_k^j+\alpha \vee_r-y_k^{j,r})\leq\mu_k^j, && \forall r\in\{1,\,\dots,\,n_v\}, (j,k\geq N_r)\in I,\\
%	&-\eta_k^j\leq Rv_k^j\leq\eta_k^j ,&&\forall(j,k\geq N_r)\in I, \\
%	& y_k^{j,r}\in\mathbb{Z}_f,&&\forall r\in\{1,\,\dots,\,n_v\}, \,(j,k\geq N_r)\in I,\\
%	&\textbf{1}^T\mu_k^j+\textbf{1}^T\eta_k^j\leq \gamma_k^j&&(j,k\geq N_r)\in I.
%	\end{align}
%\end{subequations}
}

The terminal set $\mathbb{Z}_f\oplus\mathsf{S}$  is asymptotically stable for the proposed scheme as it is shown in the following theorem. 
%SL. Here repeat what the advantage of the proposed method with respect to the one you cite. That is, the good thing is that we retain the same convergence properties but can handle something more (parametric uncertainties). isnt it?
%SS: tick
\begin{Theorem}
Suppose the assumptions \ref{as:weights}-\ref{as:3} are satisfied, then the set $\mathbb{Z}_f\oplus\mathsf{S}$ is robustly asymptotically stable for the controlled uncertain system given in~\eqref{lsys} using the proposed scheme~\eqref{mpc_generalm} \bluec{ with the implementation \eqref{MS_implement} or  \eqref{MS_implement2}.}
\end{Theorem}
%\textbf{Proof.} From~\eqref{stable1},~\eqref{stable2} and~\eqref{stable3}, we have,
%\begin{align}
%V_{N_p}^*(\ve x_0)&\le c_2 \vert z_0^{0*}(\ve x_0)\vert^2\label{step0}\\
%V_{N_p}^*(\ve x_{1})&\le c_2 \vert z_0^{0*}(\ve x_{1})\vert^2\label{step1}
%\end{align}
%SL: Is this really c_1? In the conditions the inequalities are in the other direction
%SS: Correct
\textbf{Proof.}
Since the optimal value function is established as a Lyapunov function in Lemma~\ref{lem:stab}, \bluec{ Lemma~\ref{lem:stab_1}, and Lemma~\ref{lem:stab_2} } with respect to the terminal set $\mathbb{Z}_f$, the state $z$ of \eqref{lnom} converges to the terminal set asymptotically. The state of the system \eqref{lsys} satisfies the property $x\in\{z\}\oplus\mathsf{S}$ and converges robustly asymptotically to the set $\mathbb{Z}_f\oplus\mathsf{S}$. \qed

Finite time reachability of the terminal set and robust asymptotic stability of the minimal RPI set can be proven for the proposed scheme using a dual mode control policy as proposed in~\cite{kerrigan2004feedback}. The required conditions are formalized in the following assumption.
\begin{assumption}\label{as:4}
There exists a dual mode control policy that is employed as follows:
\begin{align}\label{eq_policy}
\ve u(\ve x)=\begin{cases}
   \ve K_f\ve x, & \text{if}\, x\in\mathbb{X}_{max},\\
   \ve v+\ve K_{\mathrm{inv}}(\ve x-\ve z), & \text{otherwise},
 \end{cases}
\end{align}
where $\mathbb{X}_{max}$ is an RPI set for the asymptotically stabilizing control law $u=K_fx$ for the system~\eqref{lsys} that satisfies the conditions $\mathbb{Z}_f \oplus \mathsf{S}\subset \mathbb{X}_{max}$ and $\mathbb{X}_{max}\subseteq\mathbb{X}$  and $v$ and $z$ are optimal solutions of the proposed scheme~\eqref{mpc_generalm} with an asymptotically stabilizing feedback gain $K_{\mathrm{inv}}$. 
\end{assumption}
\begin{Theorem}
Suppose the assumptions \ref{as:weights}-\ref{as:4} are satisfied, the minimal RPI set $\mathsf{S}_{min}$ of the system~\eqref{lsys} is robustly asymptotically stable for the controlled uncertain system defined in~\eqref{lsys} employed using the dual mode control policy~\eqref{eq_policy}.
\end{Theorem}
\textbf{Proof.}
As $\mathbb{Z}_f\oplus \mathsf{S}\subset \mathbb{X}_{max}$ and $\mathbb{Z}_f\oplus \mathsf{S}$ is robustly asymptotically stable from Theorem 1, the state enters $ \mathbb{X}_{max}$ in finitely many time steps. Since the control policy is switched to $K_f x$ when $x\in\mathbb{X}_{{max}}$  and that the control law $K_fx$ is asymptotically stabilizing for the system \eqref{lsys}, the state $x$ converges to the minimal RPI set asymptotically. Hence for the uncertain system defined in~\eqref{lsys} controlled using the dual mode control policy~\eqref{eq_policy}, the minimal RPI set $\mathsf{S}_{min}$ is robustly asymptotically stable.\qed
\begin{corollary}
Suppose the assumptions \ref{as:weights}-\ref{as:4} are satisfied, the set $\mathbb{Z}_f\oplus\mathsf{S}$ can be reached in finite time steps if $\mathsf{S}_{min}\subset \mathbb{Z}_f\oplus\mathsf{S}$ holds.
\end{corollary}
\textbf{Proof.} This follows directly from the proof of Theorem 2. If the minimal RPI set $\mathsf{S}_{min}$ is contained in $\mathbb{Z}_f\oplus\mathsf{S}$ and $\mathsf{S}_{min}$ is robustly asymptotically stable, the state reaches $\mathbb{Z}_f\oplus\mathsf{S}$ in finite time steps.\qed

In the proposed approach, a multi-stage MPC solution is computed on the scenario tree for the large uncertainties with recourse, i.e. a tree of future inputs depends on the realization of the uncertainty. The affine feedback is added ``on top'' to robustify the solution against the small disturbances. The feedback gain $\ve K_{\mathrm{inv}}$ is fixed only for small disturbances and the degrees of freedom are increased using the multi-stage approach for large uncertainties resulting in an improved trade-off between optimality and complexity. Also, if the robust horizon is chosen as $N_r\geq 1$, we have different feed-forward terms at each stage in the predictions. This results in a scheme with the following advantages when compared to multi-stage MPC and tube-based MPC independently:
\begin{enumerate}
\item The growth in problem complexity is reduced when compared to a pure multi-stage approach because the small uncertainties are not considered in the scenario tree.
\item The structurally relaxed recourse which is modeled in the prediction for the realizations of the large uncertainties until robust horizon reduces the conservatism compared to pure tube-based MPC.
\item The choice of a robust horizon on the one hand limits the rapid growth of the scenario tree and on the other hand, provides increased degrees of freedom when compared to a standard tube-based MPC resulting in an improved trade-off.
\item The use of low complexity tubes is possible for less conservatism because of increased degrees of freedom in the form of feed-forward terms beyond the robust horizon. This enables the application of the approach to high dimensional systems.
\end{enumerate} 

\section{Case study}\label{sec:Results}
The example considered in this paper is a continuous stirred-tank reactor (CSTR) with a reaction scheme that is adapted from~\cite{klatt1998}. Two chemical reactions take place in the reactor:
\begin{align*}
\mathrm{A}~&\to\,\mathrm{B}~\to\,\mathrm{C}\\
2\mathrm{A}~&\to\,\mathrm{D}
\end{align*}
The linearized discrete-time model has the form given in~\eqref{lsys}, where $\ve x=[\Delta C_\mathrm{a},\,\Delta C_\mathrm{b},\,\Delta T_\mathrm{R},\Delta T_\mathrm{J}]^T$ and $\ve u=\Delta F$. $\Delta C_\mathrm{a}$ and $\Delta C_\mathrm{b}$ denotes the deviations of concentration of component A and B in mol/l , $\Delta {T}_{\mathrm{R}}$ and $\Delta {T}_{\mathrm{J}}$ denote the change in the temperature of the reactor and in the jacket temperature in $^\circ$C with respect to the equilibrium point. The input is the deviation of the feed from the equilibrium in l/h.   The uncertain model is given as: $$\ve A_{\mathrm{unc}}(d_1,d_2,d_3,d_4)= \begin{pmatrix}0.3+d_1&-0.09&-0.01&0\\0.2&0.29+d_2&0.002&0\\d_3&d_4&1.10&0.15\\0.05&0.07&0.13&0.68\end{pmatrix}.$$ The system vertex matrices are
\begin{align*}
\ve A_1&=\ve A_{\mathrm{unc}}(0.1,0.1,0.33,0.26),\\
\ve A_2&=\ve A_{\mathrm{unc}}(-0.1,-0.1,-0.33,-0.26),\\
\ve A_3&=\ve A_{\mathrm{unc}}(0.1,-0.1,0.33,-0.26),\\
\ve A_4&=\ve A_{\mathrm{unc}}(-0.1,0.1,-0.33,0.26).
\end{align*}
The input vertex matrices are $\ve B_1=\ve B_2=\ve B_3=\ve B_4=[0.1, -0.05, 0.8,  0.1]^T$. The additive uncertainty bounds on all states are $\ve w\in\ve{\mathsf W}:=\{\ve w|\,\vert\vert w\vert\vert_{\infty}\le 0.1\}$.  The state constraints are $\ve x\in\mathbb{X}:=\{x|[-5,-5,-3,-5]^T\le x\le[5,5,3,5]^T\}$ and the input bounds are given by $\ve u\in\mathbb{U}:=\{\ve u|\,\vert u\vert\le 2\}$. The control task is to take the system to a bounded set around the origin while respecting the state and input constraints. The $\ve{Q}$ matrix of the stage cost is chosen as an identity matrix and the $\ve{R}$ matrix is chosen as $0.01$. The length of the prediction horizon of is chosen as $N_p=5$.
\subsection{Details of the simulation study}
The tube-based part of the proposed scheme was implemented with three different types of tubes and different simulation studies were performed. The types of the tubes studied are given in Table~\ref{tab:schemes}.
\begin{table}
\caption{Details of the different types of tubes studied in this work.}
\label{tab:schemes}
\begin{center}
\begin{tabular}{|p{2.0cm}|p{3.0cm}|p{3cm}|p{1.5cm}|p{1.5cm}|}
\toprule
\textbf{Tube type}&\textbf{Implemented Optimization problem}& \textbf{Tube complexity}&\textbf{Recursive feasibility} & \textbf{Stability}\\
\midrule
%& \textbf{NMPC} & \textbf{NMPC} & \textbf{NMPC}\\
\midrule
 general complexity tube tube &\eqref{MS_implement} & 18 inequalities and 44 vertices& YES & YES\\
\midrule
 Homothetic tube&  \eqref{MS_implement2}& 18 inequalities and 44 vertices&\ YES &YES\\
\midrule
  Low complexity tube& \eqref{MS_implement2} with $8$ row $T$ matirx & 8 inequalities and 16 vertices&YES&YES\\
\midrule
\end{tabular}
\end{center}
\end{table}
\iffalse
\begin{comment}
\begin{enumerate}
\item general complexity tube tube
\begin{itemize}
\item The right hand side of the inequality $\tau_k^j$ associated with the tube represented in \eqref{eq_tube_ineq} are degrees of freedom in the optimization problem.
\item The $\lambda-$contractive set was obtained for the value of $\lambda=0.68$. It was chosen close to spectral radius to maximize the performance.
\item The resulting $\lambda-$contractive set is characterized by $18$ inequalities and $44$ vertices.
\item The implemented scheme provides recursive feasibility guarantees (and not stability guarantees but convergence was observed for this example).
\end{itemize}
\item  Homothetic tube
\begin{itemize}

\item Here $\tau_k^j$ is restricted by the equality  $\tau_k^j=\alpha_k^j\textbf{1}$ to characterize the tube represented in \eqref{eq_tube_ineq} and $\alpha_k^j$ are degrees for freedom in the optimization problem. 
\item The same $\lambda-$contractive as in the case of general complexity tube tube was employed.
\item The implemented scheme provides recursive feasibility and robust asymptotic stability guarantees.
    
\end{itemize}
\item Low complexity tube
\begin{itemize}
\item The tube was also implemented as  homothetic.
\item The $\lambda-$contractive set was chosen with low complexity and was characterized by $8$ inequalities and $16$ vertices.
\item The scheme provides recursive feasibility and robust asymptotic stability guarantees and is less complex compared to the implemented homothetic tube.
\end{itemize}
\end{enumerate}
\end{comment}
\fi
The additive disturbances are considered as small disturbances and an offline invariant tube $\mathsf{S}$ was obtained. The set $\ve{\mathsf S}$ is obtained by using \eqref{LP2} using the $\lambda-$contractive set for the value of $\lambda=0.68$ obtained for the system \eqref{lsys} and it is contained in the box given by $\ve{\mathsf S}\subseteq\{x|[-0.4088 , -0.5670,-0.3936,-0.3518]^T\le x\le [0.4088 , 0.5670,0.3936,0.3518]^T\}$ for the LQ-optimal feedback gain $\ve K_{\mathrm{inv}}=\ve K_{\mathrm{pred}}=\ve K_f=K=[-0.0493,-0.0004,-1.3330,-0.3485]$.  In Table~\ref{tab:performance}, the volumes and the computation times of the proposed approach for different robust horizons and the tube-based MPC approach are given. In the following, we investigate all aspects of the proposed approach.
\begin{table}
\caption{Quantitative comparison of feasible domains obtained using the proposed tube-enhanced multi-stage (TEMS) MPC scheme for varying robust horizons and the tube-based MPC.}
\label{tab:performance}
\begin{center}
\begin{tabular}{|p{1.2cm}|p{1.0cm}|p{2cm}|p{1cm}|p{1cm}|p{1cm}|p{1cm}|p{1cm}|p{1cm}|}
\toprule
\textbf{Tube type}&\textbf{Para-meters}&\textbf{Tube MPC (without the invariant tube)} & \textbf{TEMS MPC with $N_r=0$}& \textbf{TEMS MPC with $N_r=1$}& \textbf{TEMS MPC with $N_r=2$}& \textbf{TEMS MPC with $N_r=3$}& \textbf{TEMS MPC with $N_r=4$}& \textbf{TEMS MPC with $N_r=5$}\\
\midrule
%& \textbf{NMPC} & \textbf{NMPC} & \textbf{NMPC}\\
\midrule
& Volume &1197.1&1110.7&4007.6&4392.7&4570.9&4574.6&4574.6\\
\multirow{-2}{1cm}{General complexity}&Comp. time [s]&6.55&0.15&0.45&1.27&3.54&8.65&1.2\\
\midrule
& Volume &1065.2&1001.0&3820.3&4319.5&4536.6&4574.2&$\quad$as\\
\multirow{-2}{1cm}{Homo- thetic}&Comp. time [s]&5.04&0.20&0.73&2.3&7.03&14.18&above\\
\midrule
 & Volume &96.02&96.02&1415.9&3563.2&4411.3&4545.9&$\quad$as\\
\multirow{-2}{1cm}{Low complexity} &Comp. time [s]&1.1&0.07&0.28&0.82&2.8&6.5&above\\
\midrule
\end{tabular}
\end{center}
\end{table}
\subsection{The effect of the invariant tube}
If the pure multi-stage MPC is applied, it  gives rise to 64 branches per node in the scenario tree resulting in more than 100 million scenarios. The additive disturbances therefore are removed from the multi-stage part and are formulated in the (invariant) tube-based part of the scheme. Hence, $\overline{\mathsf{W}}=\{0\}$ and $\underline{\mathsf{W}}=\mathsf{W}$ for the studied example. This results in four branches at every node which is a dramatic reduction when compared to the $64$ branches required in the case of a full scenario tree for all the uncertainties.
 
\begin{figure}%
    \centering
    \subfloat[ Projection on $\Delta C_a$, $\Delta C_b$ and $\Delta T_{\mathrm{R}}$.]{{\includegraphics[width=7cm]{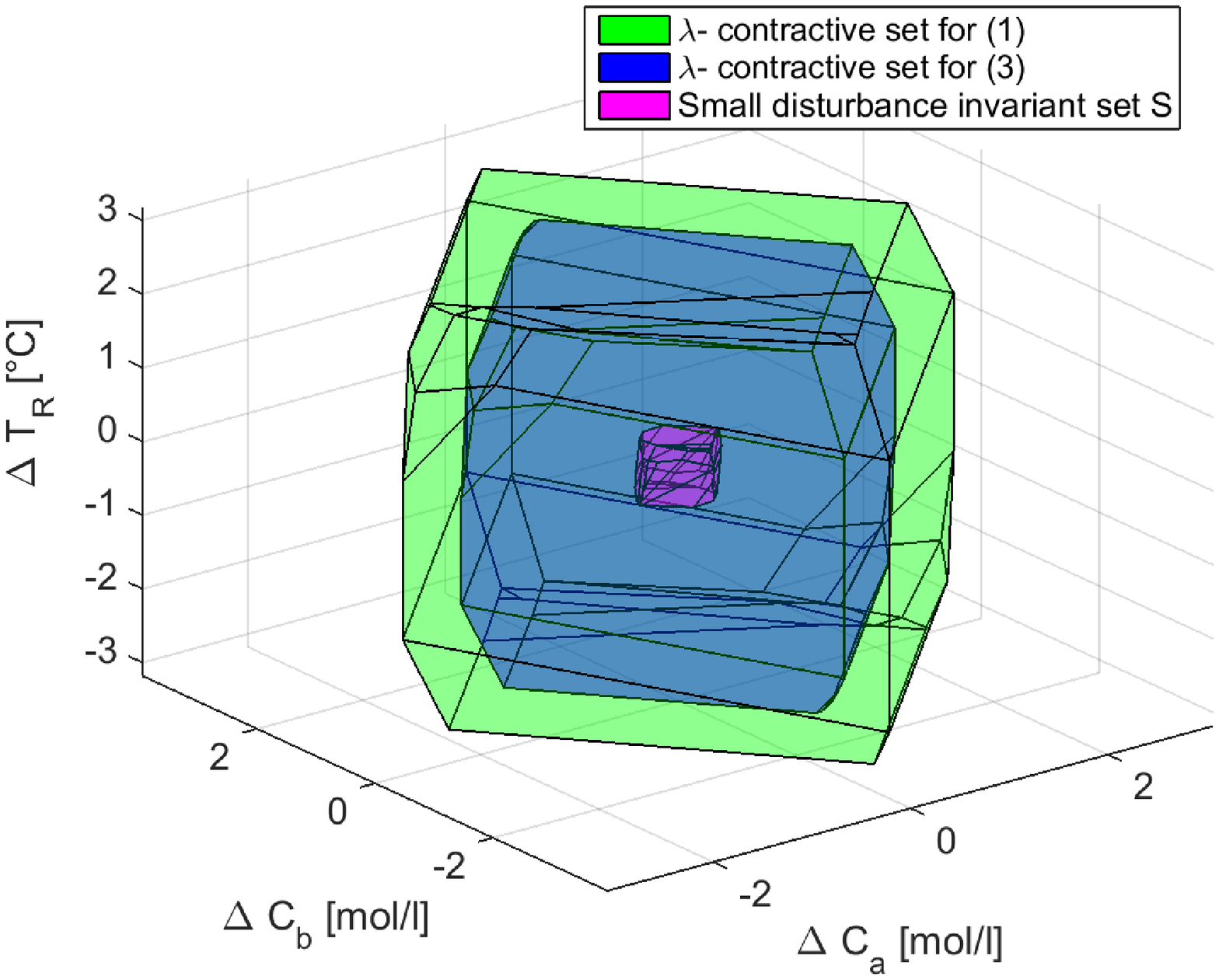}}}%
    \subfloat[ Projection on $\Delta C_b$, $\Delta T_{\mathrm{R}}$ and $\Delta T_{\mathrm{J}}$.]{{\includegraphics[width=7cm]{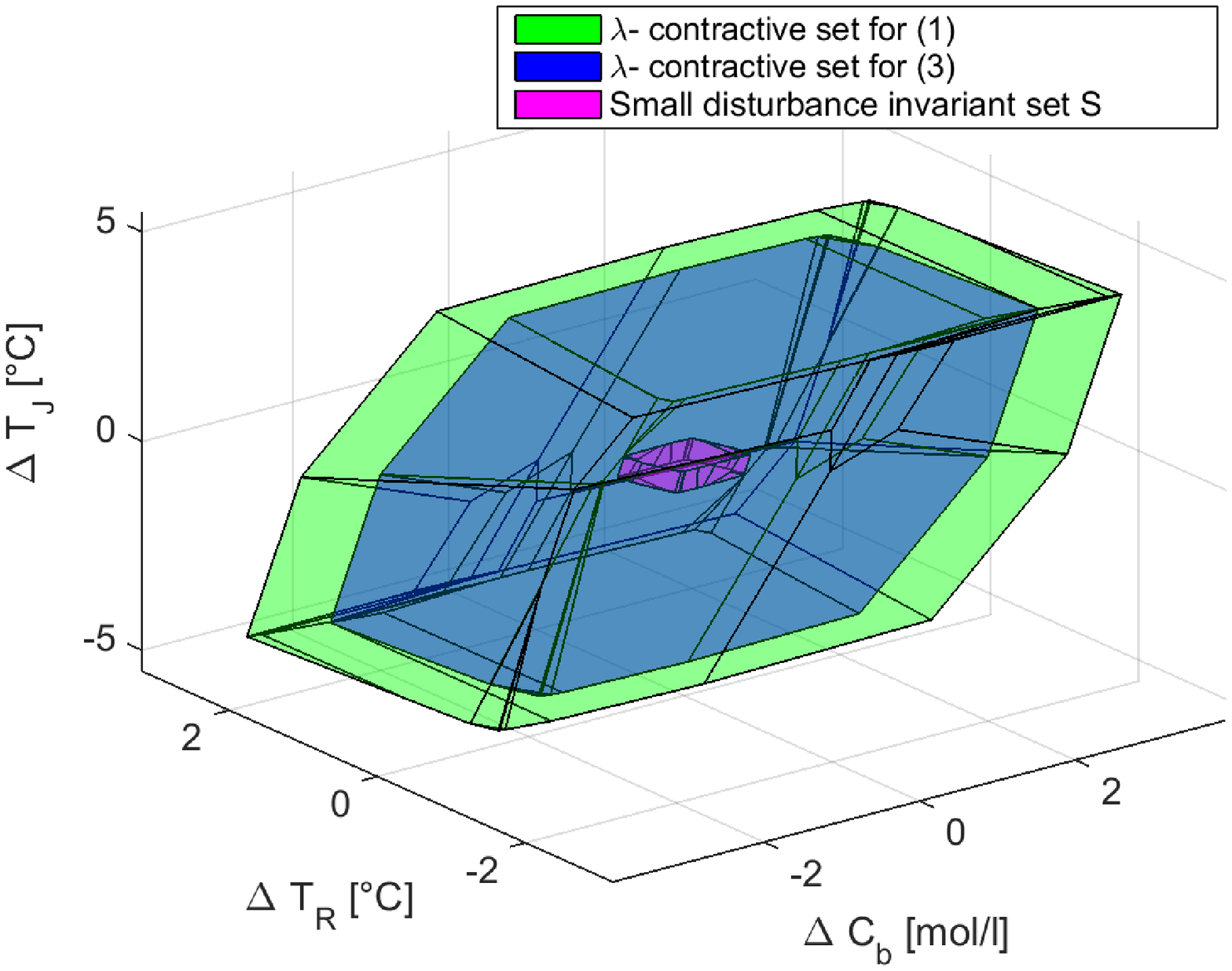} }}%
    \caption{Plot of $\lambda-$ contractive sets and the disturbance invariant set $\mathsf{S}$}%
    \label{fig_lambda_sets}%
\end{figure}

 For comparison purposes, tube-based MPC was implented for the three different types of tubes considered in the proposed framework.  The $\lambda-$contractive set resulted in $32$ inequalities for the system \eqref{lsys} compared to  $18$ inequalities for the system \eqref{lnom}. The contractive sets were computed using the multi-parametric toolbox~\cite{MPT3}. Since the additive disturbances are not considered in \eqref{lnom}, the resulting complexity of the $\lambda-$contractive set is different.  The $3$D projections of the $\lambda-$ contractive sets and the small disturbances invariant set $\mathsf{S}$ are shown in Figure \ref{fig_lambda_sets}.
\begin{figure}%
    \centering
    \subfloat[ Projection on $\Delta C_a$, $\Delta C_b$ and $\Delta T_{\mathrm{R}}$.]{{\includegraphics[width=7cm]{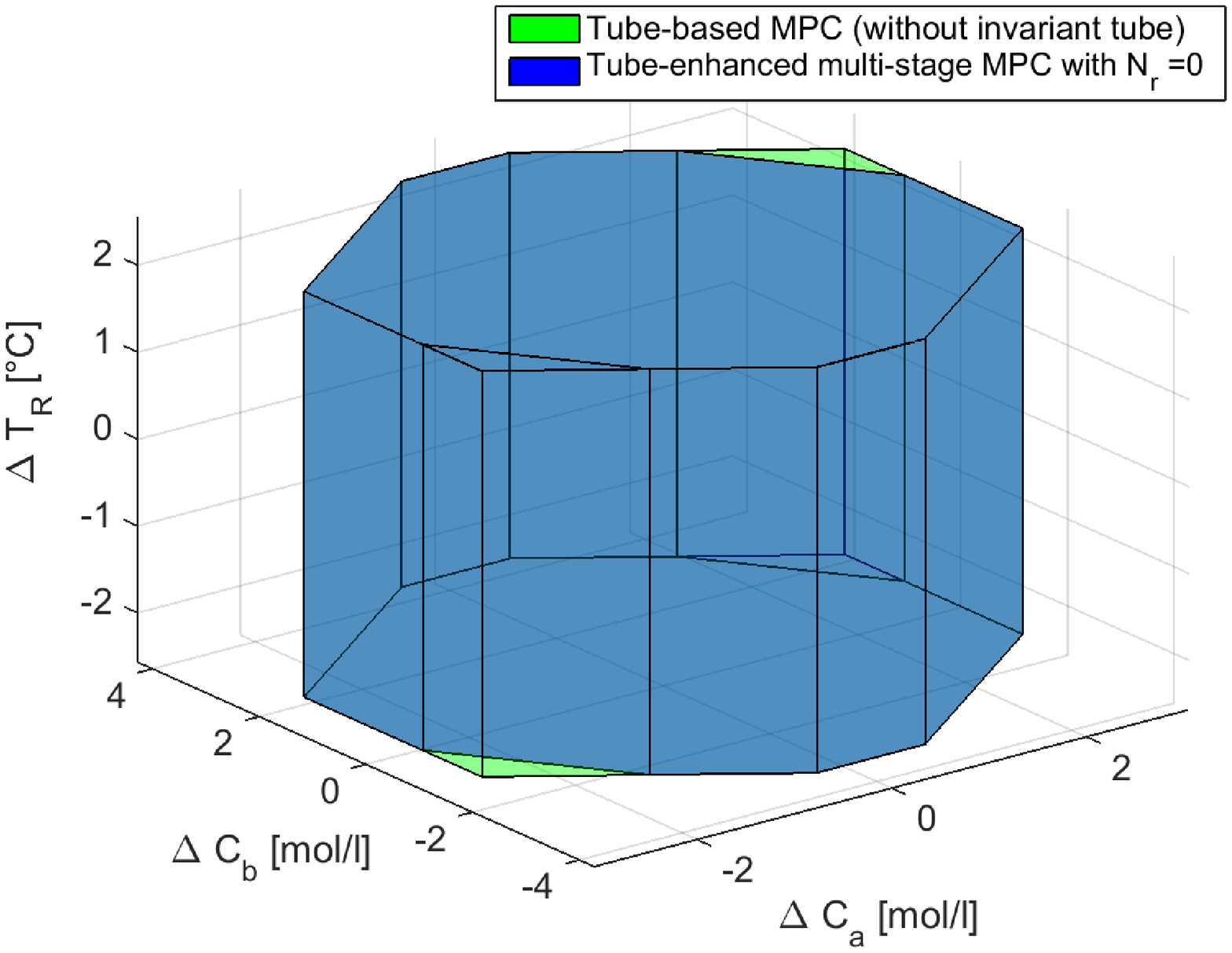}}}%
    \subfloat[ Projection on $\Delta C_b$, $\Delta T_{\mathrm{R}}$ and $\Delta T_{\mathrm{J}}$.]{{\includegraphics[width=7cm]{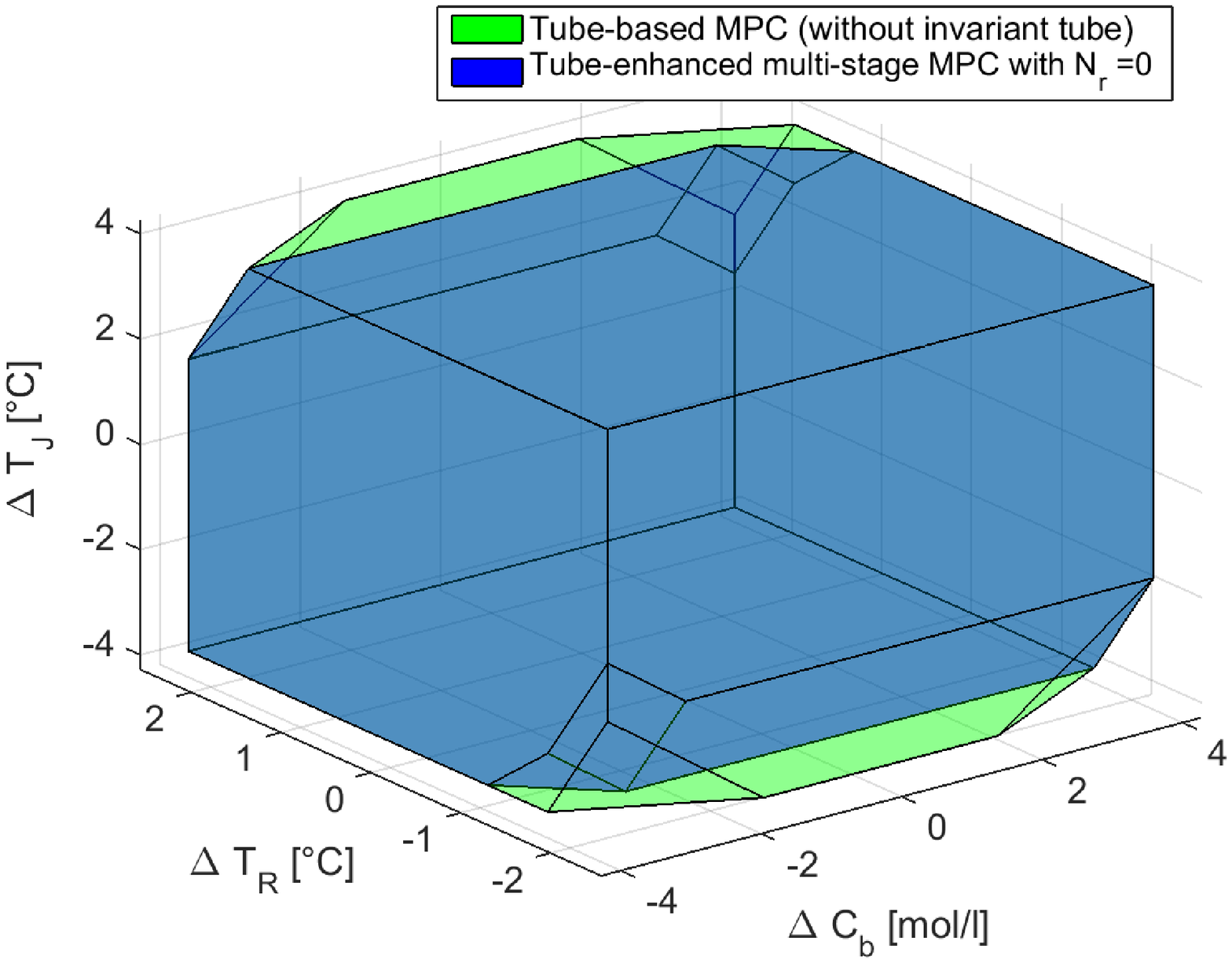} }}%
    \caption{Comparison of the feasible domains obtained by the tube-based NMPC (without the invariant tube $\mathsf{S}$) and the proposed scheme with $N_r=0$ .}%
    \label{fig_compare_tube}%
\end{figure}

 If the tube-based MPC~\cite{Fleming2015} is employed without the classification of uncertainties proposed in this paper, the tube at every time step is characterized by $32$ inequalities in the case of homothetic and general complexity tube tubes and the number of constraints increases with the number of additive and parametric uncertainties. In the studied example, we must consider $4$ vertex matrices and $16$ vertices of the additive disturbance set. Hence, to formulate the propagation of tubes, $32\times4\times16=2048$ constraints are required per prediction step. In contrast, for the proposed scheme with robust horizon $N_r=0$, only  $18\times4=72$ constraints are required to formulate the propagation of the tubes. This is because the small uncertainties are not considered both in the computation of contractive sets and in the online problem. Instead, a suitable back-off is obtained by making use of the disturbance invariant set $\mathsf{S}$. If the low complexity tube is employed, the complexity of the tube-based MPC scheme without the invariant tube is $16$ times larger than the proposed scheme with $N_r=0$ because of the $16$ vertices of the additive disturbance bounds.
 
 The feasible domains of the tube-based MPC scheme (without the invariant tube) and the proposed tube-enhanced multi-stage MPC scheme with $N_r=0$ obtained using the general complexity tube predicted tubes are shown in Figure~\ref{fig_compare_tube}. It can be seen that the feasible region of the tube-based MPC scheme that does not employ the invariant tube is larger than the feasible region of the proposed scheme with $N_r=0$. The volume of the feasible domain of tube-based MPC scheme (without the invariant tube) is 1197.1 whereas the proposed scheme with $N_r=0$ results in a volume of 1110.7 (approx. 7\% smaller). However, the computation time of the proposed approach with $N_r=0$ is only 0.15 seconds compared to the other scheme which has a computation time of 6.55 seconds. If the homothetic tubes are employed, the feasible domain of the proposed scheme is $6\%$ smaller while the computation time of the proposed approach is approximately $96\%$ smaller (for exact values refer to Table~\ref{tab:performance}). In the case of low complexity tube, there is no difference in the volumes of feasible domains observed.
  
 With this example, it can be seen that the invariant set can introduce a certain conservatism in the closed-loop. However, it reduces significantly the computational complexity and improves the computation time of the approach largely (up to 98\% reduction). Thanks to the use an invariant tube for small uncertainties, an important reduction in computation time can be expected at the cost of only minor additional conservativeness.
  
 \subsection{The effect of the robust horizon}
 \begin{figure}%
    \centering
    \subfloat[ Projection on $\Delta C_a$, $\Delta C_b$ and $\Delta T_{\mathrm{R}}$.]{{\includegraphics[width=7cm]{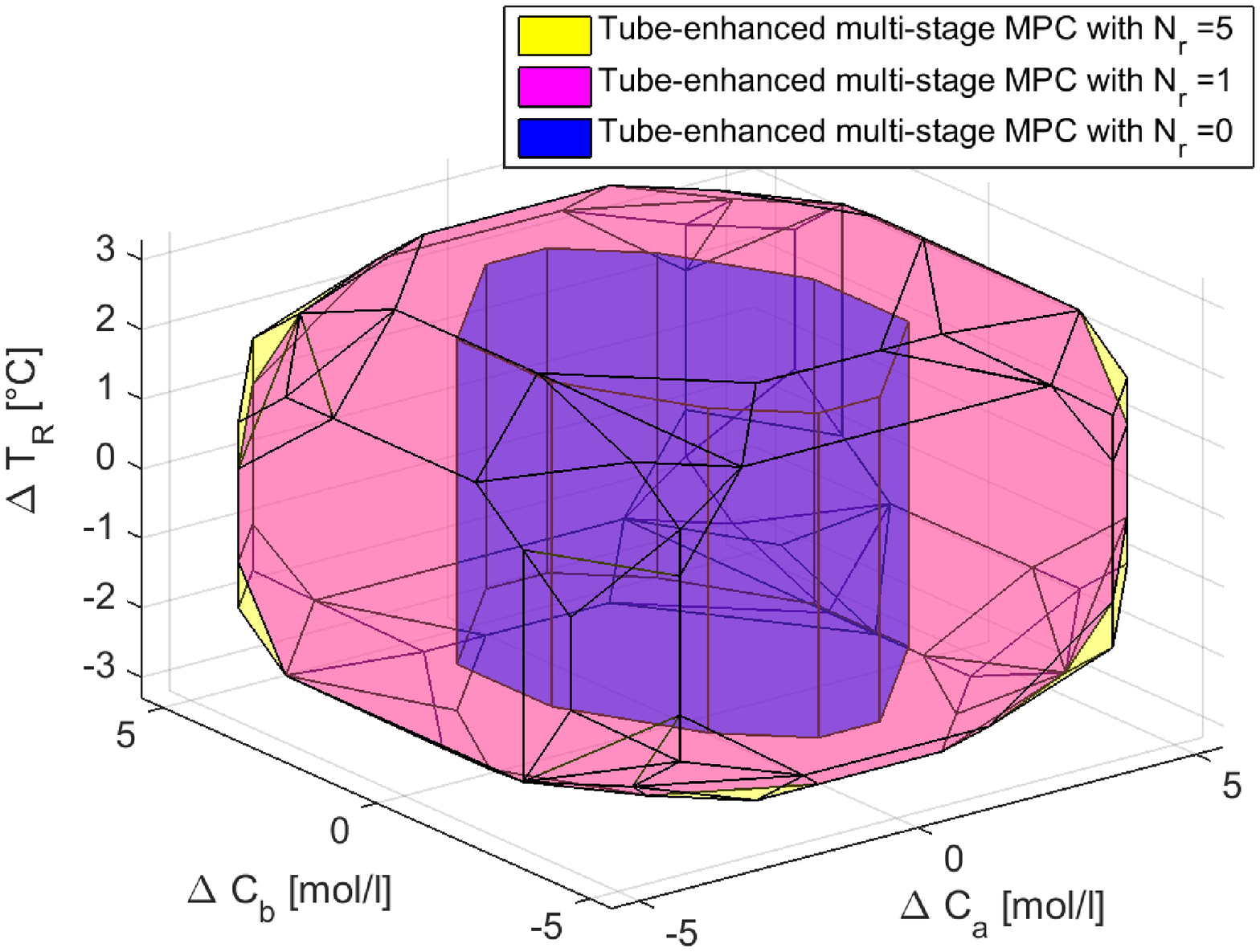}}}%
    \subfloat[ Projection on $\Delta C_b$, $\Delta T_{\mathrm{R}}$ and $\Delta T_{\mathrm{J}}$.]{{\includegraphics[width=7cm]{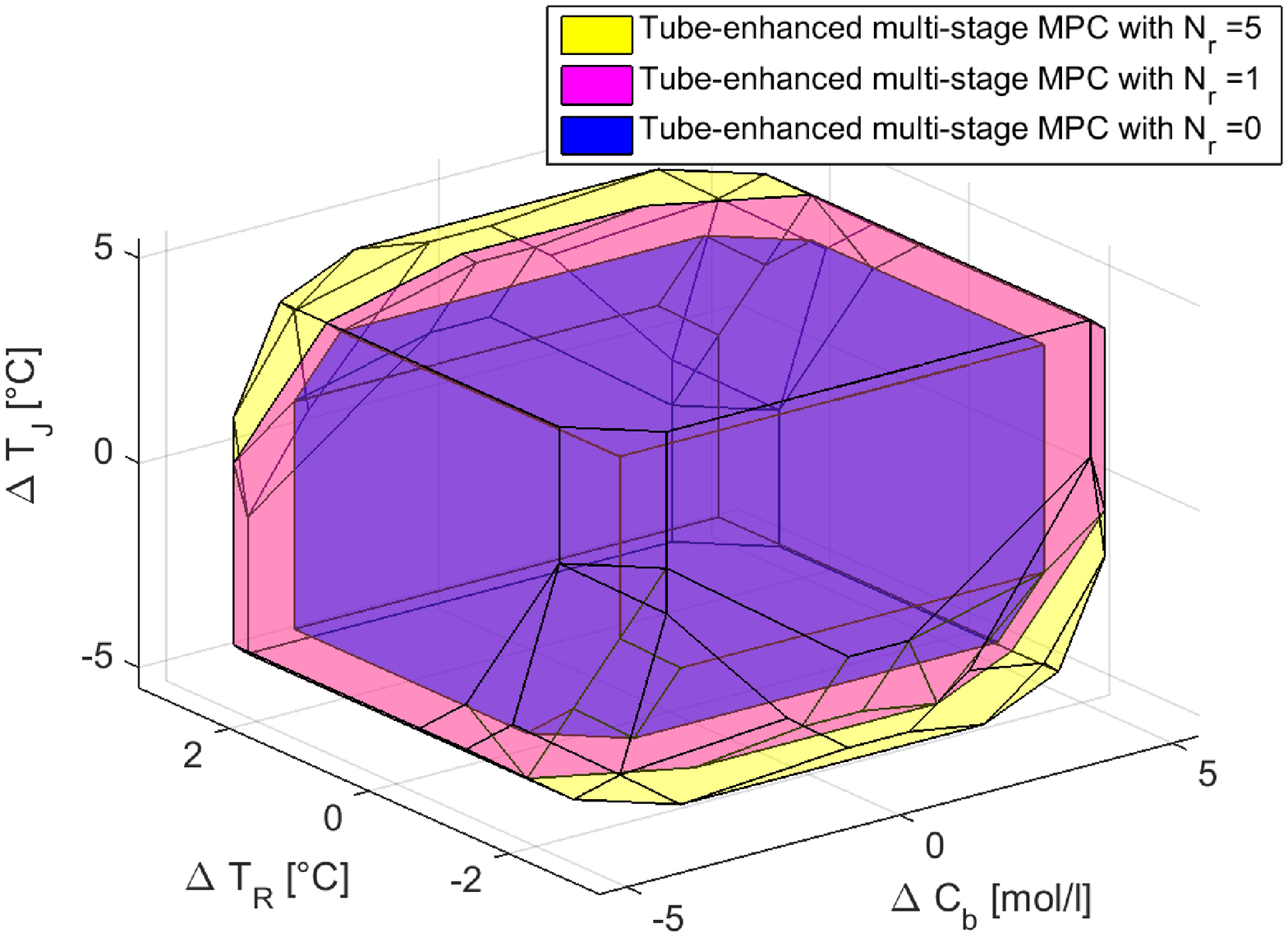} }}%
    \caption{Comparison of the feasible domains obtained using the proposed scheme that uses general complexity tubes for varying robust horizons.}%
    \label{fig_compare_Std}%
\end{figure}
\begin{figure}%
    \centering
    \subfloat[ Projection on $\Delta C_a$, $\Delta C_b$ and $\Delta T_{\mathrm{R}}$.]{{\includegraphics[width=7cm]{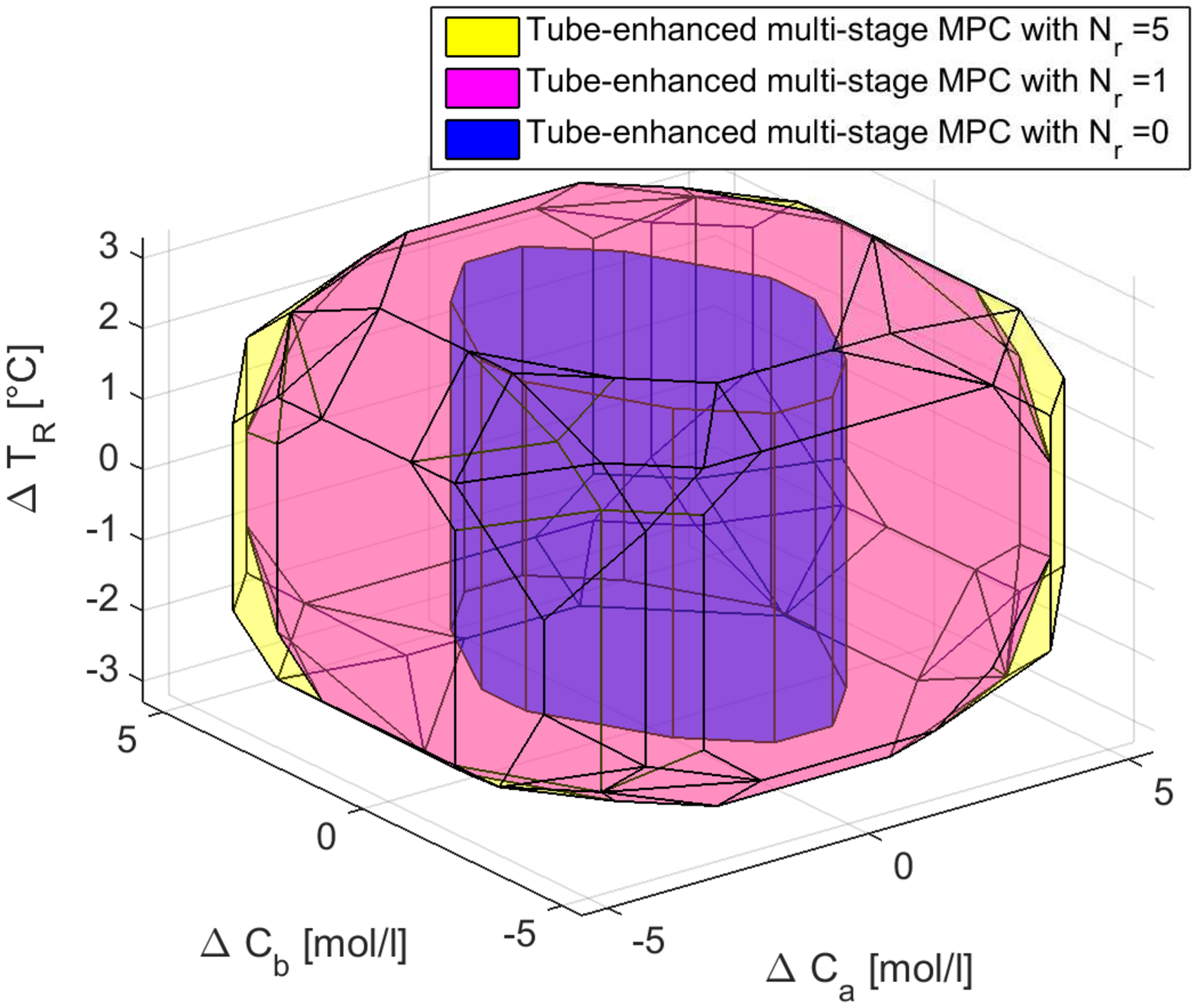}}}%
    \subfloat[ Projection on $\Delta C_b$, $\Delta T_{\mathrm{R}}$ and $\Delta T_{\mathrm{J}}$.]{{\includegraphics[width=7cm]{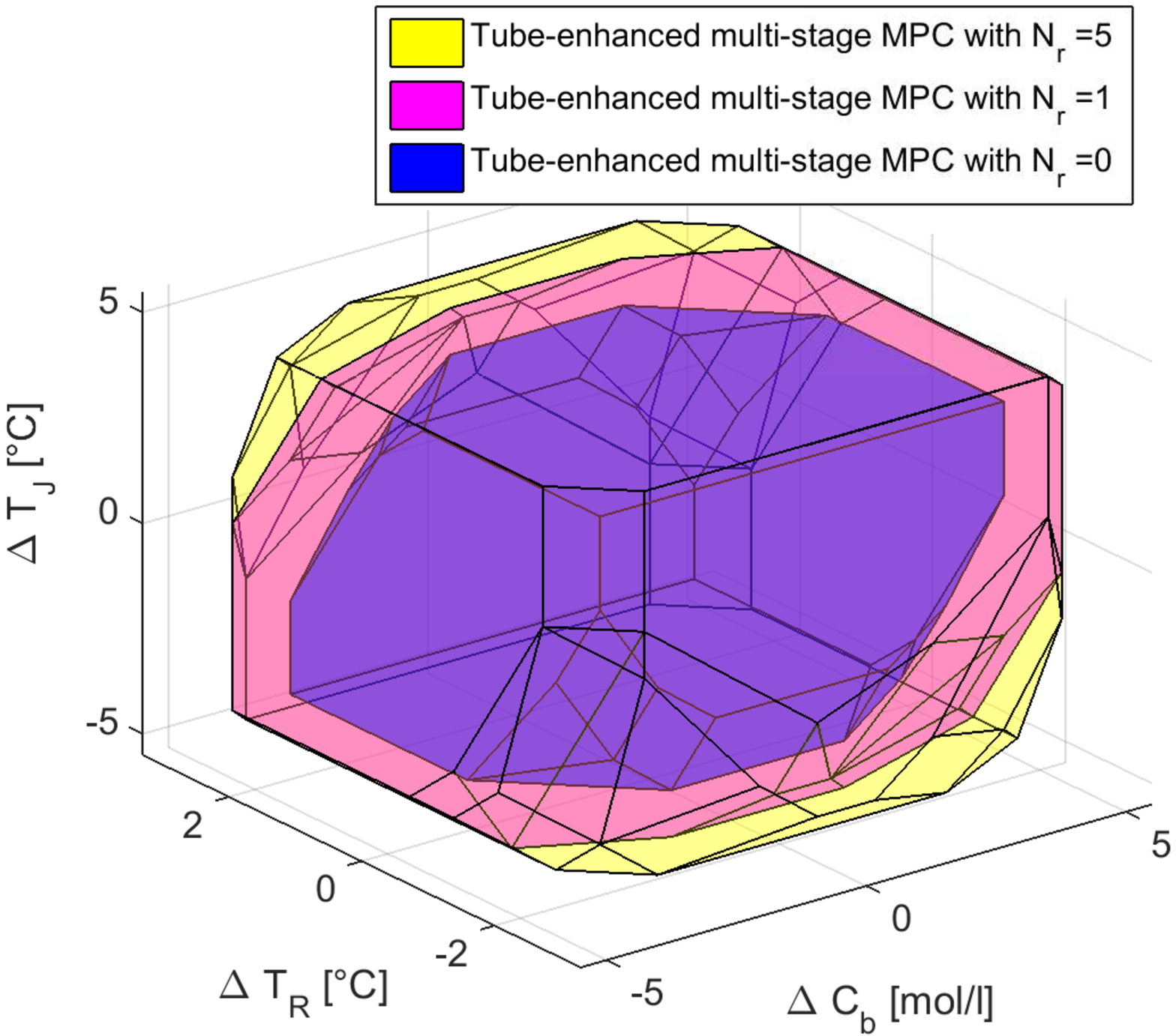} }}%
    \caption{Comparison of the feasible domains obtained using the proposed scheme that uses homethetic tubes for varying robust horizons.}%
    \label{fig_compare_Std2}%
\end{figure}

\begin{figure}
    \centering

    \subfloat[ Projection on $\Delta C_a$, $\Delta C_b$ and $\Delta T_{\mathrm{R}}$.]{{\includegraphics[width=7cm]{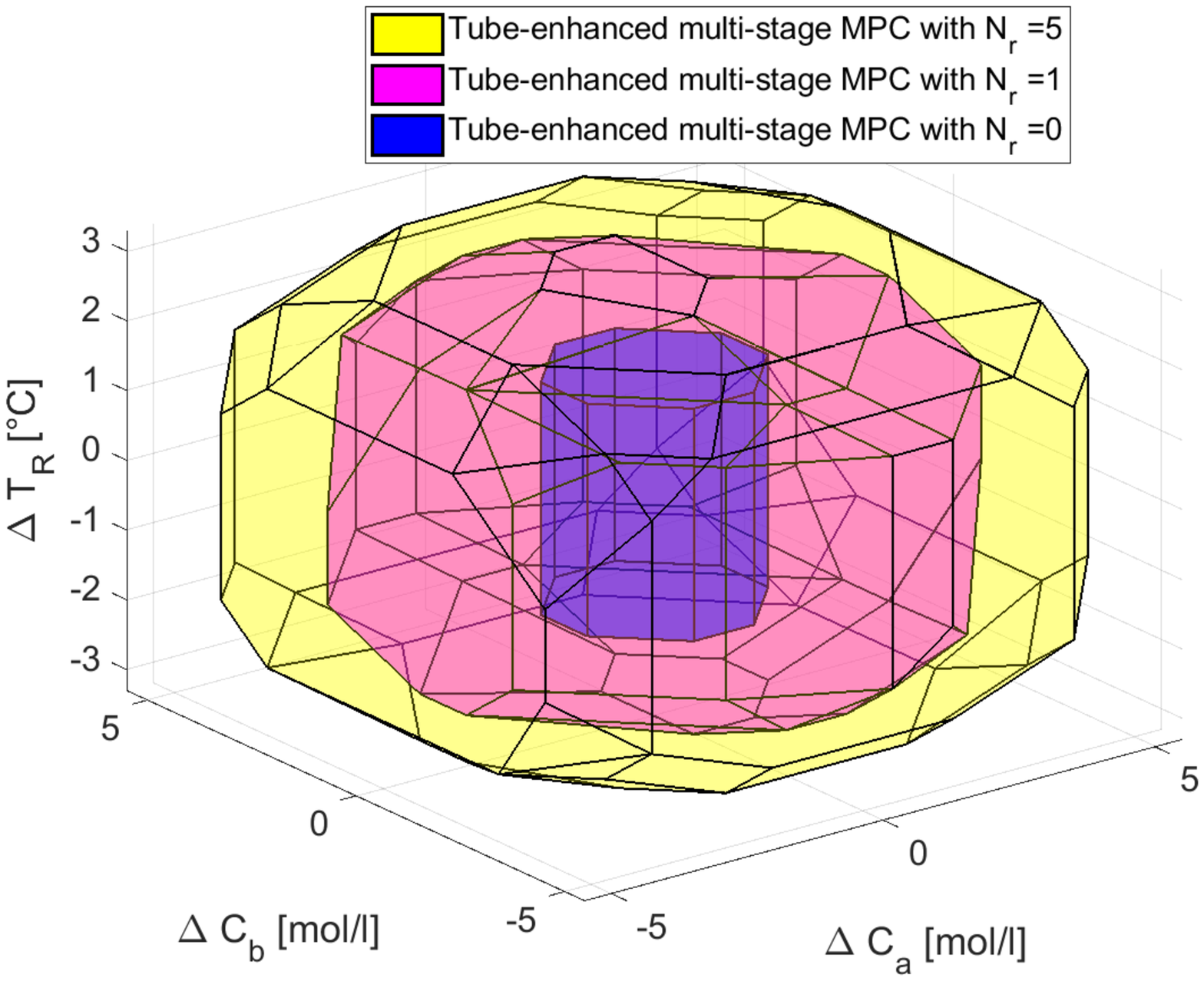}} }
    \subfloat[ Projection on $\Delta C_b$, $\Delta T_{\mathrm{R}}$ and $\Delta T_{\mathrm{J}}$.]{{\includegraphics[width=7cm]{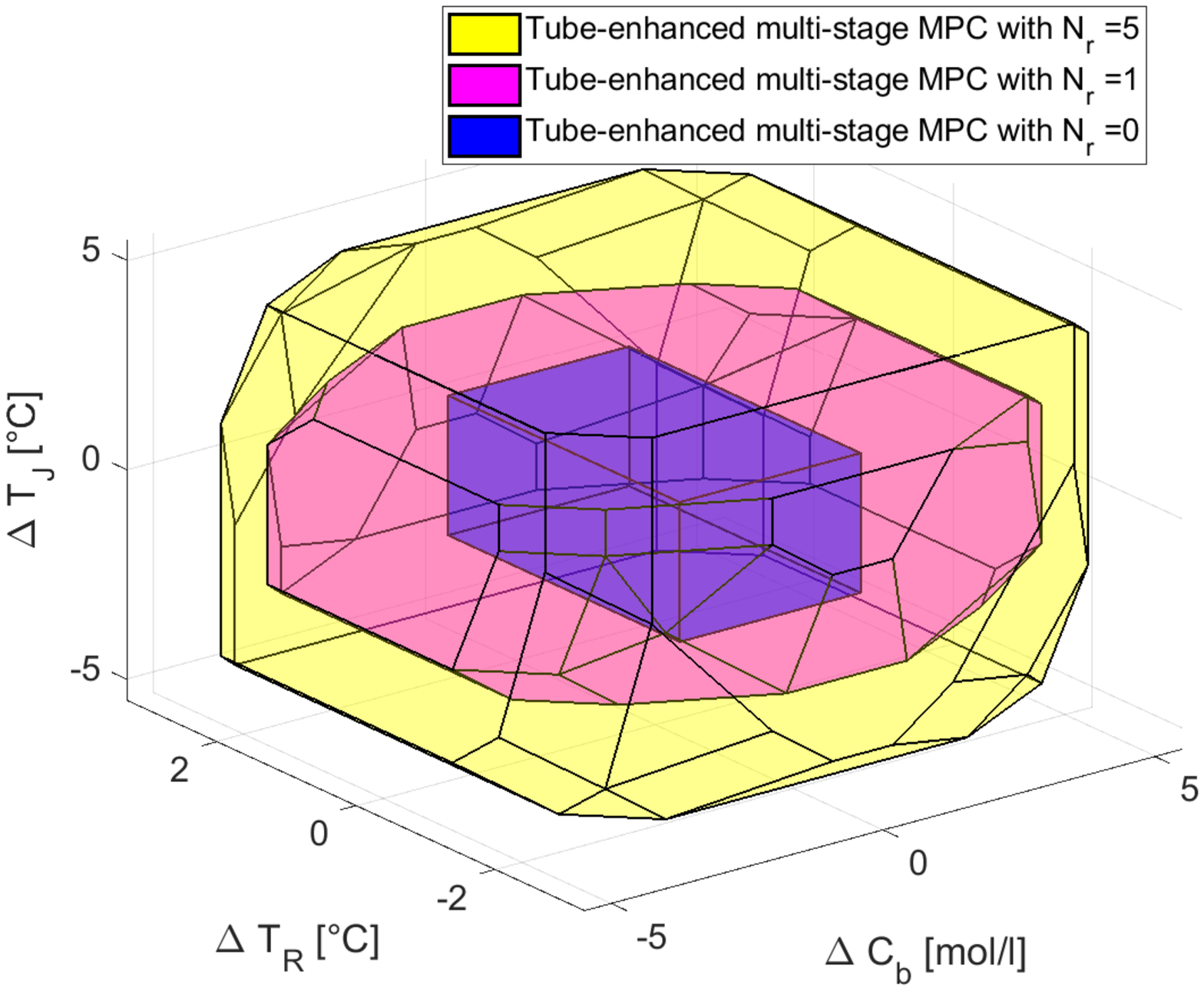} }}%
    \caption{Comparison of the feasible domains obtained using the proposed scheme that uses low complexity homethetic tubes for varying robust horizons.}%
    \label{fig_compare_Std3}%
\end{figure}

If the proposed tube-enhanced multi-stage approach is employed to handle large uncertainties, the number of scenarios considered by the optimization problem increases with the length of  robust horizon. Total number of scenarios in the problem is determined by $N_r$. For $N_r=1$, the problem has only four scenarios and for $N_r=5$, the problem has $4^5=1024$ scenarios. First, we discuss the  effect of the types of the predicted tubes that are employed and then summarize the observations. 

\subsubsection{General complexity tube}
 The scheme with a robust horizon $N_r=4$ has the same feasible domain as the scheme with full robust horizon. When the robust horizon decreases further, the volume of the feasible domain decreases monotonically. The comparison of the feasible domains of the proposed scheme with different robust horizons that uses general complexity tubes are shown in Figure~\ref{fig_compare_Std}. The proposed scheme with $N_r=1$ has a volume that is approximately 12\% smaller than the volume of the full horizon case. The volumes of the feasible domains of the tube-based MPC implemented without the invariant tube and the proposed scheme with $N_r=0$ are significantly smaller (approx. 74\% smaller and 76\% smaller).  In the tube-based MPC scheme, the optimization problem has one feed-forward term per prediction step as degrees of freedom. Whereas, in the cased of $N_r=1$, there are four feed-forward terms optimized at every stage. This improves the degrees of freedom of the controller and results in an improved performance.  The polytopes are plotted with the help of the multi-parametric toolbox~\cite{MPT3}.
 
 The computation times of the scheme with different robust horizons however does not show a uniform trend. The computation times of robust horizons $N_r=0$ and $N_r=1$ are smaller than the full robust horizon. However, the schemes with the robust horizons 2 to 4 have computation times larger than the scheme with full robust horizon. This is because of the difference in complexity associated with the tree and the tube. The scheme with $N_r=2$ has $16$ scenarios. However, from the second prediction step, the propagation of tubes is characterized by $72$ inequalities and this is formulated for all $16$ scenarios requiring $72\times16=1152$  inequalities after the second prediction step. This leads to an increased computational effort when compared to a full tree with $4\times4=16$ equality constraints per node (though exponentially increasing every stage).
 \subsubsection{Homothetic tube}
The trends in the volumes of the feasible domain and the computation times are similar to the general complexity tube case, if the homothetic tubes are employed in the predictions. However, the scheme results in an increased conservatism compared to the general complexity tube case. This is expected because the shape of the predicted sets is restricted. The proposed scheme with $N_r=1$ has a volume of the feasible domain that is smaller by approximately $17\%$ in this case compared to the full robust horizon case. The feasible domains of the proposed scheme implemented with homothetic tubes are given in Figure~\ref{fig_compare_Std2}. There is also an increase in computation times when compared to the general complexity tube case for a fixed horizon. This is because of the increase in the number of constraints due to $44$ vertices of the tube considered to obtain the tight upper bound of the stage cost. This leads to a proportional increase in computational cost. 
\subsubsection{Low complexity tube}
The computation times are smaller in this case as expected because the tube is represented using the minimal number of inequalities. There is however large conservatism as a result. When $N_r=0$ is employed, the volume of the feasible region is only $96.02$ which is more than $10$ times smaller than for the homothetic tube case. Similar reductions in the feasible domains can be observed across all robust horizons. The feasible domains of the proposed scheme implemented with  the Low complexity tube are given  in Figure~\ref{fig_compare_Std3}. An interesting point to note here is that when $N_r=1$ is applied, the volume of the feasible domain is much smaller for the same robust horizon than when general complexity/homothetic tubes are employed. Despite this reduction, the volume is larger than the best feasible domain observed for the tube-based MPC. This clearly demonstrates the advantages of the proposed approach. For robust horizon $N_r=2$, the feasible domain is larger and the computational cost is still lower than tube-based MPC. Hence the scheme can be used to control high dimensional systems, where the pure tube-based scheme is either highly conservative (in the case of using a Low complexity tube) or intractable (in the case of a higher complexity tube). The proposed scheme offers an alternative with an improved trade-off between optimality and complexity.

\bluec{
The computation time and the feasible domain for the case $N_r=N_p=5$ does not depend on the type of tube employed because there is no tube-based part in the predictions of  the optimization problems \eqref{MS_implement} and \eqref{MS_implement2} for this case.}

\subsubsection{Summary}
Below are the summary of the observations:
\begin{enumerate}
\item The volumes of the feasible domains of the proposed scheme increases monotonically with the increase in the robust horizon.  For a given robust horizon, the volume of the feasible domain of the proposed scheme that employs the Low complexity tube is lower and that of the general complexity tube is higher. The volume of the scheme that employs a homothetic tube is in between but it is closer to the general complexity tube than to the Low complexity tube.
\item The proposed scheme with $N_r=1$ has a larger volume of the feasible domain and a smaller computational effort than the tube-based MPC that is employed without the invariant tube. Even the proposed scheme with $N_r=1$ that employs the Low complexity tube has a larger volume than the tube-based MPC that employs a general complexity tube.
\item Though the proposed scheme with Low complexity tube shows conservatism with respect to the scheme with full robust horizon, it gives the best computation times. In addition, the proposed scheme employed with Low complexity tubes with  $N_r\geq 1$ shows better performance than the tube-based MPC scheme with complex/general complexity tubes.
\item The scheme offers flexibility with respect to the choice of tube and robust horizon. For example, the proposed scheme that employs a homothetic tube with $N_r=1$ has a feasible volume and computation times comparable to that of the proposed scheme with Low complexity tube with $N_r=2$. The scheme offers a wide variety of options to achieve a desired trade-off between optimality and complexity than the existing robust MPC schemes.
\end{enumerate}

\begin{figure}\psfrag{Time}{Time Steps}
\begin{center}
\includegraphics[width=0.8\columnwidth]{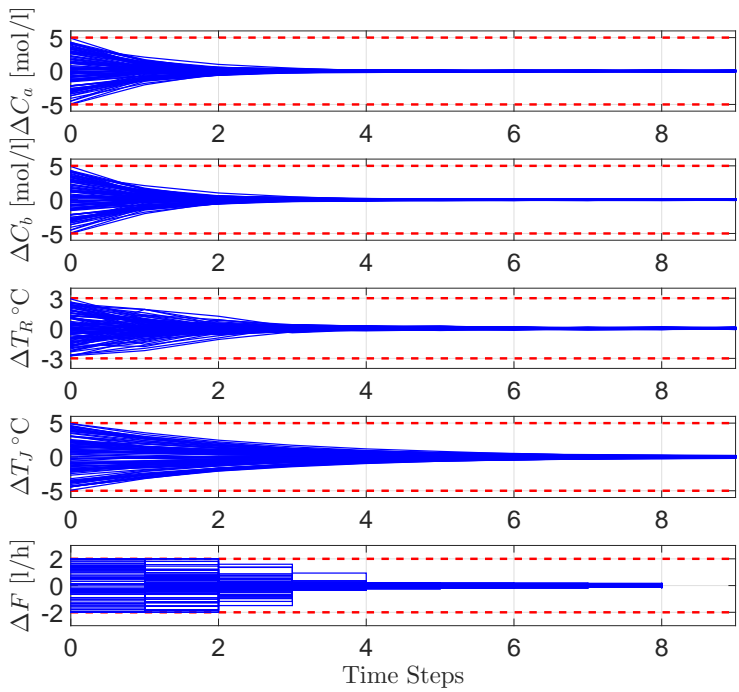}
\caption{Trajectories of $\Delta C_{\mathrm{a}},\,\Delta C_{\mathrm{b}},\,\Delta T_{\mathrm{R}},\, \Delta F$ obtained using 100 simulation runs.}  % width is 8.4 cm.
\label{fig_plot_100}                                 % Size the figures
\end{center}                                 % accordingly.
\end{figure}

For this example, $N_r=1$ using a homothetic/general complexity tube, $N_r=2$ using Low complexity tube and $N_r=5$ are possible choices to obtain good trade-offs overall in optimality and computational complexity. The scheme with $N_r=5$ gives the best performance for a reasonable computational time. Though the volume of the feasible domain of $N_r=1$ using general complexity tube is approximately 12\% smaller, the computation time is reduced by 62\%. Hence $N_r=1$ using general complexity/homothetic tube is a good choice for this example. The closed-loop state and input trajectories of the proposed scheme employed using a general complexity tube with $N_r=1$ for random initial conditions and random realizations of the uncertainties for $100$ simulation runs is shown in Figure~\ref{fig_plot_100}.
\subsubsection{The effect of the prediction horizon}
\begin{figure}%
    \centering
    \subfloat[ Y-axis in Linear scale ]{{\includegraphics[width=7cm]{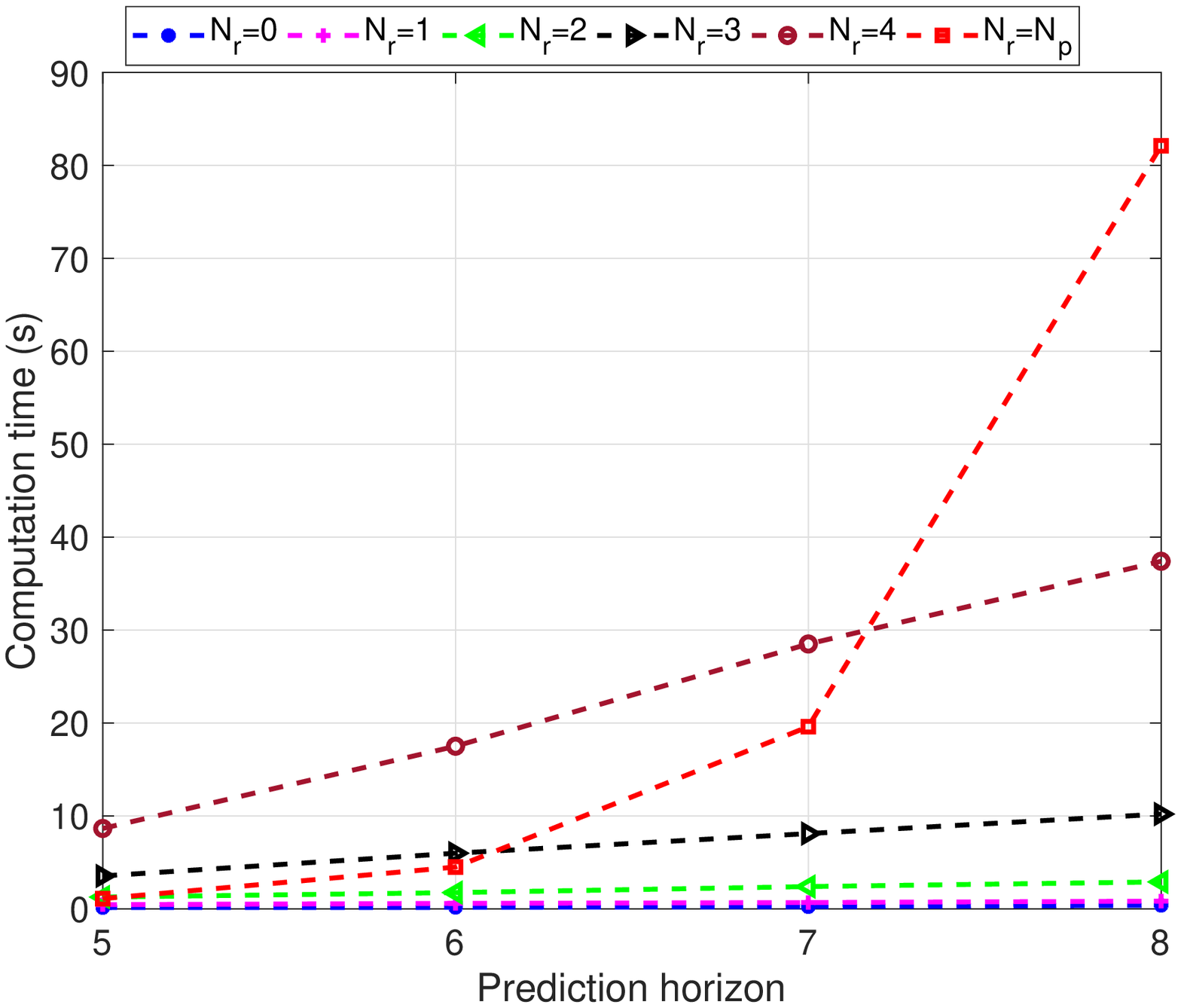}}}%
    \subfloat[ Y-axis in  Logarithmic  scale]{{\includegraphics[width=7cm]{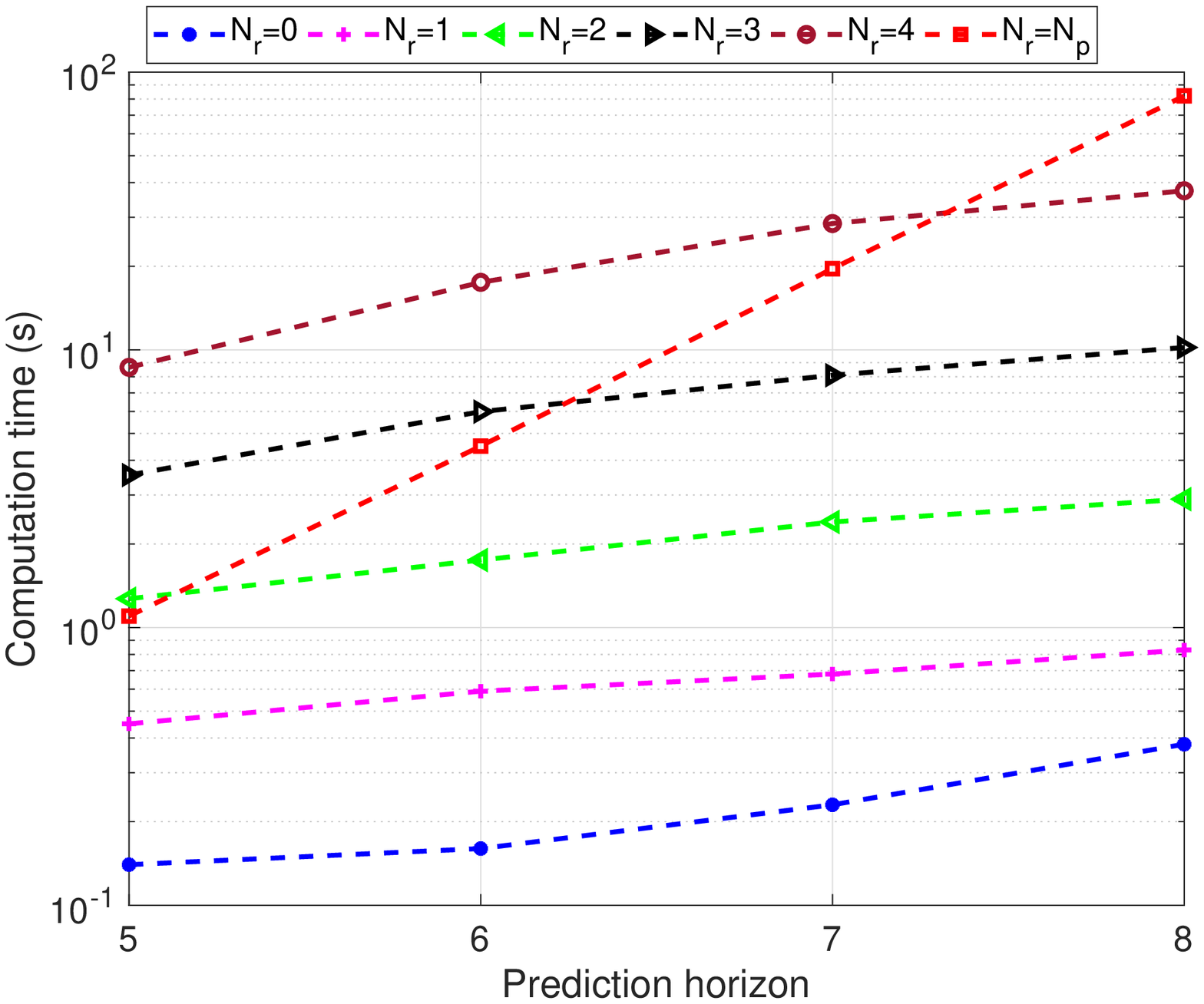} }}%
    \caption{Trend of computation times of the proposed scheme for varying different robust horizons}%
    \label{fig_computation_time}%
\end{figure}

The growth of the problem complexity with respect to the prediction horizon is analyzed by comparing the average computation times of the schemes with different robust horizons. The results are plotted in Figure~\ref{fig_computation_time}. It can be seen that the computational cost increases exponentially if a full robust horizon is used and is significantly larger than for the other robust horizons considered for $N_p=8$. The computation times of the schemes with $N_r<N_p$ grows linearly in complexity with respect to the prediction horizon. However, the slope  is seen increasing when the robust horizon increases. The proposed scheme with robust horizon $N_r=1$ has a computation time of less than one second  and the scheme with $N_r=2$ has a computation time of less than $3$ seconds. The variations in the times are much smaller than in the full robust horizon case. As the horizon grows larger, the computational advantages of the proposed scheme increase. 

\section{Conclusion}
In this paper, we have shown that the combination of multi-stage and tube-based model predictive control schemes offers a flexible framework to manage the trade-off between the performance and computational complexity of the robust scheme. The proposed method uses a tube-based method to handle uncertainties that are small or occur far in the prediction (after the robust horizon), while the multi-stage approach handles the significant and immediate uncertainties to increase the performance.  The stability of the scheme for linear systems with parametric and additive disturbances was demonstrated for any choice of robust horizon, including the pure multi-stage case. Simulation results show that the proposed method provides flexibility to obtain a good trade-off between complexity and performance. 
\bibliography{references_all}

\appendix

\end{document}